\documentclass[onecolumn,authoryear]{els-mrw} 

\usepackage{amsmath,amssymb,amsfonts,amsthm,makeidx,graphicx}
\usepackage{txfonts}
\usepackage{helvet}

\usepackage[utf8]{inputenc}
\usepackage[normalem]{ulem}
\usepackage{color}
\usepackage{xspace}

\newcommand{\ud}{\mathrm{d}}

\newcommand{\pT}{p_{\rm T}}
\newcommand{\mub}{\ensuremath{\mu_{B}}\xspace}
\newcommand{\dndy}{\ensuremath{\ud N/\ud y}\xspace}

\newcommand{\jpsi}{\ensuremath{\mathrm{J}/\psi}\xspace}
\newcommand{\psip}{\psi(2{\rm S})}

\newcommand{\dd}{\mathrm{d}}

\newcommand{\meanNp}{\ensuremath{\langle N_{part}\rangle}}
\newcommand{\sqrtsNN}{\sqrt{s_{\mathrm{NN}}}}
\newcommand{\qqbar}{\ensuremath{\mathrm{q\bar{q}}}}
\newcommand{\ccbar}{\ensuremath{\mathrm{c\bar{c}}}}

\renewcommand\sout{\bgroup \color{blue} \ULdepth=-.5ex \ULset}

\graphicspath{{figs/}}

\begin{document}

\chapter{Experimental exploration of the QCD phase diagram}\label{tit}

\author[1]{Anton Andronic}%
\author[2,3,4]{Peter Braun-Munzinger}%
\author[5]{Krzysztof Redlich}
\author[3,4]{Johanna  Stachel}

\address[1]{\orgname{Universit\"at M\"unster}, \orgdiv{Institut f\"ur Kernphysik}, \orgaddress{Wilhelm-Klemm-Str. 9, 48149 M\"unster, Germany}}
\address[2]{\orgname{GSI Helmholtzzentrum f\"ur Schwerionenforschung},  \orgdiv{Research Division and EMMI}, \orgaddress{Planckstr. 1, 64291 Darmstadt, Germany}}
\address[3]{\orgname{Universit\"at Heidelberg}, \orgdiv{Physikalisches Institut}, \orgaddress{69120 Heidelberg, Germany}}
\address[4]{\orgname{Central China Normal University}, \orgdiv{Institute of Particle Physics and Key Laboratory of Quark and Lepton Physics (MOE)}, \orgaddress{Wuhan 430079, China}}
\address[5]{\orgname{University of Wroc\l aw}, \orgdiv{Institute of Theoretical Physics}, \orgaddress{50-204 Wroc\l aw, Poland}}


\maketitle

\begin{glossary}[Glossary]
\term{Statistical Hadronization Model} describes hadron production in a canonical or grand canonical ensemble (also called hadron resonance gas model) \\
\term{Chemical freeze-out} The stage in a nucleus-nucleus collision at which the abundances of hadron species are fixed (frozen in) \\
\term{Centrality} A method to sort nucleus-nucleus collisions based on the geometric overlap of the colliding nuclei \\
\term{Deconfinement} A state of quarks and gluons no longer confined within hadrons (of size about 1 fm = 10$^{-15}$ m)
\end{glossary}

\begin{glossary}[Nomenclature]
\begin{tabular}{@{}lp{34pc}@{}}
AGS & The Alternating Gradient Synchrotron accelerator at Brookhaven National Laboratory, USA\\
LHC & The CERN Large Hadron Collider \\
QCD & Quantum Chromodynamics, the quantum field theory of the strong interaction\\
QGP & Quark-Gluon Plasma, the deconfined state of matter with quarks and gluons as constituents, assumed to have existed in the early universe. \\
LQCD & Lattice QCD, a method to solve the equations of QCD by discretizing space and time. \\
RHIC & Relativistic Heavy-Ion Collider at Brookhaven National Laboratory, USA\\
SHM & Statistical Hadronization Model\\
SHMc & Statistical Hadronization Model for Charm Hadrons\\
SIS & Schwerionensynchrotron at GSI Darmstadt, Germany\\
SPS & The Super-Proton Synchrotron at CERN\\
$T_c$ & Pseudo-critical (crossover) temperature \\
$T_{CF}$ & Temperature at chemical freeze-out\\
$\mu_{B}$ & Baryo-chemical potential\\
\end{tabular}
\end{glossary}

\section*{Key points / Objectives:}
\begin{itemize}
\item We address the experimental exploration of hot and dense QCD matter as produced in high-energy heavy-ion collisions, with focus on hadron yields and their relevance for the delineation of the QCD phase diagram, discussing briefly  related aspects and approaches.
\item The global characteristics of the experimental data over a broad range of collision energies is discussed.
\item The model used to interpret the measurements, the SHM, is outlined; SHM allows the characterization of the chemical freeze-out stage in terms of the thermodynamic quantities, energy and entropy densities and pressure, and concerning the meson, baryon, strangeness and charm content of the hadron gas.
\item The main objective is to provide evidence for the presence of a phase transition between hadronic matter and the QGP.
\end{itemize}

\begin{abstract}[Abstract]
The different phases of strongly interacting matter are governed by Quantum Chromodynamics (QCD). At high temperature and/or density a deconfined, chirally symmetric phase of quarks and gluons is expected to govern the nature of strongly interacting matter, the so-called quark-gluon plasma. Here we review what is known about the existence of this exotic form of matter from an experimental point of view and confront the results with QCD predictions based on 'Lattice QCD', where QCD is discretized on a space-time lattice and solved numerically in a Monte Carlo approach. The form of the presentation is explicitly pedagogical but the results include even very recent developments. Our research is based on the interpretation of hadron production data from relativistic nucleus--nucleus
collisions over a wide energy range, with the aim to make progress in the understanding of the QCD phase, and of phenomena like deconfinement and hadronization.
\end{abstract}

\section{Introduction} \label{sect:intro} 

Quarks and gluons are the fundamental building blocks of the matter around us. But Quantum Chromodynamics, the theory of the strong interaction, implies that they can never be observed as free particles, a phenomenon called confinement. However, if one heats strongly interacting matter to temperatures as existed in the first 10 microseconds of the Big Bang, one expects (\cite{Itoh:1970uw,Collins:1974ky,Cabibbo:1975ig,Chapline:1976gy}) that confinement breaks down and quarks and gluons  can move freely inside Big Bang matter. Recent experiments at large  accelerators  at the US Brookhaven Laboratory and in Europe at CERN have provided strong evidence that such Big Bang matter can be produced in the laboratory by colliding large nuclei at ultra-relativistic energies. The matter formed in such collisions is akin to Big Bang matter but consists of little drops the size of a few times the size of a large atomic nucleus filled with tens of thousands of quarks and gluons, a state called Quark-Guon Plasma (QGP).

Quarks come with different masses. The lightest quarks called $u$ and $d$ quarks each carry baryon number of $\frac{1}{3}$ and are nearly massless with masses of a few MeV. The matter around us consists of baryons with 3 quarks: the proton has a $(uud)$ composition, the neutron in turn is $(udd)$. The third in the league of light quarks is the strange quark s. It carries a mass of approximately 100 MeV and the lightest strange baryon is the $\Lambda$ baryon with  $(uds)$ composition. In addition to baryons, the hadronic world consists of mesons with a quark-antiquark structure and hence vanishing baryon number. The lightest mesons are the pions and kaons, the positive pion is of $u \bar d$ type, the lightest kaon of $u \bar s$
type. While quarks have fractional charges, baryons and mesons have integer charges. The world of strongly interaction particles is the made of baryons and mesons, jointly called hadrons, the baryons with half-integer spin, the mesons with integer spin. For details on the quark model of hadrons, see ~\cite{ParticleDataGroup:2024cfk}. 

Up to the present, there exist 6 different quarks. In addition to the light quarks $(u,d,s)$ there are the heavy quarks charm $c$ and beauty $b$ with masses of approximately and 1.3 and 4.2 GeV, respectively, and the corresponding charm and beauty hadrons. The heaviest quark is the top quark with approximate mass of 173 GeV, this quark decays too rapidly to form any hadrons. Altogether there are currently more than 600 different hadrons identified to-date, and more states are expected. For details see ~\cite{ParticleDataGroup:2024cfk}.

Massless quarks have their spin either aligned (right-handed chirality) or counter-aligned (left-handed chirality) to their momentum and chirality is then a conserved quantity. Since the light quark masses nearly vanish, the Langrangian of Quantum Chromodynamics (QCD) exhibits approximate chiral symmetry. This fundamental symmetry is spontaneously broken in hadronic and nuclear matter. For a rather elementary discussion see ~\cite{Koch:1997ei}.

At conditions very similar to where deconfinement takes place, also the chiral symmetry is restored. Whether the two transitions take place at the same temperature and baryo-chemical potential (a measure of the net baryon density, i.e. the difference between densities of baryons and antibaryons) is still under investigations, see~\cite{Borsanyi:2025lim}. In any case, two decades of detailed lattice QCD (LQCD) studies, see~\cite{Karsch:2022opd,Borsanyi:2025ttb} have led to the firm prediction that, beyond the phase transition line, the matter is in a deconfined, chirally symmetric state, the Quark-Gluon Plasma (QGP). The name QGP was coined nearly 50 years ago by~\cite{Shuryak:1978ij}. With these theoretical studies the critical energy density above which a QGP can exist was  determined as $\epsilon = 0.4 $\,GeV/fm$^3$. 

Throughout this article,  temperatures will be expressed in units of the Boltzmann constant (i.e. $k_B = 1$). We also use, as is common in nuclear and particle physics, natural units, i.e. $\hbar = c = 1$.

This new form of matter is expected to have existed in the early universe between the electroweak phase transition at picoseconds after the Big Bang and about 10 microseconds later (\cite{Boyanovsky:2006bf}). It is  currently studied experimentally in a world-wide effort involving 4 large international collaborations at the LHC accelerator via collisions of nuclei at high energies (\cite{Busza:2018rrf,Braun-Munzinger:2015hba,Braun-Munzinger:2026krf}).

The main idea behind these experiments is that collisions of heavy nuclei such as Au or Pb at high energies liberate the quarks and gluons initially confined to the nucleons and thereby form hot and dense 'fireballs' over large space-time volumes. How to determine the energy density in such fireballs was first demonstrated more then 40 years ago with an ingeniously simple argument by J. D. Bjorken. For a simple derivation and experimental results for energy densities achieved in relativistic nuclear collisions see section ~\ref{sect:global} below.  From these measurements it is clear that, in relativistic nuclear collisions, energy densities are achieved exceeding by more than one order of magnitude the critical energy density discussed above. Hence the necessary condition for QGP formation is fulfilled in such collisions: A hot and dense fireball is formed beyond the QCD phase boundary. Subsequently the fireball expands hydrodynamically and thereby cools (\cite{Braun-Munzinger:2015hba}) until the phase boundary is reached. 

There, hadron formation (hadronization) takes place. This non-perturbative process is approximated assuming full chemical equilibrium at the phase boundary. In the region around the critical temperature the density is rapidly reduced due to the large reduction in degrees of freedom (colored quarks and gluons converting to colorless hadrons). Thereby, the systems falls out of equilibrium and hadron yields are frozen in. This process is in thermodynamic language called 'Chemical Freeze-out' and is addressed phenomenologically within the statistical hadronization model (SHM)  with chemical freeze-out temperature $T_{CF}$, see ~\cite{Braun-Munzinger:2003pwq,Andronic:2005yp,
Andronic:2017pug} and an according chemical potential as discussed in section ~\ref{sect:lightq}. 

At very high energy (LHC) baryons and antibaryons are produced in equal proportions and all chemical potentials vanish. The transition from QGP to hadronic matter is found under these conditions to be of rapid crossover type. The value of the pseudo-critical temperature $T_c$ for the chiral crossover transition at vanishing \mub is currently calculated in LQCD by two collaborations to be 156.5$\pm 1.5$ MeV ~(\cite{Bazavov:2018mes}) and 158.0$\pm 0.6$ MeV ~(\cite{Borsanyi:2020fev}), in complete agreement within the calculation uncertainties. LQCD results also quantify a small decrease of $T_c$ with increasing \mub as long as $\mub \lesssim 450$ MeV ~(\cite{Bonati:2018nut,Bazavov:2018mes,Borsanyi:2020fev}). Within this parameter range the chiral transition is still of crossover type (\cite{Aoki:2006we}). The temperature for the deconfinement transition is more difficult to evaluate in LQCD due to the lack of an order parameter at finite quark masses. A recent extrapolation of the static quark entropy (\cite{Borsanyi:2024xrx}) puts the deconfinement transition line very close to that for the chiral transition. 

Arguments for the possible presence of new phases near the phase boundary have recently been presented in ~\cite{Glozman:2022zpy} and in ~\cite{McLerran:2026dio}. However, we are not aware of evidence from LQCD for such phases.  Recently, there are increasingly firm theoretical expectations about the presence of a  critical end point in the QCD phase diagram at values of $\mu_B$ around 600 MeV. This is currently discussed intensely and addressed experimentally and theoretically, see very recent reviews~\cite{Fischer:2026uni,Braun-Munzinger:2026krf}.
For more general reviews on the QCD phase diagram see~\cite{BraunMunzinger:2008tz,Fukushima:2010bq,Weise:2012yv,Harris:2023tti,Borsanyi:2025ttb,Fukushima:2025ujk}.

In the following we will, in  section~\ref{sect:global}, first briefly summarize the status of global observables in relativistic nuclear  collisions. This section is based in part on ~\cite{Braun-Munzinger:2025mud}. The main part of this review deals with hadron production and what has been learned from that on the QCD phase diagram. We first introduce in section~\ref{sect:model} the Statistical Hadronization Model SHM. In section~\ref{sect:lightq} a more detailed discussion is presented of the SHM  for hadrons composed of light (u,d,s) quarks. Importantly, this contains a comparison of SHM predictions to results from hadron measurements over a wide range of energies. Next, in section~\ref{sect:heavyq}, we will introduce  a novel concept how to deal with hadrons containing heavy quarks (charm, beauty). This section will then focus on heavy quark hadronization. The observation of deconfinement in the charm-quark sector will be discussed, as well as the  production of exotica such as multi-charm hadrons and nuclei containing charm quarks. This section is based, in part, on section 3 of \cite{Braun-Munzinger:2026krf}.  We conclude, in section~\ref{sect:future}, with a brief summary of new experimental projects and  a brief outlook towards the coming ten years.

In addition to this focused, pedagogical review there are several general reviews of the physics of the QGP and of results from relativistic nuclear collisions and other QGP related research areas, see~\cite{Braun-Munzinger:2015hba,Busza:2018rrf,Harris:2023tti,Fukushima:2025ujk}. Recent reviews on specific observables to characterize the dense QCD matter (QGP) are available on: collective flow and vorticity for the determination of  the Equation of State (EoS)(~\cite{Sorensen:2023zkk});
chiral symmetry restoration and the temperature of the fireball via dilepton production~(\cite{Rapp:1999ej,Salabura:2020tou,Bailhache:2025kwa});
heavy-quark (charm and bottom) diffusion coefficients(~\cite{He:2022ywp,Apolinario:2022vzg});
quarkonium dissociation and (re)generation (\cite{Rothkopf:2019ipj,Andronic:2025jbp});
parton energy loss (\cite{Wang:2025lct}); and
event-by-event fluctuations of baryon number~(\cite{Braun-Munzinger:2026krf}).

{\section{Global observables in relativistic nuclear  collisions and QGP properties} \label{sect:global}

Experimentally, the regime of the QCD phase transition is accessible by investigating collisions of heavy nuclei at high energy. Here, the hot fireball produced in the collisions is very nearly baryon-free, containing in addition to thousands of mesons, equal numbers of baryons and anti-baryons, implying that the baryon chemical potential \mub vanishes. For experimental values on the number of baryons and on \mub see ~section  \ref{sect:lightq}. The properties of this fireball are essentially determined by strongly interacting particles, i.e. mesons and baryons. Photons and lepton-pairs are, of course also produced in the collisions, but leave the fireball as electro-weak particles without interactions. It was conjectured already in~\cite{Shuryak:1978ij} that, in such hadronic collisions, after some time local thermal equilibrium is established and the properties of the system (fireball) are determined by a single parameter, the temperature $T$, depending on time and spatial coordinates. This is exactly the regime probed by collisions of nuclei at the Large Hadron Collider (LHC), as will be outlined in the following section ~\ref{sect:lightq}. The nuclear stopping region corresponding to finite and  large \mub is accessed by nuclear collisions at lower energies.

In the early phase of the collision, the incoming nuclei lose a large fraction of their energy, leading to the creation of a hot fireball characterized by an energy density $\epsilon$ and a temperature $T$. The concomitant deceleration of the nucleons in the colliding nuclei, called stopping, is characterized by an average rapidity\footnote{Rapidity is defined as: $y=\frac{1}{2}\ln\frac{E+p_{\mathrm L}}{E-p_{\mathrm L}}=\mathrm{tanh}^{-1}(\beta_{\mathrm L})$, where $p_{\mathrm L}$ and $\beta_{\mathrm L}$ are the longitudinal (in beam direction) momentum and velocity (in units of $c$), respectively; $E=\sqrt{m^2+p^2}$ is the particle's total energy. Midrapidity denotes the rapidity of the center-of-mass, c.m. ($y=0$ in the usual case of $y$ being evaluated in c.m.).} shift $\Delta y$ = -ln($E/E_0$) with nucleon energies $E_0$ and $E$ before and after the collision. Quantitative information is contained in the experimentally measured net-proton rapidity distributions (i.e. the difference between proton and anti-proton rapidity distributions). These distributions are summarized for collision energies from the SPS to RHIC energy range in~\cite{Braun-Munzinger:2020jbk}. Including also the AGS data it emerges that the rapidity shift saturates at approximately two units from $\sqrt{s_{NN}}\approx$ 17.3 GeV upwards, implying a fractional energy loss of $1-{\rm exp}(-\Delta y) \approx 86\%$. In fact, the same rapidity shift was already determined for p--nucleus collisions at Fermi National Accelerator Laboratory for 200 GeV proton momentum~\cite{Abe:1988hq} compared to about one unit for pp collisions. With increasing collision energy, the target and projectile rapidity ranges are more separated, leaving at central rapidity with a small or even zero net-baryon density, and universal fragmentation regions forward and backward following the concept of limiting fragmentation~(\cite{Benecke:1969sh}).

The large energy loss (rapidity shift) of the incident nucleons leads to high energy densities at central rapidity, i.e., in the center of the fireball. These initial energy densities can be estimated using the Bjorken model (\cite{Bjorken:1982qr}):
\begin{equation}
\label{eq-Bj}
    \epsilon_{BJ} = \frac{1}{A_T\tau_{0}}\frac{\ud E_{T}}{\ud \eta}\frac{\ud\eta}{\ud y},
\end{equation}
where $A_T$ is the transverse overlap area of two nuclei and the kinetic equilibration time $\tau_0$. Eq.~\ref{eq-Bj} is typically evaluated at a time $\tau_{0}$ = 1 fm and the resulting energy densities are displayed in Table~\ref{tab:Edensity} for central\footnote{Centrality is estimated based on experimental data and the Glauber model~(\cite{dEnterria:2020dwq}), a geometric nuclear overlap model, and expressed either as a fraction in the total geometric cross section (starting with 0 \% for the most central collisions) or as the corresponding average number of nucleons, \meanNp, participating in the collision.} Au--Au and Pb--Pb collisions at the different collision energies. For central Pb--Pb collisions ($A_T \simeq$ 150 fm$^2$) at $\sqrt{s_{NN}}$ = 2.76 TeV this yields an energy density of about 14 GeV/${\rm fm^3}$~(\cite{CMS:2012krf}), more than a factor of 30 above the critical energy density for the chiral crossover phase transition as determined in LQCD calculations. In fact, for all collision energies shown, the initial energy density significantly exceeds the critical value quoted above, indicating that the matter in the fireball consists of  colored quarks and gluons (QGP) rather than hadrons. The corresponding initial tem\-pe\-ra\-tures can be computed using the energy density of an ideal relativistic gas of quarks and gluons with two quark flavors, $\epsilon = 37\frac{\pi^{2}}{30}T^{4}$, yielding e.g. $T\approx$ 307 MeV for LHC energy. Temperature values for lower collision energies are also shown\footnote{The values reported in the table are all for vanishing chemical potentials. We have evaluated the differences if one assumes values for chemical potentials as determined at chemical freeze-out as shown below. The resulting temperature values differ by less than 5\%  from those reported in Table~\ref{tab:Edensity}. Owing to the proportionality of energy density to the fourth power of temperature, inclusion of a bag pressure only mildly changes the calculated temperature values.} in Table~\ref{tab:Edensity}.

\begin{table}[htb]
\centering
\vspace{-2mm}
\begin{tabular}{| c | c | c | c | c |}
 \hline
 & $\sqrt{s_{NN}}$  & $\ud E_{T}/\ud \eta$ & $\epsilon_{BJ}$ & $T$ \\ [1ex]
 & [GeV] & [GeV] & [GeV/fm$^{3}$] & [GeV] \\ [1ex]
 \hline
AGS & 4.8 & 200 & 1.9 & 0.180 \\  [1ex]
 \hline
SPS & 17.2 & 400 & 3.5 & 0.212 \\ [1ex]
\hline
RHIC & 200 & 600 & 5.5 & 0.239 \\ [1ex]
\hline
LHC & 2760 & 2000 & 14.5 & 0.307 \\ [1ex]
\hline
\end{tabular}
\vspace*{2mm}
\caption{Collision energy per colliding nucleon pair, measured transverse energy pseudo-rapidity density at midrapidity~(\cite{E814E877:1993rlr,WA98:2000mvt,PHENIX:2015tbb,CMS:2012krf,ALICE:2016igk}), energy density, and initial temperature estimated as described in the text for central Pb--Pb and Au--Au collisions at different accelerators.}
\label{tab:Edensity}
\end{table}

Depending on energy, collisions of heavy ions populate different regimes falling into two categories: (i) the stopping or high net baryon density region reached at $\sqrt{s_{NN}} \approx$ 3-20 GeV and (ii) the transparency or net-baryon-free region reached at higher collision energies. The net-baryon-free QGP presumably existed in the early universe after the electro-weak phase transition and up to about 10 microseconds after the Big Bang. This corresponds to the time of the QCD phase transition, where the strongly interacting constituents, i.e. the quarks and gluons, were converted into hadrons. For a brief but concise description, see section 22.3 of ~(\cite{ParticleDataGroup:2024cfk}). In the QGP of the early universe, particles interacting via the strong and electro-weak force are part of the system, while an accelerator-made QGP mostly contains strongly interacting particles. On the other hand, a baryon-rich QGP may be produced in neutron star mergers or could exist, at very low temperatures, in the center of neutron stars~(\cite{Bauswein:2018bma,Baym:2019iky,Gorda:2022jvk}). 

Another important difference between the `laboratory-created' QGP and the QGP phase in the early universe is that, after the QCD phase transition, the `laboratory-created' QGP falls out of equilibrium and never returns to it. This is denoted as chemical freeze-out. In contradistinction, the standard cosmological model phase, which after hadronization contains hadrons, leptons, photons and neutrinos, immediately returns to equilibrium and stays there until neutrino-freeze-out at a time of about 1 s after the Big Bang. For a detailed description of this phase see ~\cite{Rafelski:2023emw}. 

\begin{figure}[hbt]
\begin{tabular}{cc}  \begin{minipage}{.47\textwidth}
{\includegraphics[width=1.\textwidth]{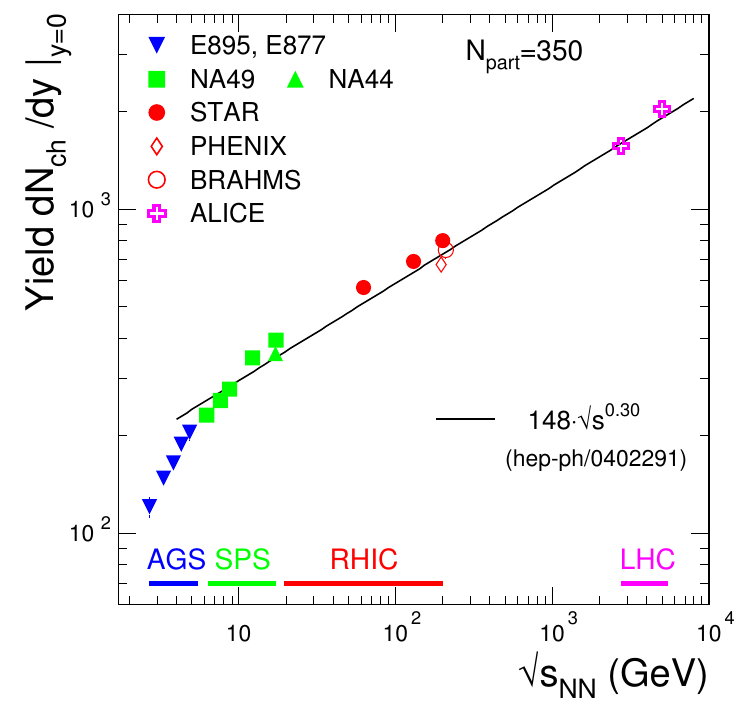}}
\end{minipage} & \begin{minipage}{.5\textwidth}
{\includegraphics[width=1.04\textwidth]{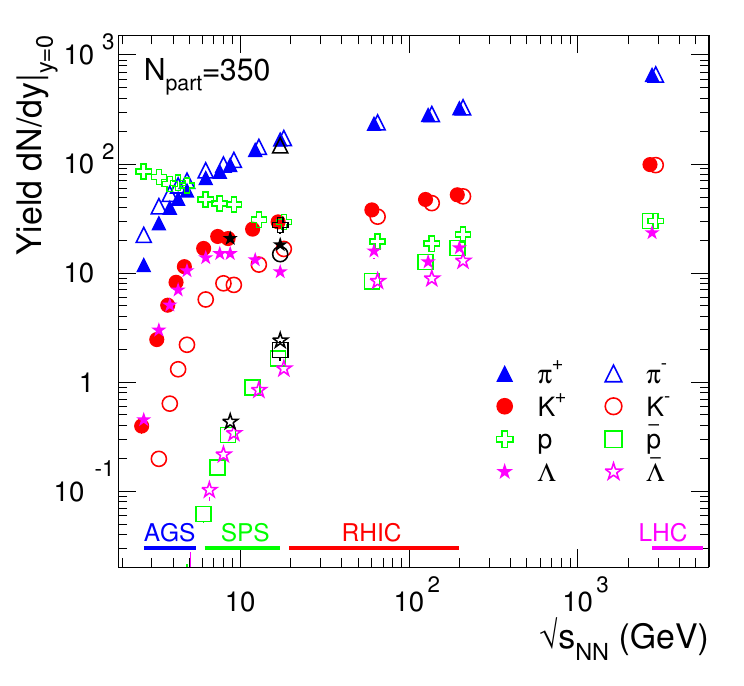}}
\end{minipage} \end{tabular}
\caption{Collision energy dependence of rapidity densities at midrapidity ($y=0$) of charged particles (left) and of identified hadrons (right) for central collisions, corresponding to a (average) number of participating nucleons $N_{part}$=350. The energy ranges of the various accelerators are indicated; the RHIC energy range does in fact extend down to the AGS range. Figures taken (adapted) from \cite{Andronic:2014zha}.
}\label{fig:dndy}
\end{figure}

In Fig.~\ref{fig:dndy} (left) the collision energy dependence of the measured charged-particle rapidity density $\ud N_{\mathrm{ch}}/\ud y$ is shown. The data are for midrapidity (where particles are emitted preferentially in the transverse direction). A strong increase of $\ud N_{\mathrm{ch}}/\ud y$ is observed for lower collision energies of up to $\sqrt{s_{NN}}\simeq$5 GeV, followed by a milder increase for the higher energies, where the trend is well described by the functional $(\sqrt{s_{NN}})^{0.3}$ dependence.

In Fig.~\ref{fig:dndy} (right) the collision energy dependence of selected identified hadron yields at midrapidity is shown. The data shown in the figure and the additional data shown later on are from measurements spanning more than 30 years, by experiments at 
the SIS: FOPI~(\cite{FOPI:2010xrt}), HADES~(\cite{HADES:2017jgz,HADES:2025qum}),
AGS: E895 ~(\cite{Klay:2001tf,Klay:2003zf,Pinkenburg:2001fj}), 
E866/E917 ~(\cite{Ahle:1999uy,Ahle:2000wq}, E891 \cite{Ahmad:1998sg});
the SPS: NA49 ~(\cite{Afanasiev:2002mx,Alt:2007aa,Alt:2005gr,Alt:2008qm}), 
NA44 ~(\cite{Bearden:2002ib}),
NA57 ~(\cite{Antinori:2004ee});
RHIC: STAR ~(\cite{Adler:2002uv,Adams:2006ke,Abelev:2008ab,Abelev:2009bw,Aggarwal:2010ig,STAR:2019bjj,STAR:2019sjh,STAR:2021hyx,STAR:2023uxk,STAR:2024znc}), 
BRAHMS ~(\cite{Arsene:2005mr}), 
PHENIX ~(\cite{Adler:2003cb});
and the LHC: 
ALICE ~(\cite{Abelev:2012wca,Abelev:2013vea,Abelev:2013xaa,ABELEV:2013zaa,Abelev:2014uua,Adam:2015yta,Adam:2015vda,ALICE:2016fbt,ALICE:2024koa,ALICE:2024djx}).
The monotonic decrease of the proton yield as a function of energy indicates that fewer and fewer of the nucleons (or their valence $u,d$ quarks) in the colliding nuclei are ``stopped'' in the fireball at midrapidity. An onset of meson production is seen, with the kaons (most massive and containing a strange quark) produced less abundantly than pions.
The  difference in production yields of $\pi^+$ and $\pi^-$  at low energies reflects the isospin composition of the fireball. 
The difference between $K^+$ and $K^-$ meson and $\Lambda$ and $\bar{\Lambda}$ hyperon yields is determined by the quark content of the hadrons, $K^+$($u\bar{s}$), $K^-$($\bar{u}s$) $\Lambda$($uds$), $\bar{\Lambda}$($\bar{u}\bar{d}\bar{s}$).  
The availability in the fireball of valence $u$, $d$ quarks from colliding nucleons stopped in the fireball leads to a preferential production of hadrons carrying those quarks.
These differences vanish gradually for higher energies, where the quarks are mostly newly created and the hadron production yields exhibit a clear mass ordering.

\section{The statistical hadronization model} \label{sect:model} 

One of the consequences of confinement in QCD is that isolated,  free quarks have never been detected, even in collisions at the highest energies: physical observables require a representation in terms of hadronic states. Indeed, as has been noted in the context of QCD thermodynamics (see, e.g., ~\cite{Bazavov:2017dus} and refs. therein), as long as the temperature stays below $T_c$,  the corresponding partition function $Z^{GC}$ can be very well approximated within the framework of the hadron resonance gas (HRG) {which is a mixture of ideal gases of all stable hadrons and resonances}.

{In the HRG} comprised of $N$ particle species, all thermodynamic quantities can be obtained from the grand-canonical partition function that can be expressed as the product of N individual partition functions $Z_i$  for particle i. {Thus, for a multi-component hadron gas in} volume $V$ at temperature $T$ the grand canonical partition function {is given by:} 
\begin{equation}
\ln Z^{GC}(T,V,\mu_B,\mu_{Q},\mu_S,\mu_C) =\sum_{i=1}^N{{Vg_i}\over {2\pi^2}}\int_0^\infty  p^2\ud p \ln [1\pm  \exp (-(E_i-\mu_i)/T)]^{\pm 1} 
\label{eq:partition_fun}
\end{equation}
with $+$ for fermions and $-$ for bosons, where $g_i=(2J_i+1)$ is the spin degeneracy factor, while  $E_i =\sqrt {p^2+m_i^2}$ is the energy of particle $i$, and the chemical potential is
$\mu_i = \mu_B B_i+\mu_{I_3} I_{3,i}+\mu_S S_i+\mu_C C_i$. The chemical potentials, $\mu_B, \mu_{I_3}, \mu_S, \mu_C$, ensure conservation (on average) of baryon, isospin, strangeness, and charm quantum numbers, and the $X_i$ denote the corresponding quantum numbers carried by particle $i$.
{Employed but not shown explicitly in Eq.~\ref{eq:partition_fun} is an additional integration over the Breit-Wigner mass distribution of resonances. 
\par\noindent
{From the partition function (\ref{eq:partition_fun}), the particle multiplicity of species $i$, the entropy density, pressure and energy density are obtained by differentiation:}
\begin{equation}
\langle N_i \rangle^t =T{{\partial \ln Z^{GC}}\over {\partial\mu_i}},~ ~~s={1\over V}{{\partial (T \ln Z^{GC})}\over {\partial T}}, ~~~P=T {{\partial ( \ln Z^{GC})}\over {\partial V}},~~~
\epsilon={{T^2}\over V} {{\partial ( \ln Z^{GC})}\over {\partial T}}+ {1\over V}\sum_{i=1}^N\mu_i \langle N_i\rangle^{t}.
\label{eq:definition}
\end{equation}
The above equation of state represents a non-interacting gas of a particle mixture. However, as has been pointed out in  ~\cite{Beth:1937zz}, the inclusion of attractive interactions is approximately obtained when the sum implied in ~(\ref{eq:partition_fun}) contains not only stable particles but also the very many resonances in the complete hadron spectrum. A fully consistent approach to interactions among hadrons is an implementation of the S-matrix formulation of statistical mechanics
where the thermodynamic potential is linked with the scattering phase-shifts, see e.g.  (\cite{Andronic:2018qqt}, \cite{Blaschke:2025qvv}). In that way, one includes not only attraction but also repulsion and non-resonant components in hadron-hadron interactions, see ~\cite{Cleymans:2020fsc}. For an interesting new contribution, see  also ~\cite{Yasui:2026vve}. However, the absence of experimental phase shifts for many different hadron-hadron scattering combinations currently prevents a universal application of this method.

One notes that, due to the contribution of resonances, the total average number of particle species $i$ is calculated in the HRG from:
\begin{equation}
\label{yieldR}
\langle N_i\rangle
= \langle N_i\rangle^t + \sum_r \Gamma(r\rightarrow i) \langle N_r\rangle^{t}. 
\end{equation}
The first term is the thermal average number of particles $i$. The second term describes the overall contribution from resonances decaying to particle $i$ with the corresponding decay branching,  $\Gamma(r\to i)$.

If, in high-energy heavy-ion collisions, the observed hadrons in the final state originate from the hadronizing QGP, then they should be of thermal origin with respect to the HRG partition function (\ref{eq:partition_fun}) as it describes the LQCD thermodynamics in the hadronic phase. Furthermore, the particle composition of the fireball should be consistent with (\ref{yieldR}).  

To verify the above, one needs, as input for the calculations, knowledge of the complete hadron spectrum, and the default is what is listed by the PDG; see ~\cite{ParticleDataGroup:2024cfk}. 
Furthermore, three initial conditions in heavy-ion collisions help to fix ($\mu_{I_3},\mu_S,\mu_C$):
i)  the third component of isospin, $\langle I_3\rangle=\sum_i \langle N_i\rangle  I_3$ to net baryon number, $\langle B\rangle=\sum_i \langle N_i\rangle  B_i$ conservation:  $\langle I_3\rangle/\langle B\rangle=(Z-N)/2A$ with $Z$ and $A$ being  the atomic and mass number of colliding nuclei, respectively, and $N=A-Z$ the neutron number;
ii) vanishing net initial strangeness: $\sum_i \langle N_i\rangle  S_i = 0$; iii) vanishing net initial charm content: $\sum_i{ \langle N_i\rangle}  C_i = 0$.
Consequently, when describing particle multiplicity in heavy-ion collisions within the above HRG model, which in the literature is referred to as the Statistical Hadronisation Model (SHM), one needs to fix only three parameters: the volume of the fireball and its temperature, as well as the value of the baryo-chemical potential. 

The SHM introduced above is formulated in the Grand Canonical (GC) ensemble with respect to charge conservation laws. The conservation of baryon number, strangeness, electric charge and charm holds on average and is controlled by the corresponding chemical potentials.   
It is already well established, however, that such a GC model can be successfully applied to heavy ion collisions if the number $\langle N_Q\rangle$  of produced hadrons with a given charge (conserved quantum number) $Q$ is sufficiently large, in practice, when $\langle N_Q\rangle\gtrsim10$. In the opposite limit,  particularly if  $\langle N_Q\rangle < |Q_p|$,  where $Q_p$ is the charge carried by a particle $p$, the thermal description requires exact implementation of charge conservation, which is introduced in the canonical, C-ensemble, see e.g.~\cite{Hagedorn:1984uy,Hamieh:2000tk,Braun-Munzinger:2003pwq}. This is particularly the case when applying the thermal model to particle production in central heavy-ion collisions at $\sqrt{ s_{NN}}\lesssim5$ GeV or at higher energies in peripheral collisions, as well as when applying SHM to pp collisions~(\cite{Cleymans:2020fsc}). 
The yields of hadrons with a given charge calculated in the C-ensemble for that charge are suppressed relative to the values obtained in the GC-ensemble. In a simplified treatment, applied here explicitly for strangeness, the density of species $i$, carrying net-strangeness $S$, $n_{i,S}$, is, in the C-ensemble, calculable from the respective density in the GC-ensemble as: 
\begin{equation}
n_{i,S}^{C}\simeq  n_{i,S}^{GC}\cdot \frac{I_n(x)}{I_0(x)},
\label{eq:ce}
\end{equation}
where the argument of the modified Bessel functions of order $n=|S|$ is $x=2\sqrt {N_SN_{\overline{S}}}$, with 
$N_S$ and $N_{\overline{S}}$ the total yields of hadrons carrying strangeness and anti-strangeness quantum number, respectively. 
The volume needed to calculate these yields, the strangeness-canonical volume, $V_{sc}$, interpreted as a strangeness-correlation volume, is a parameter of the model, which needs to be constrained with the data. This was recently accomplished using multi-strange hyperon data in pp and p-Pb collisions at the LHC~\cite{Cleymans:2020fsc}, but remains to be done systematically for AA collisions at low energies, where multi-strange hyperon data are scarce. While a first constraint was achieved with recent STAR data~(\cite{STAR:2021hyx}), we have here parametrized $V_{sc}$ as a function of $\sqrt{s_{NN}}$ adopting a proportionality with the beam rapidity in the center-of-mass; at lower energies this parametrization is similar to one based on the charged-particle multiplicities, Fig.~\ref{fig:dndy} (left) and  has currently a $\pm30$\% uncertainty. We note that the exact form of the $V_{sc}$ parametrization is not influencing the results. What the data clearly constrain is the increase of $V_{sc}$ with $\sqrt{s_{NN}}$ and the absolute values; for instance at $\sqrt{s_{NN}}=3$~GeV $V_{sc}=172\pm52$~fm$^3$ ($r_{sc}=3.4^{+0.3}_{-0.4}$~fm, close to the constraints achieved by STAR,~\cite{STAR:2021hyx}), which is much smaller than the fireball volume.
The ratio of Bessel functions in Eq.~\ref{eq:ce}, which explicitly quantifies the canonical suppression, is shown in Fig.~\ref{fig:edep-ratios} in the next section (see Eq.~\ref{eq:balance} and (\cite{Braun-Munzinger:2024ybd}) for the case of charm at large energies).

The HRG model formulated in the C-ensemble has provided an instrumental framework for the centrality and system-size dependence of particle production, particularly for strangeness production and suppression, see e.g.~\cite{Cleymans:2020fsc} and references therein.  The characteristic prediction of the HRG model in the C-ensemble was an increasing suppression of strange particle yields per pion with decreasing collision energy and collision centrality, as well as with increasing strangeness content of the particle, see~\cite{Hamieh:2000tk,Braun-Munzinger:2003pwq,Cleymans:2020fsc}. Such a pattern of suppression of (multi-)strange hadrons with multiplicity was indeed observed by the ALICE Collaboration at the LHC~(\cite{ALICE:2017jyt}) and is qualitatively similar to what has been first measured by the WA97 and NA57 Collaborations in Pb-Pb collisions at SPS energies (\cite{Antinori:2004ee}). 

\section{Statistical hadronization of light quarks}  \label{sect:lightq} 

In the SHM, the thermal parameters at chemical freeze-out, $T_{CF}$, $\mub$, and $V_{\Delta y= 1}$, are determined from a fit to the experimental data. 
For the most-central (0-10\%) Pb--Pb collisions at the LHC, the best description of the ALICE data (see ~\cite{Acharya:2017bso} and ref. therein) on yields of particles in one unit of rapidity at midrapidity is obtained with $T_{CF}=156.6\pm 1.7$~MeV, $\mu_B=0.7\pm 3.8$~MeV, and $V_{\Delta y= 1}=4175\pm 380$~fm$^3$~(\cite{Andronic:2017pug,Andronic:2018qqt}), see  Fig.~\ref{fig:Fit}.
The standard deviations quoted here are due exclusively to experimental uncertainties and do not reflect the systematic uncertainties associated with the model implementation. Further investigations have led to an order of magnitude improvement in the precision of $\mub =0.7\pm0.45$ MeV, see ~\cite{ALICE:2023ulv}, demonstrating that the central region at LHC energies is essentially baryon-free.

\begin{figure}[hbt]
\begin{tabular}{cc}  \begin{minipage}{.56\textwidth}
 \hspace{-.2cm}  \includegraphics[width=1.25\textwidth]{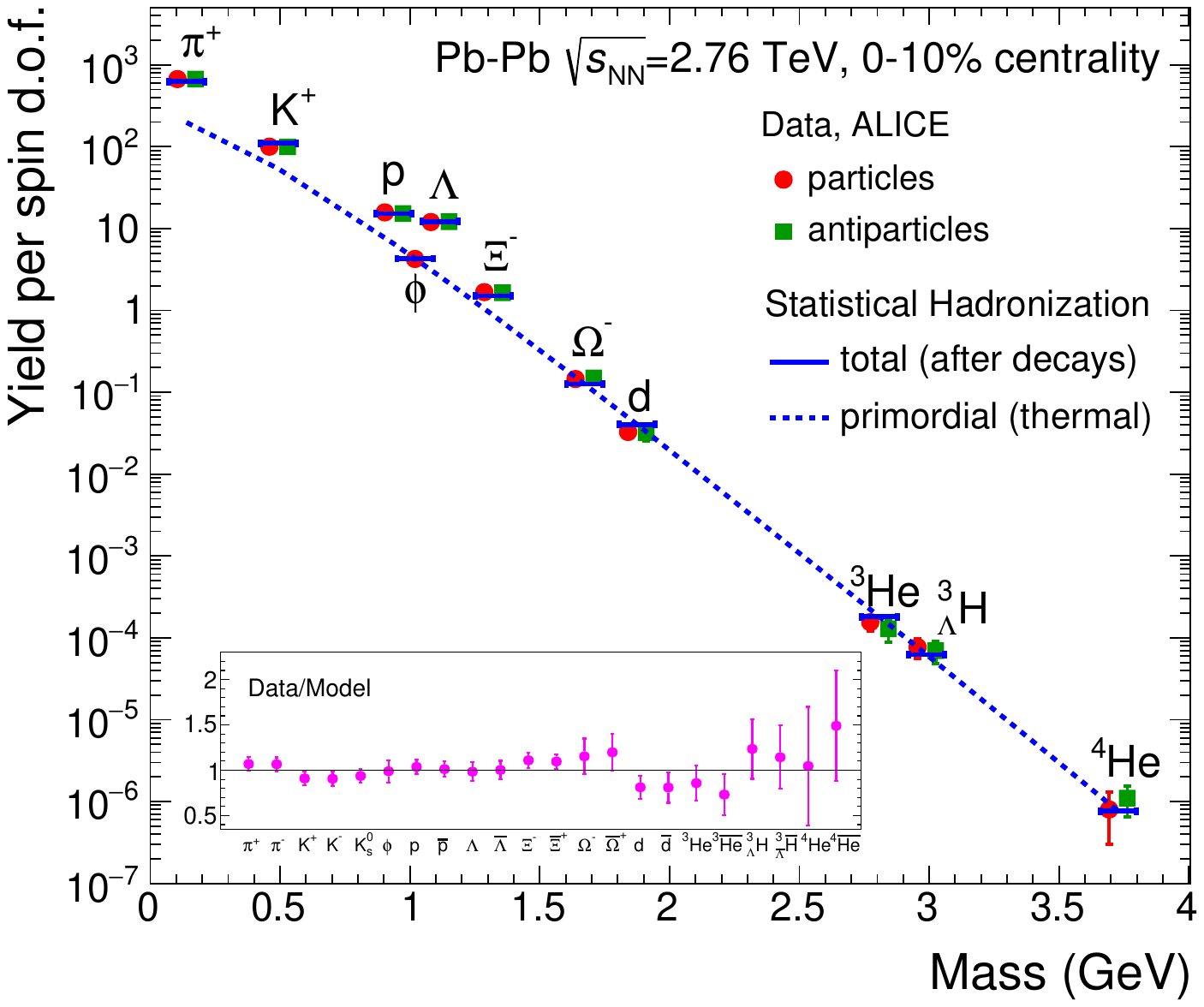}
\end{minipage} & \begin{minipage}{.4\textwidth}
\caption{Mass dependence of hadron yields divided by the spin degeneracy factor ($2J+1$), obtained using the SHM best fit in comparison to the ALICE data. The SHM results are shown for both the ``total'' particle yields, including strong and electromagnetic decays from excited resonances, and for the ``primordial'' thermal yields only. 
The insert shows the ratio data to model.
}\label{fig:Fit}
\end{minipage} \end{tabular}
\end{figure}

The results shown in Fig. \ref{fig:Fit} demonstrate that a very good agreement is obtained between the measured particle yields and SHM results over nine orders of magnitude in abundance values, encompassing strange and non-strange mesons, baryons, including strange and multiply-strange hyperons, as well as light nuclei and hypernuclei and their anti-particles.
The initially observed overprediction of about 17\% of the data compared to the model for proton and anti-proton yields (a deviation of 2.7$\sigma$) is entirely accounted for by the S-matrix treatment of the pion-nucleon interactions ~(\cite{Andronic:2018qqt}), leading to an excellent fit with a $\chi^2_{red} = 16.9/19$. 
Furthermore, it was recently shown that the addition, compared to what is listed by PDG ~(\cite{ParticleDataGroup:2024cfk}), of about 500 new (mostly baryonic) states predicted by LQCD and the quark model does lead to a strong deterioration of the fit, while a restoration of the good fit quality at no change of the thermal parameters is observed when the S-matrix treatment is employed as well for this expanded hadron spectrum, see ~\cite{Andronic:2020iyg}.

\begin{figure}[htb] \begin{tabular}{cc}  \begin{minipage}{.49\textwidth}
 \includegraphics[width=.96\textwidth]{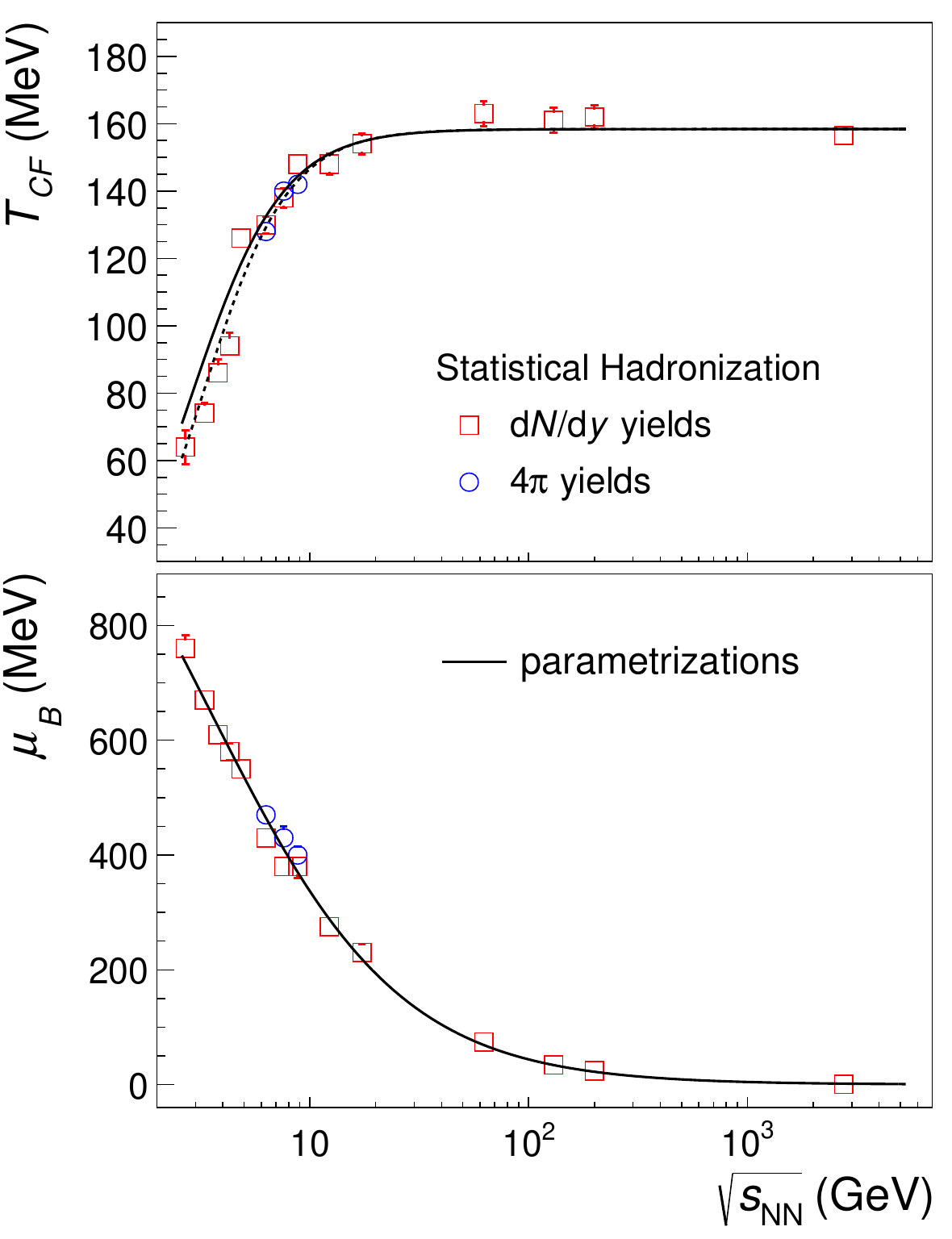} 
\end{minipage} & \begin{minipage}{.47\textwidth}
 \includegraphics[width=1.\textwidth]{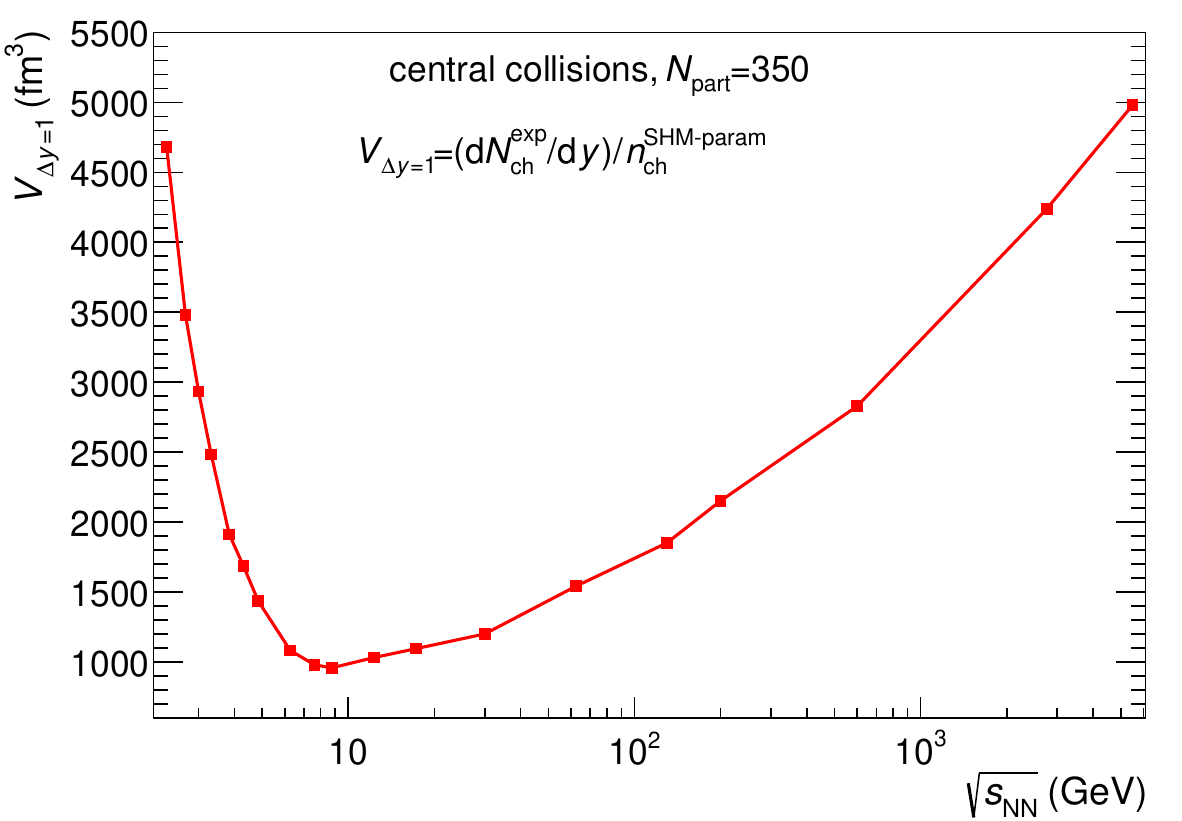} 

\hspace{.5cm}\caption{Left panel: energy dependence of the chemical freeze-out parameters $T_{CF}$  and \mub. The results are obtained from SHM analyses of hadron yields (at midrapidity, \dndy, and in full phase space, $4\pi$) for central collisions at different collision energies.
The parametrization for the temperature is slightly updated for the low energy domain to account for the current strangeness canonical treatment, see text (the previous parametrization~(\cite{Andronic:2017pug}) is shown for comparison as the dashed line). Right panel: energy dependence of the fireball volume $V_{\Delta y=1}$.} 
\label{fig:edep}
\end{minipage}
\end{tabular}
\end{figure}

We note that the yields of the measured lightest mesons and baryons ($\pi,K,p,\Lambda$) are substantially increased relative to their primordial thermal production by the resonance decay contributions (for pions, e.g., the decay contribution amounts to 70\% of the total yield). For the subset of light nuclei, the SHM predictions are, however, not affected by resonance decays.
For these nuclei, due to their large masses, a small variation in temperature leads to a large variation of the yield, resulting in a relatively precise determination of the freeze-out temperature $T_{nuclei} = 159 \pm 5$ MeV, well consistent with the value of $T_{CF}$ extracted above for all hadronic states.

The rapidity densities of light (anti-)nuclei and hypernuclei were actually predicted ~(\cite{Andronic:2010qu}), based on the systematics of hadron production at lower energies. It is nevertheless remarkable that such loosely bound objects (the deuteron binding energy is 2.2 MeV, much less than $T_{CF} \approx T_c  \approx 157$ MeV) are produced with temperatures very close to that of the phase boundary at LHC energy.
The detailed production mechanism for loosely bound states remains, however,  an open question (see the recent review ~(\cite{Braun-Munzinger:2018hat})). One possibility is that such objects, at QGP hadronization, are produced as compact, colorless droplets of quark matter with quantum numbers of the final-state hadrons~(\cite{Andronic:2017pug}), see the discussion below.

The thermal nature of particle production in ultra-relativistic nuclear collisions has been experimentally verified not only at LHC energy, but also at the lower energies of the RHIC, SPS and AGS accelerators. The essential difference is that, at these lower energies, the matter anti-matter symmetry observed at the LHC is lifted, implying substantial values of the chemical potentials. %
An overview of the SHM results for all investigated collision energies is shown in Fig.~\ref{fig:edep}.
While \mub decreases smoothly with increasing collision energy, the dependence of $T_{CF}$ on energy exhibits a striking feature: it increases with increasing energy from about 60 MeV to a saturation for $\sqrt{s_{\rm NN}} > 20$ GeV, of about 157 MeV. 
The saturation of $T_{CF}$ observed in Fig.~\ref{fig:edep} lends support to the earlier proposal~(\cite{BraunMunzinger:1998cg,Stock:1999hm,BraunMunzinger:2003zz}) that, at least at high energies, the chemical freeze-out temperature is very close to the QCD hadronization temperature ~(\cite{Andronic:2008gu}), implying a direct connection between data from relativistic nuclear collisions and the QCD phase boundary. Hagedorn noted long ago~(\cite{Hagedorn:1965st}) that hadronic matter cannot be heated beyond a certain limit due to the divergence of the number of states,
but the saturation observed here indicates a different boundary, i.e. the chiral phase transition.

\begin{figure}[htb]
\begin{tabular}{cc}  \begin{minipage}{.48\textwidth}
\includegraphics[width=.9\textwidth,height=.7\textwidth]{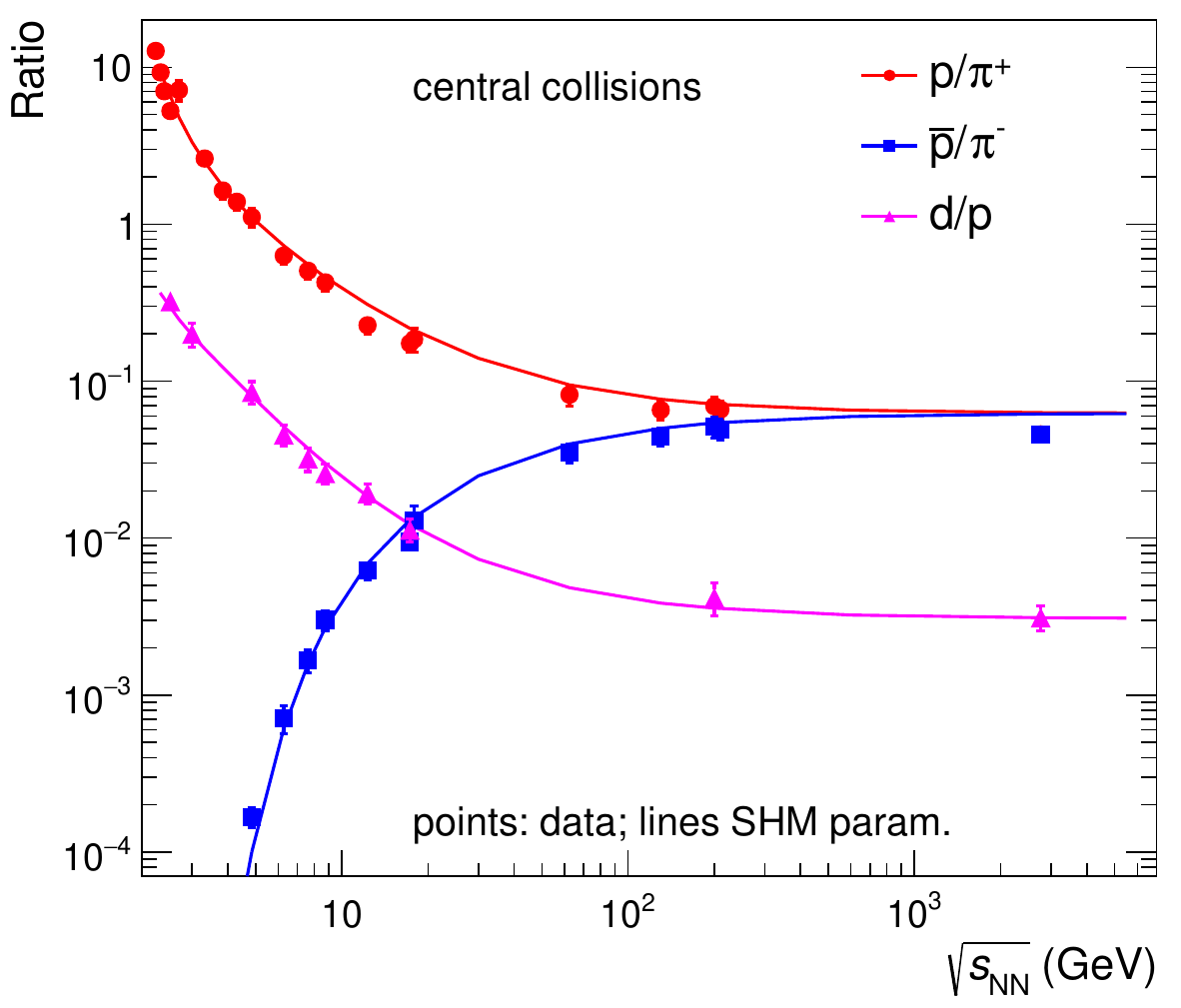} 
\end{minipage} & \begin{minipage}{.48\textwidth}
\includegraphics[width=.9\textwidth,height=.7\textwidth]{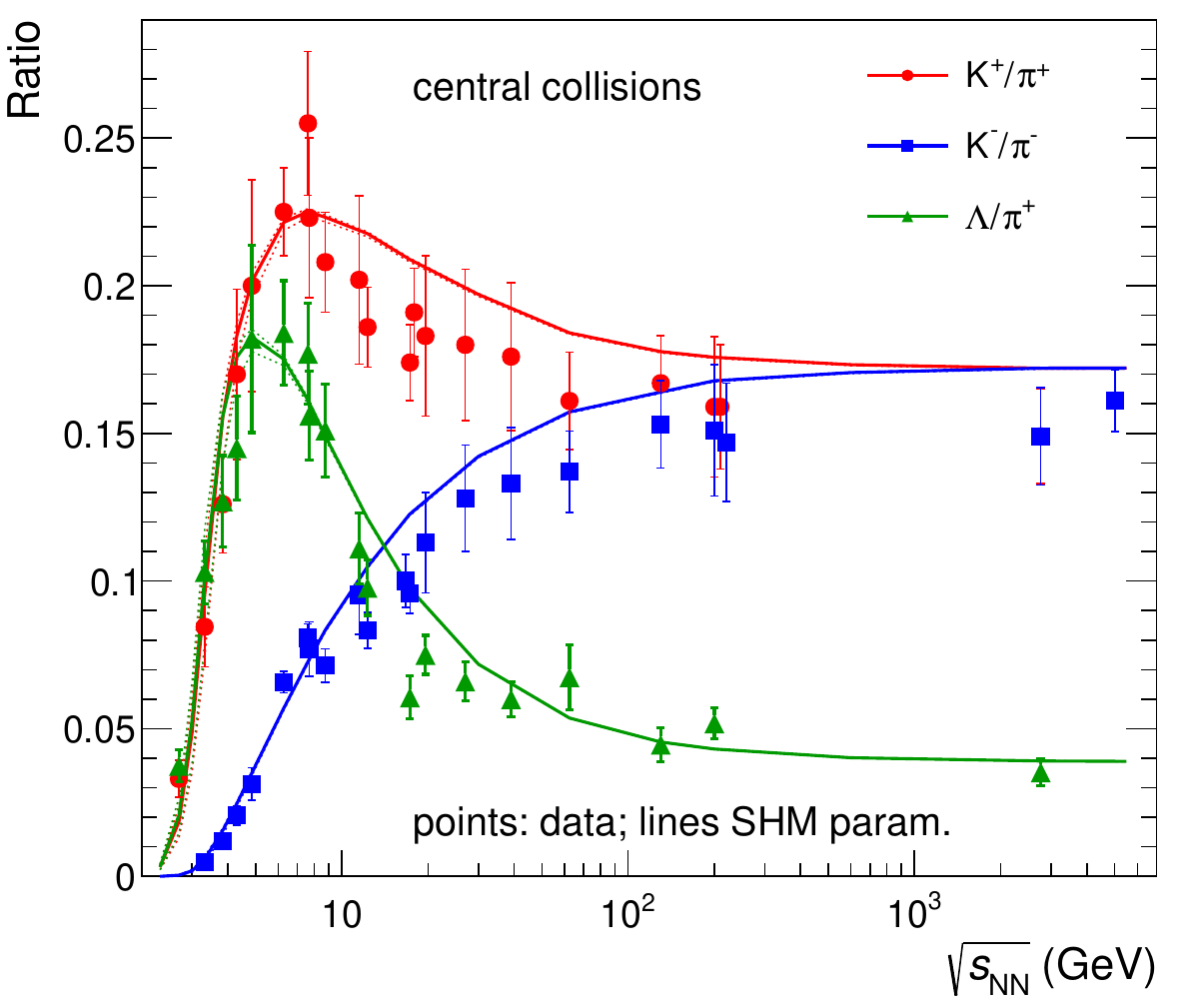}
\end{minipage}  \\
 \begin{minipage}{.48\textwidth}
\includegraphics[width=.9\textwidth,height=.7\textwidth]{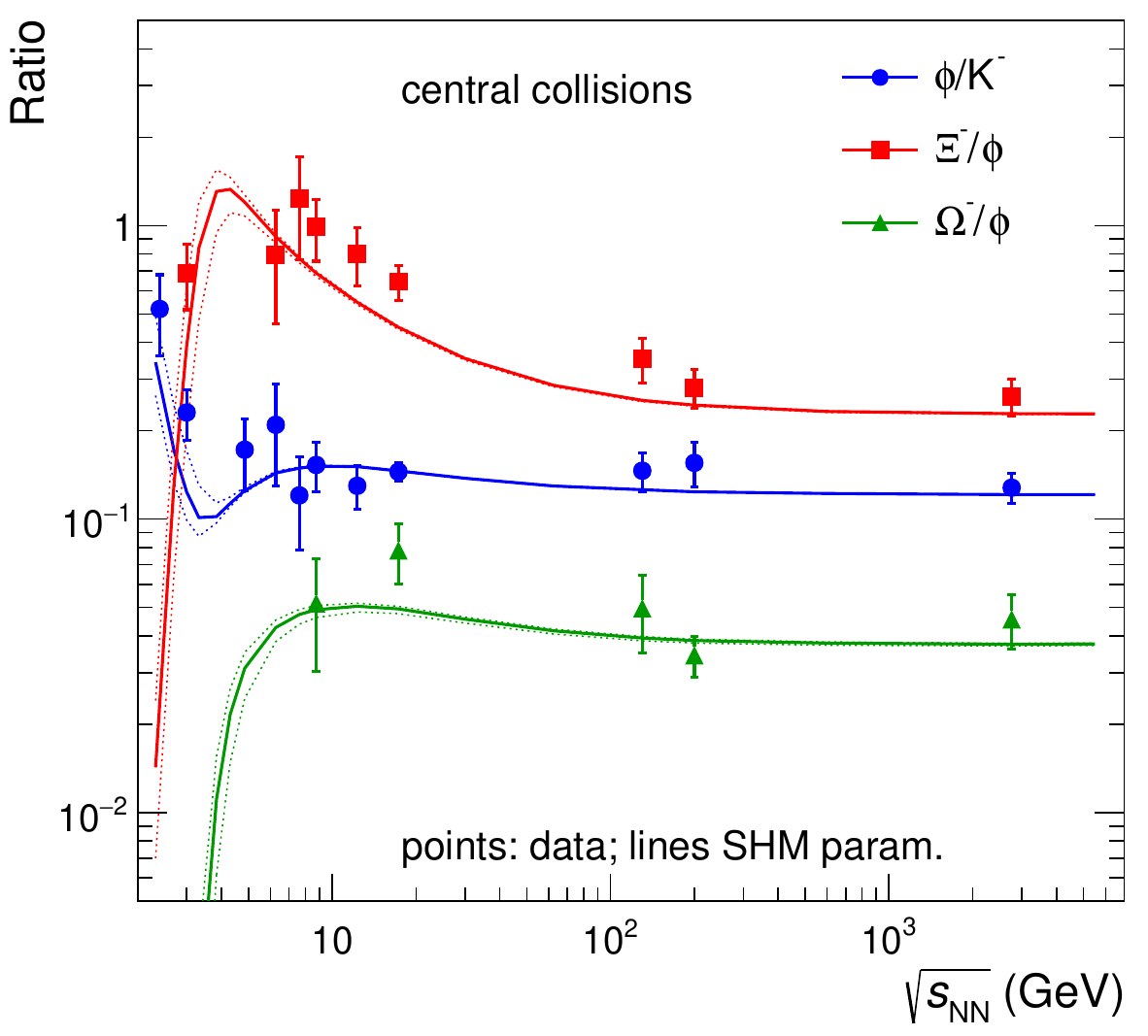} 
\end{minipage} & \begin{minipage}{.48\textwidth}
\includegraphics[width=.9\textwidth]{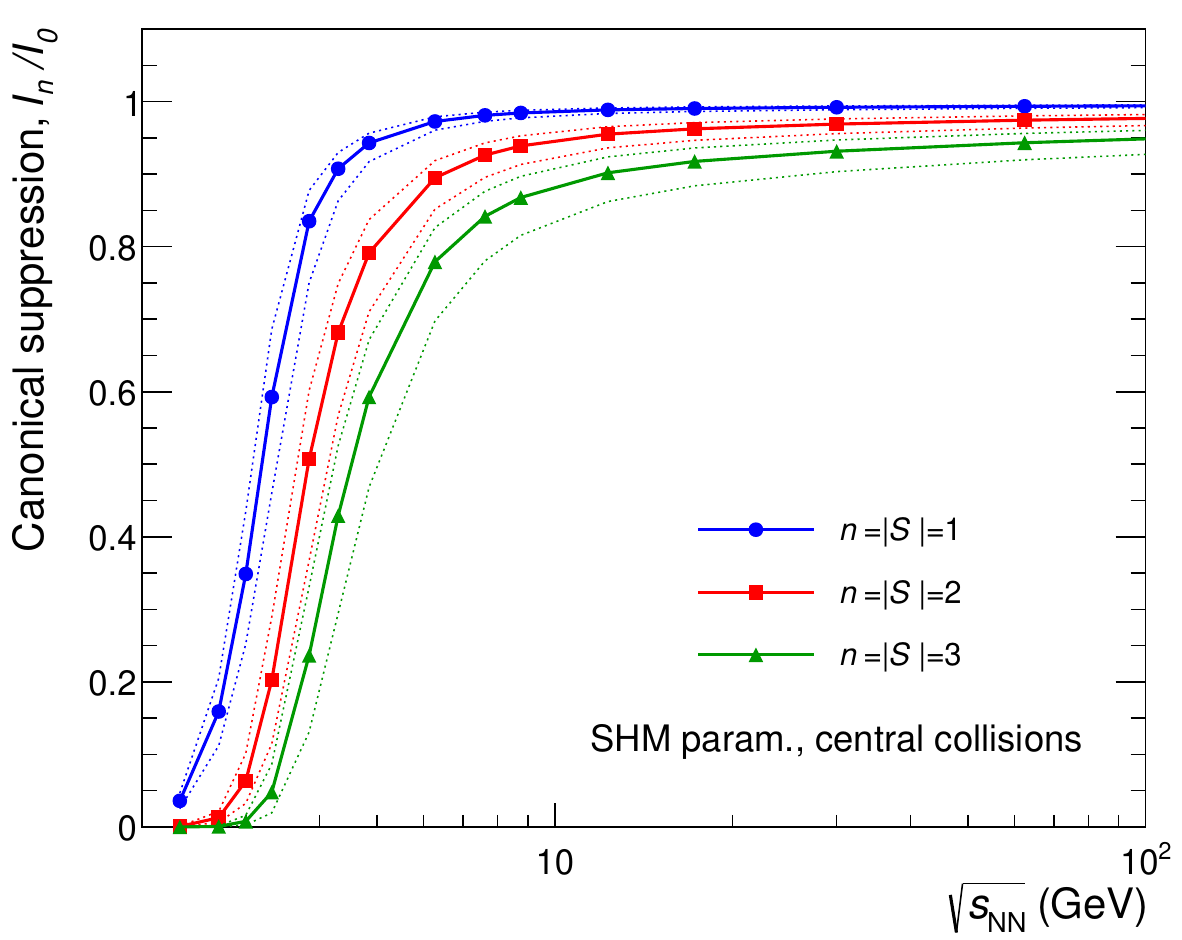} 
\end{minipage}
\end{tabular}
\caption{Collision energy dependence of the relative abundance of selected hadron species. The data, where the error bars represent the total statistical and systematic uncertainties, are compared to SHM predictions for the parametrized dependence of $T_{CF}$ and $\mu_B$ with energy. The dotted band, visible at the lower energies, quantifies the $\pm30$\% uncertainty of the strangeness-canonical volume $V_{cs}$. The canonical suppression, ratio of Bessel functions $I_n/I_0$, see Eq.~\ref{eq:ce}, is shown in the lower right panel, where the $\pm30$\% uncertainty of $V_{cs}$ is shown as dotted lines.}
\label{fig:edep-ratios}
\end{figure}

The parametrizations shown for $T_{CF}$ and $\mu_B$ in Fig.~\ref{fig:edep} (left) are:
\begin{equation}
T_{CF}={T_{CF}^{lim}}/(1+\exp(2.20-\ln(\sqrt{s_{\rm NN}})/0.48)), \quad
\mub ={a}/(1+0.288\sqrt{s_{\rm NN}}),
\label{eq:param}
\end{equation}
with $\sqrt{s_{\rm NN}}$ in GeV, with a 'limiting temperature' $T_{CF}^{lim}=156.6\pm 1.7$ MeV fixed to the temperature obtained from the fit of the LHC data and $a=1307.5$ MeV.
The parametrization for $T_{CF}$ is slightly updated for the low energy domain with respect to the previous parametrization~(\cite{Andronic:2017pug} to account for a re-evaluation of the strangeness canonical treatment, as imposed by the recent data at lower energies~(\cite{STAR:2021hyx}). 
The third parameter of the SHM, the fireball volume, can be determined as a fit parameter. Here, the volume corresponding to one unit of rapidity (at midrapidity, $y=0$), $V_{\Delta y=1}$ was determined for the calculations using the parametrizations for $T_{CF}$ and $\mu_B$, Eq.~\ref{eq:param}, via the charged-particle multiplicity densities at midrapidity, $\ud N_{ch}/\ud y$, Fig.~\ref{fig:dndy} (left): $V_{\Delta y=1}=(N_{ch}/\ud y)/n_{ch}^{\mathrm SHM-param}$.
The non-monotonic dependence of $V_{\Delta y=1}$ may to some extent be caused by the strong energy dependence of the width of the rapidity distributions of hadrons, leading to a larger effective fraction of the fireball for one unit of rapidity considered here. The larger baryon stopping at lower energies plays a role too.

To illustrate how well the thermal description of particle production in central nuclear collisions works we show in Fig.~\ref{fig:edep-ratios} and Fig.~\ref{fig:edep-yields} the energy dependence of the relative abundance of several hadron species and absolute yields $\ud N/\ud y$, respectively, along with SHM predictions using the parametrized dependence of $T_{CF}$ and $\mu_B$ with energy.
We note that the S-matrix correction, which at the LHC energies leads to a decrease of the proton yield in the SHM by about 17\%, is not applied for the curves in Fig.~\ref{fig:edep-ratios} and Fig.~\ref{fig:edep-yields}.
The model-data agreement is overall very good.
In particular, the maxima  (occurring at slightly different c.m. energies) in the $K^+/\pi^+$ and $\Lambda/\pi^+$ ratios are naturally explained~(\cite{Andronic:2008gu}) due the energy dependence of $T_{CF}$ and $\mu_B$, as shown in the left panel of Fig.~\ref{fig:edep}, and  strangeness conservation. The rather steep increase of strangeness production is well reproduced within the framework of canonical thermodynamics and only very indirectly and in a 'limiting temperature' sense related to a 'rapid onset of deconfinement' as argued in ~(\cite{Gazdzicki:1998vd}). Deuterons and the multi-strange hyperons $\Xi^-$ and $\Omega^-$ are also well reproduced. 
For the strangeness-carrying hadrons the uncertainty of the strangeness-canonical volume is visible at the lower energies, where the canonical suppression is a significant effect, see the lower right plot in Fig.~\ref{fig:edep-ratios}; this uncertainty will be reduced with the measurements expected in the coming years, which would also put to test the interesting predictions of SHM for the non-monotonic evolution of the ratios $\phi$/K$^-$ and $\Xi^{-}/\phi$ at lower energies, seen in the lower left panel in Fig.~\ref{fig:edep-ratios}.

\begin{figure}[htb]
\includegraphics[width=.49\textwidth]{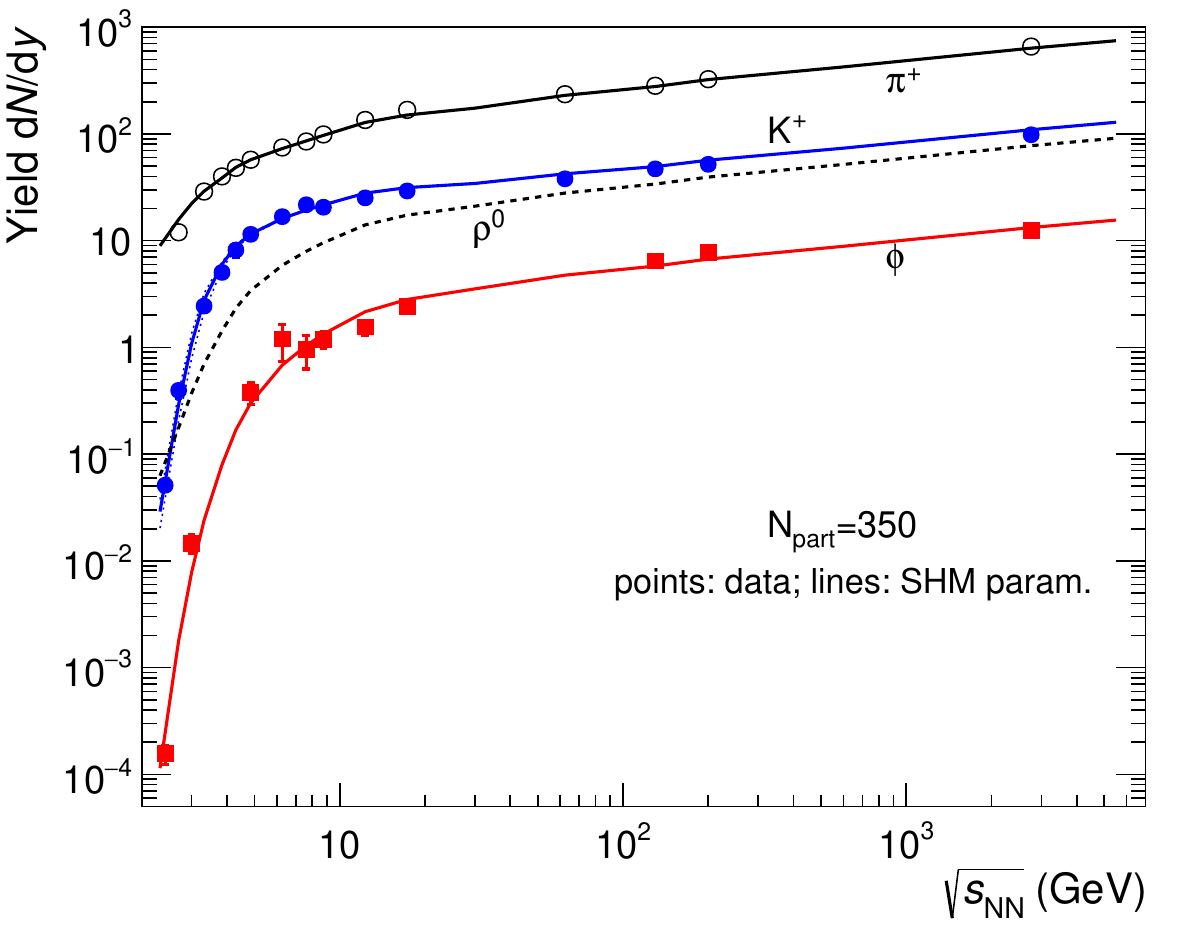} \hspace{.3cm} 
\includegraphics[width=.49\textwidth]
{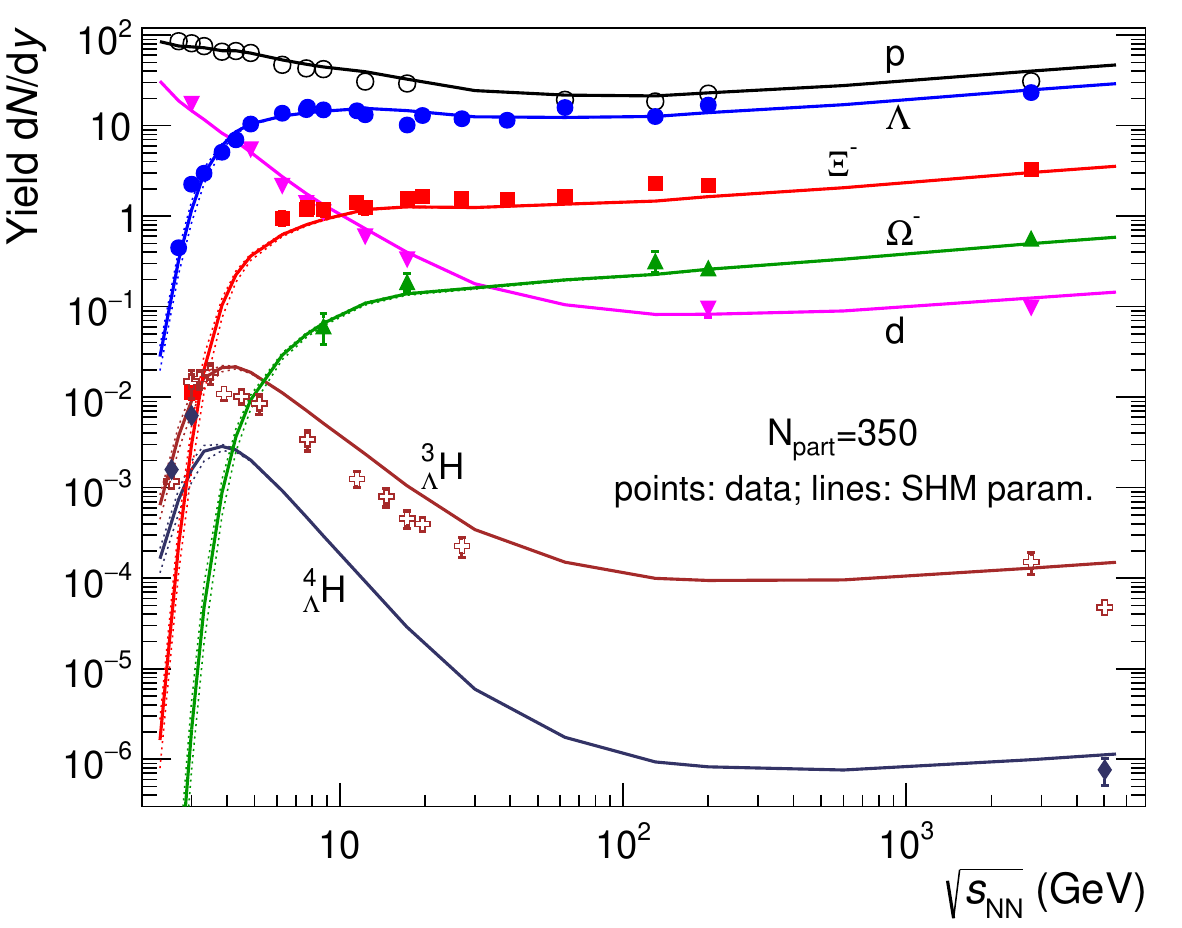}
\caption{As Fig.~\ref{fig:edep-ratios}, but for yields of selected mesons (left) and baryon and hypernuclei (right) species. Note the performed $N_{part}$-scaling of the data to correspond to $N_{part}=350$. In the left panel we show for comparison the predicted yields of the $\rho^0$ mesons.
}
\label{fig:edep-yields}
\end{figure}

The remarkable description of the absolute yields seen in Fig.~\ref{fig:edep-yields} implies a universal hadron production mechanism at chemical freeze-out, despite vastly different fireball properties at the lowest and highest ends of the collision range.
The production of hypernuclei, which is significantly enhanced at lower energies as a result of the large baryo-chemical potential, shows interesting features. The recent hypertriton data from HADES in Ag-Ag collisions at 2.55 GeV~(\cite{HADES:2025qum}), STAR in central Au-Au collisions over a broad range of energies in the BES program~(\cite{STAR:2026khp}), and ALICE in central Pb-Pb collisions at the LHC~(\cite{ALICE:2015oer,ALICE:2024koa} are described by the model with a mixed degree of success, good at the lowest energies and overestimated otherwise.
The $^4_\Lambda$H hypernucleus data (\cite{STAR:2021orx,ALICE:2024djx,HADES:2025qum}) is underpredicted at low energies and well described at the LHC. This lack of consistency between data and model description is under scrutiny. We note in this context that the efficiency for the detection of loosely bound states such as the hyper-triton is notoriously difficult to determine with precision.

\begin{figure}[htb]
\begin{tabular}{cc}  \begin{minipage}{.48\textwidth}
 \hspace{-.4cm}    \includegraphics[width=1.04\textwidth]{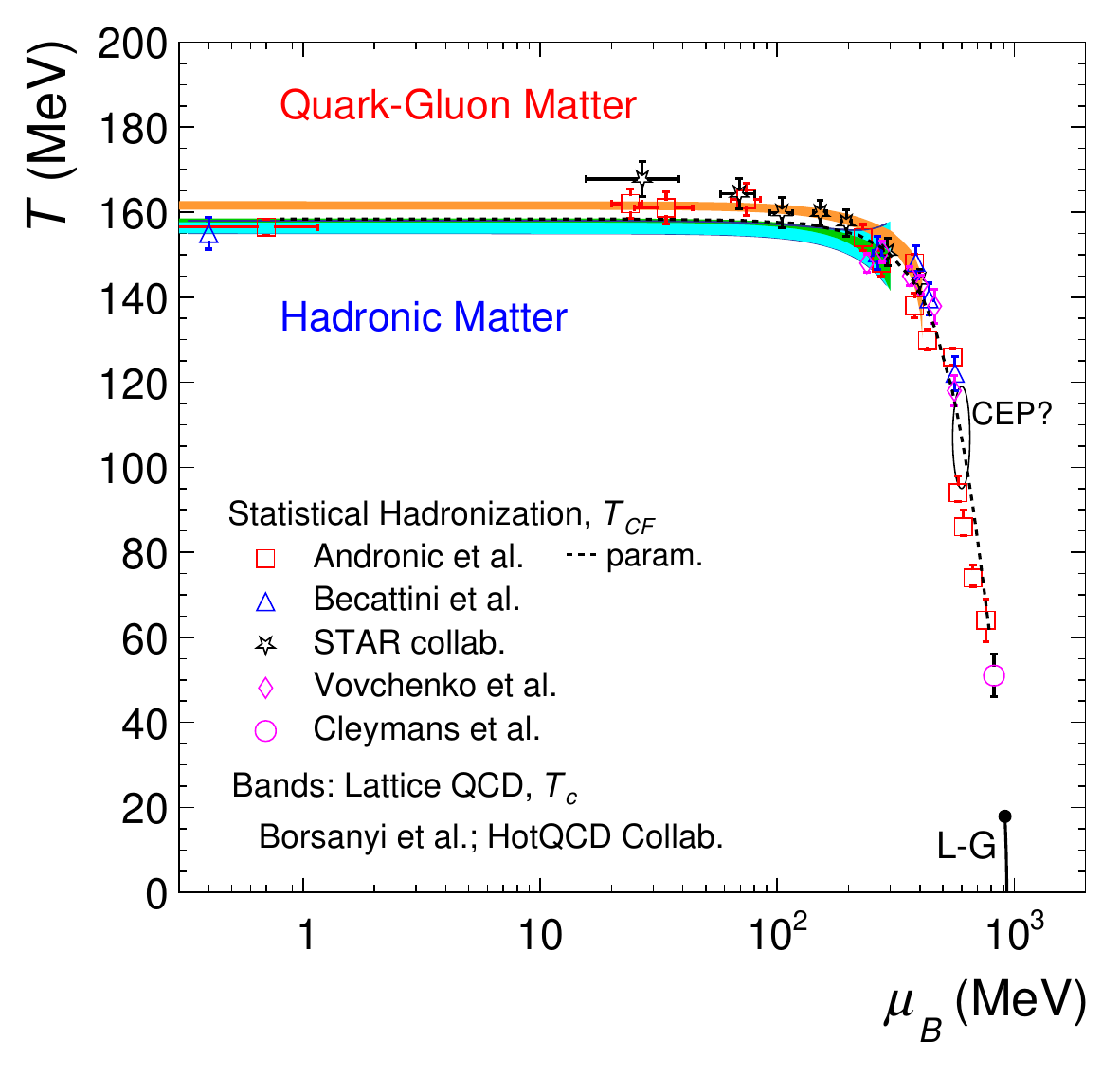}
\end{minipage} & \begin{minipage}{.47\textwidth}
\caption{Phase diagram of strongly interacting matter constructed from chemical freeze-out points for central collisions at different energies. Points are extracted from experimental data sets in our own (squares) and other similar SHM analyses by ~\cite{Cleymans:1998yb,Vovchenko:2015idt,Becattini:2016xct,Adamczyk:2017iwn}. They are compared to predictions from LQCD shown as bands. The turquoise band represents the chiral transition as computed by ~(\cite{Bazavov:2018mes,Borsanyi:2020fev}).
The orange band corresponds to the deconfinement transition~(\cite{Borsanyi:2024xrx}), in this case not yet extrapolated to the continuum. For the location of a possible critical endpoint (CEP) see~(\cite{Fischer:2026uni}). For the liquid-gas phase transition (L-G) see~(\cite{Kaiser:2026msy}).
}
\label{fig:t-mu}
\end{minipage}\end{tabular}
\end{figure}

Since the statistical hadronization analysis at each collision energy yields a pair of ($T_{CF}$,\mub) values, these points can be entered into the phase diagram of QCD, shown in Fig.~\ref{fig:t-mu}. At high collision energies the points are very close to the pseudo-critical line for the chiral phase transition (and also the deconfinement transition) as computed by ~(\cite{Bazavov:2018mes,Borsanyi:2020fev,Borsanyi:2024xrx}). The points at low temperature and high $\mu_B$ seem to converge towards the value for ground state nuclear matter ($\mub\simeq930$~MeV). This region in the phase diagram is likely far away from the QCD phase boundary  as noted by ~\cite{Floerchinger:2012xd}.
Electroweak observables like dileptons (\cite{Rapp:1999ej,Salabura:2020tou,Bailhache:2025kwa}) and jets (\cite{Wang:2025lct}) probe the earlier and hotter phases of the collision, prior to chemical freeze-out.

Fits of $\pT$ spectra of hadrons give access to the kinetic freeze-out temperature, which is in the range 100--135 MeV for central collisions for a broad range of collision energies. A second fit parameter is the average collective expansion velocity, which is in the range 0.50-0.65$c$, see a compilation in~\cite{STAR:2025xxf}.
At RHIC and the LHC, the description of hadron multiplicities, $\pT$ spectra and collective flow through hydrodynamics by~\cite{Bernhard:2019bmu,Schenke:2020mbo,Gardim:2019xjs} allows the extraction of the dynamical evolution of the QGP temperature, leading to initial values as high as 800 MeV, at the LHC~(\cite{ALICE:2022wpn}). The hydrodynamic simulations allow, in addition, the extraction of the ratio of shear viscosity to entropy density.

\begin{figure}[hbt]
\begin{tabular}{cc}  \begin{minipage}{.5\textwidth}
{\includegraphics[width=.9\textwidth]{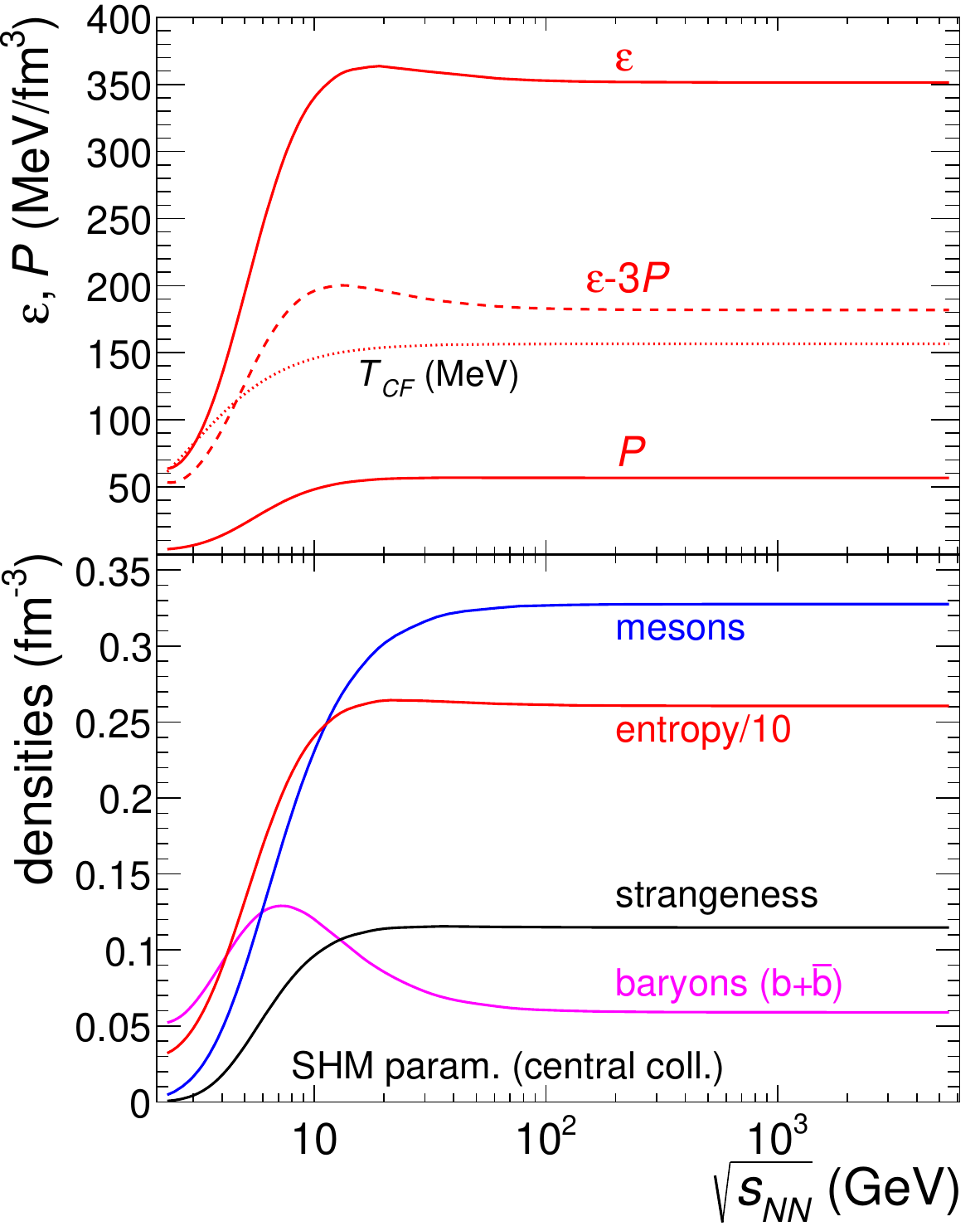}}
\end{minipage} & \begin{minipage}{.46\textwidth}
{\includegraphics[width=.85\textwidth]{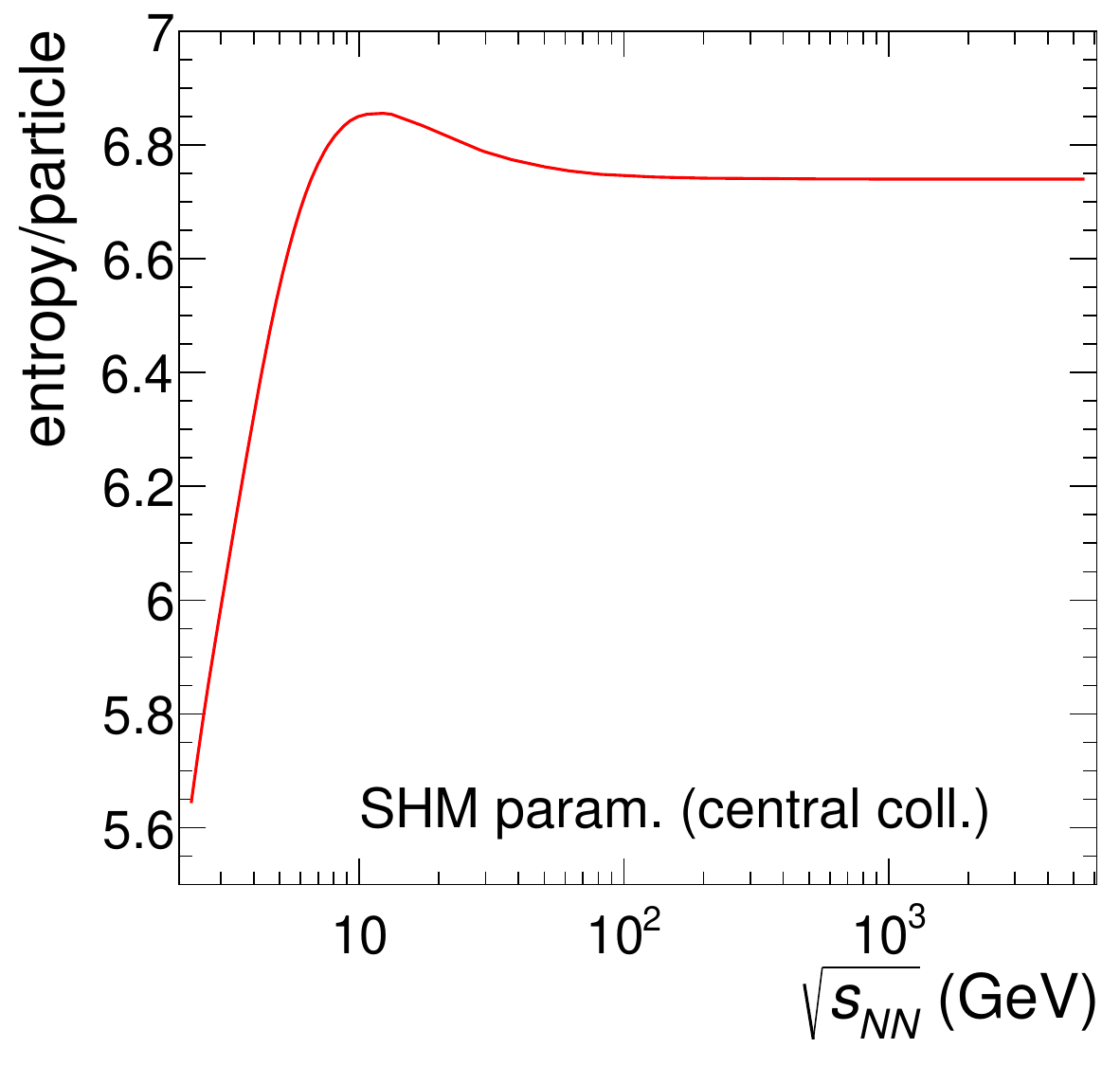}}
 \vskip -0.0 cm\caption{Left panel: collision energy dependence of energy density $\varepsilon$, pressure $P$, and Equation of State, $\varepsilon -3P$ (upper panel) and entropy, meson, baryon, and strangeness densities (lower panel) following the SHM parametrization of the chemical freeze-out.
 The energy dependence of the parametrized chemical freeze-out temperature $T_{CF}$ is overlaid for comparison. For the baryon and strangeness densities particles and antiparticles are added. The entropy per particle is shown on the right panel.
} \label{fig:therm2}
\end{minipage} \end{tabular}
\end{figure}

Employing the SHM parametrization of chemical freeze-out as a function of collision energy for central collisions, the thermodynamic quantities energy density $\varepsilon$, pressure $P$ and entropy density $s$ are shown in Fig.~\ref{fig:therm2} (left) alongside the densities of baryons (particles and antiparticles) and mesons. The Equation of State, $\varepsilon -3P$, is also included.
The main trends are determined by the collision energy dependence of $T_{CF}$, but the baryo-chemical potential plays a role too. In particular, a prominent maximum is observed at $\sqrtsNN\simeq7$\,GeV for the density of baryons and anti-baryons (at these energies dominated by the baryons). This appears to cause a very shallow maximum in the energy density and entropy density, visible at slightly larger collision energies.
A more prominent maximum is observed in the energy dependence of the entropy per particle, Fig.~\ref{fig:therm2} (right).

\begin{figure}[hbt]
\begin{tabular}{cc}  \begin{minipage}{.5\textwidth}
{\includegraphics[width=.9\textwidth]{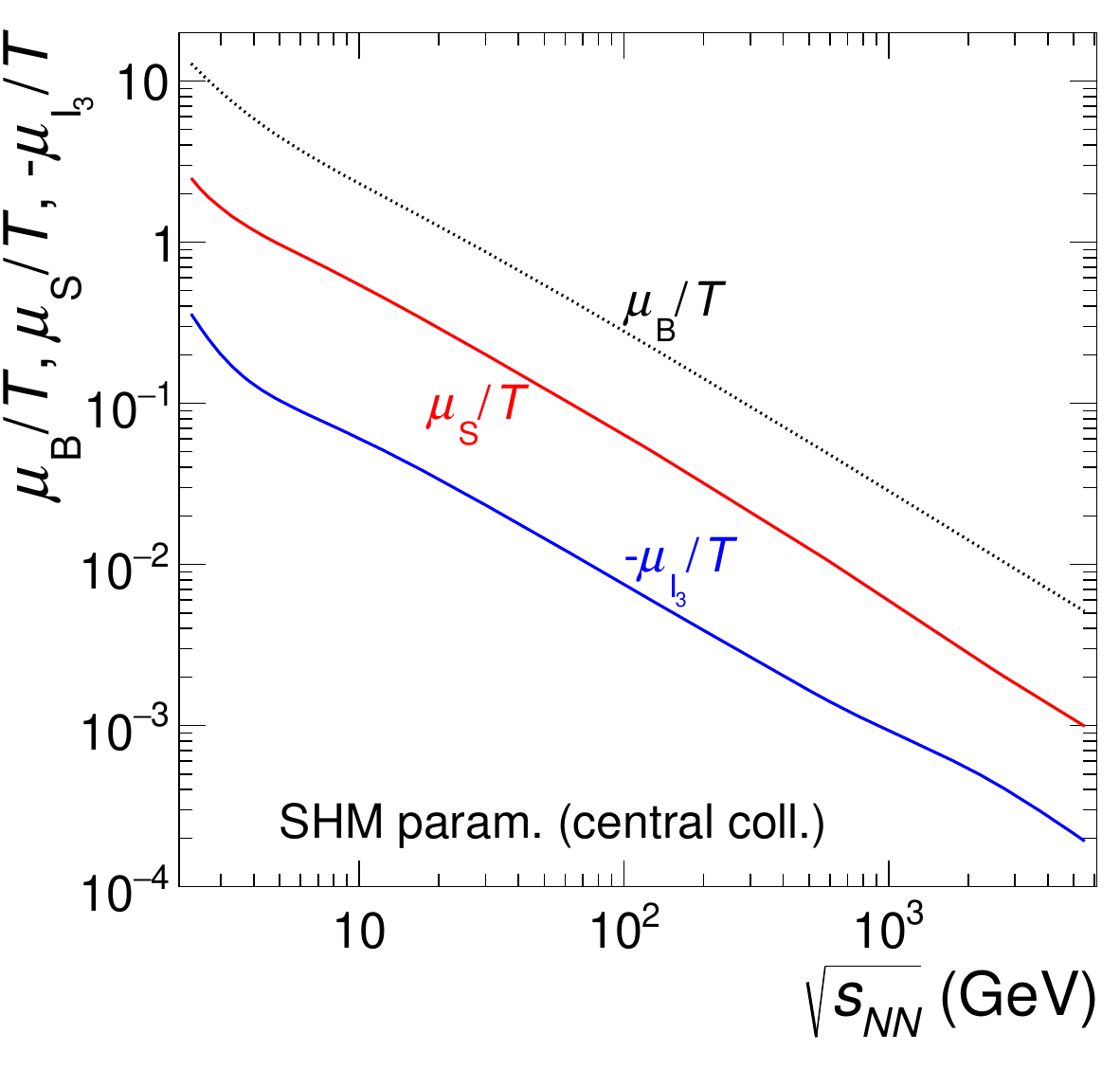}}
\end{minipage} & \begin{minipage}{.46\textwidth}
 \caption{The ratios $\mu_B/T$, $\mu_S/T$,  and $-\mu_{I_3}/T$ as a function of the collision energy. We recall that only $\mu_B$ is a fit parameter of SHM, while $\mu_S$ and $\mu_{I_3}$ are constrained by conservation laws, see Section~\ref{sect:model}.
} \label{fig:therm3}
\end{minipage} \end{tabular}
\end{figure}

In Fig.~\ref{fig:therm3} the ratios of the baryo-chemical, strangeness, and isospin chemical potentials to the temperature, $\mu_B/T$ , $\mu_S/T$, and $-\mu_{I_3}/T$, are shown, exhibiting a steep and monotonic decrease with the collision energy.

\section{Extension to hadrons with heavy quarks}   \label{sect:heavyq} 

Experiments at the RHIC and mainly LHC colliders have recently provided rather detailed information on the production of  hadrons containing charm or beauty quarks. They can be composed of states like charmonia with quark composition $c\bar{c}$ and bottomonia  with $b\bar{b}$ corresponding to hidden charm and beauty hadrons. In addition,  there are  open charm and beauty hadrons such as D mesons with $d\bar{u}$ and B mesons  with $b\bar{d}$  composition  as well as baryons like the $\Lambda_c$ containing $udc$ quarks, see ~\cite{ParticleDataGroup:2024cfk}.

From experiments with Pb--Pb collisions there is good evidence, mainly from results obtained at the CERN LHC~(\cite{ALICE:2021rxa,Andronic:2021erx,Andronic:2019wva}), that D mesons produced in such collisions closely follow the anisotropic expansion observed in such collisions, implying that  charm quarks reach a large degree of thermal equilibrium, although charm quarks in the system are chemically far out of equilibrium. This is supported by heavy quark diffusion coefficients obtained from LQCD calculations~(\cite{Altenkort:2020fgs,Altenkort:2023oms}). A further strong indication for equilibration of charm quarks is the fact that the charmonium state J/$\psi$   and  the charm baryon $\Lambda_c$ also participate in this collective,  hydrodynamic expansion ~(\cite{ALICE:2013xna,He:2021zej, ALICE:2026zcz}).

Models have been developed to provide a framework for  the understanding of the production me\-cha\-nism of charmed hadrons for collision systems ranging from pp to Pb--Pb.  Direct calculations based on QCD are typically applicable only for production of hadrons at large transverse momenta, where perturbation theory can be applied. For a survey see ~\cite{Cacciari:2012ny}. Since QGP studies with hadrons containing heavy quarks usually rely also on the total charm production in relativistic nuclear collisions, this implies measurements also at low transverse momenta, where the perturbative approach fails.

An alternative approach is to use a phenomenological approach based on the coalescence model. For a survey see ~\cite{Minissale:2020bif}. In fact, a number of quark coalescence mo\-dels have been developed ~(\cite{Cho:2019lxb,ExHIC:2017smd,Zhou:2014kka,Greco:2003vf}). This could be one possible way to study the dependence of production yields on hadron size to make progress in the understanding of the important open question whether the many recently observed exotic charm states  are compact multi-quark states or rather hadronic molecules (see ~\cite{Aarts:2016hap,Maiani:2022psl} and refs. cited there). The many conceptual difficulties with this approach include that energy conservation is not included in modeling of the coalescence process. Furthermore, the produced charm hadrons are color neutral, while charm quarks carry the color degree of freedom. Consequently, color neutralization at hadronization process is an issue that requires further assumptions, see ~\cite{Song:2021mvc}.

A parameter-free approach, named SHMc, has been proposed more than 25 years ago  ~\cite{Braun-Munzinger:2000csl}. This implies an extension of the SHM to incorporate charm quarks by treating them as impurities that thermalize in the hot fireball. In this approach the charm quarks are, because of their large mass ($m_c \approx 1.2$ GeV), not produced thermally but their  total number is fixed as external parameter by measurement of the total charm production cross section for the system under consideration, see also below. This was developed further in ~(\cite{Andronic:2003zv,Andronic:2006ky,Andronic:2017pug}) eventually to include all hadrons with hidden and open charm.  

Charm quark production in this approach takes place in initial hard collisions. The produced charm quarks then thermalize in the hot fireball, but their number is conserved during the evolution of the fireball~(\cite{Andronic:2006ky}) since the charm quark annihilation cross section is small.  This new approach requires the introduction of a charm fugacity $g_c$, introduced in~\cite{Braun-Munzinger:2000csl,Andronic:2021erx}. The value of $g_c$ is not arbitrary but has to be experimentally determined by measurement of the total charm cross section. For central Pb--Pb collisions at LHC energy, $g_c \simeq 31.5$~(\cite{Andronic:2021erx}). The charmed hadrons are, in the SHMc, all formed at the phase boundary, i.e. at hadronization, in the same way as all (u,d,s) hadrons, of course with the boundary condition that all charm quarks present in the QGP materialize in hadrons (as warranted by $g_c$).

In this approach, the knowledge of the inclusive heavy \qqbar\ production cross-section along with the chemical freeze-out (hadronization) temperature $T_{CF}= 156.6$~MeV obtained from the analysis of the yields of hadrons composed of light valence quarks  as in ~\cite{Andronic:2017pug}, is sufficient to determine the total ($\pT$-integrated) yield of all hadrons containing heavy quarks in ultra-relativistic nuclear collisions.
This is based on the charm balance equation introduced in ~\cite{Braun-Munzinger:2000csl}. Using thermal weights charm conservation this equation relates, via the fugacity factor $g_c$, the initial inclusive charm production yields to the yields of hadronic states:
\begin{equation}
    N_{\ccbar} = \frac{1}{2} g_c V \sum_{i} n^{{\rm th}}_i \frac{I_1(N_{c}^{tot})} {I_0(N_{c}^{tot})} \,
    + \, g_c^2 V \sum_{j} n^{{\rm th}}_j \, + \, \frac{1}{2} g_c^2 V \sum_{k} n^{{\rm th}}_k \frac{I_2(N_{c}^{tot})} {I_0(N_{c}^{tot})},
  \label{eq:balance}
\end{equation}
where $N_{\ccbar}\equiv \dd N_{\ccbar}/\dd y$ denotes the rapidity density of charm quark pairs produced in initial hard collisions and the (grand-canonical) thermal densities for open and hidden charm hadrons are given by $n_{i,j,k}^{{\rm th}}$. The index $i$ runs over all open charm states with one valence charm or anti-charm quark ($\mathrm{D}, \mathrm{D}_s, \Lambda_c, \Xi_c, \Omega_c$ and antiparticles), the index $j$ over all quarkonium states ($\jpsi, \chi_c, \psip$), and the index $k$ over open charm states with two charm or anti-charm quarks ($\Xi_{cc}, \Omega_{cc}$ and antiparticles). 
The fugacity factor $g_c$ is obtained by solving {Equation~\ref{eq:balance}} for a given collision centrality class and enters in the model predictions of the yields of hadrons with charm quarks and antiquarks linearly for single-charm hadrons and quadratically for charmonia and doubly-charmed baryons.
The ratio of the modified Bessel functions, $I_\alpha/I_0$, is a (canonical) correction for the exact conservation of charm~(\cite{Cleymans:1990mn,Gorenstein:2000ck}). The argument,  
$N_{c}^{tot}$, is the total open charm content (particles and antiparticles), consequently containing, besides the thermal densities of charmed hadrons and the volume, the $g_c$ factor.
The thermal densities are computed in the grand canonical ensemble using the latest version of the SHMc~(\cite{Andronic:2017pug,Andronic:2019wva}), with the chemical freeze-out temperature $T_{CF}=156.6$ MeV. The fireball volume per unit rapidity at midrapidity is $V = 4997 \pm 455$\,{fm}$^3$  for the most central 10\% Pb-Pb collisions at LHC energy $\sqrtsNN$ = 5.02 TeV. In this case, based on the measured average value $N_{\ccbar} \simeq 16$ for one unit of rapidity~(\cite{ALICE:2022wpn}), $g_c\simeq 31.5$. While the thermal densities do not vary with centrality of the collision, $N_{\ccbar}$ and $V$ in {Equation~\ref{eq:balance}} are centrality-dependent and scale with the number of nucleon-nucleon collisions, $N_{coll}$, and number of participating nucleons, $N_{part}$, respectively. This leads to a quasi-linear dependence of $g_c$ on $N_{part}$.
Thermal charm production as well as charm quark-antiquark annihilation in Pb-Pb collisions are neglected, as they were estimated to be very small at the LHC energies and negligible for lower energies as demonstrated in~(\cite{Andronic:2006ky,Braun-Munzinger:2000uqj}).

With the assumption of the kinetic  freeze-out for hadrons with charm quarks taking place also at the QCD phase boundary and employing hydrodynamics, the transverse momentum distributions can be calculated as well as established in~\cite{Andronic:2023ioz}. A corona contribution from the dilute periphery of the fireball is added, both for the total and the $\pT$-differential yields, based on measurements in pp collisions.

\begin{figure}[hbt]
\begin{tabular}{cc}  \begin{minipage}{.5\textwidth}
{\includegraphics[width=.98\textwidth]{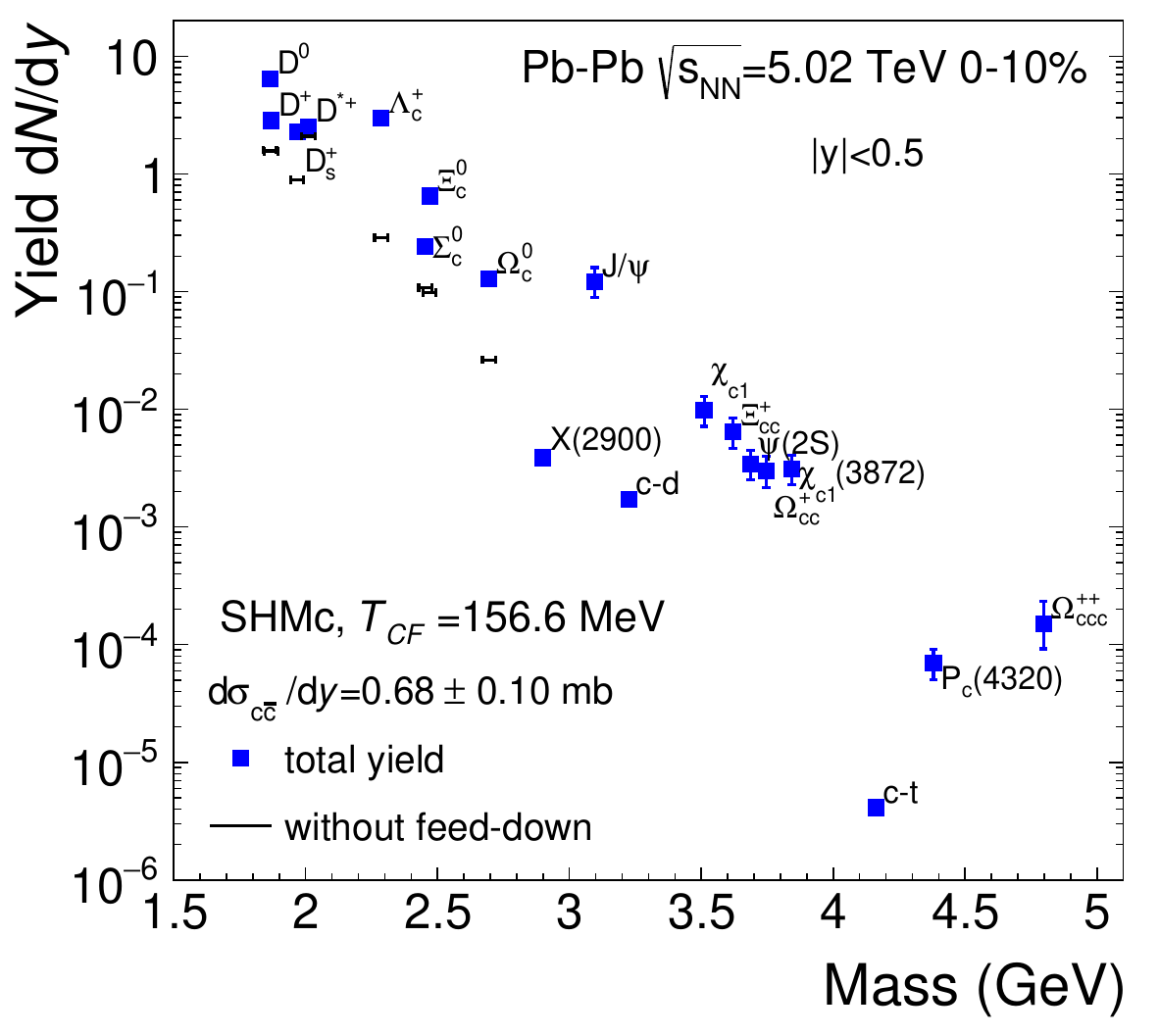}}
\end{minipage} & \begin{minipage}{.5\textwidth}
{\includegraphics[width=.94\textwidth]{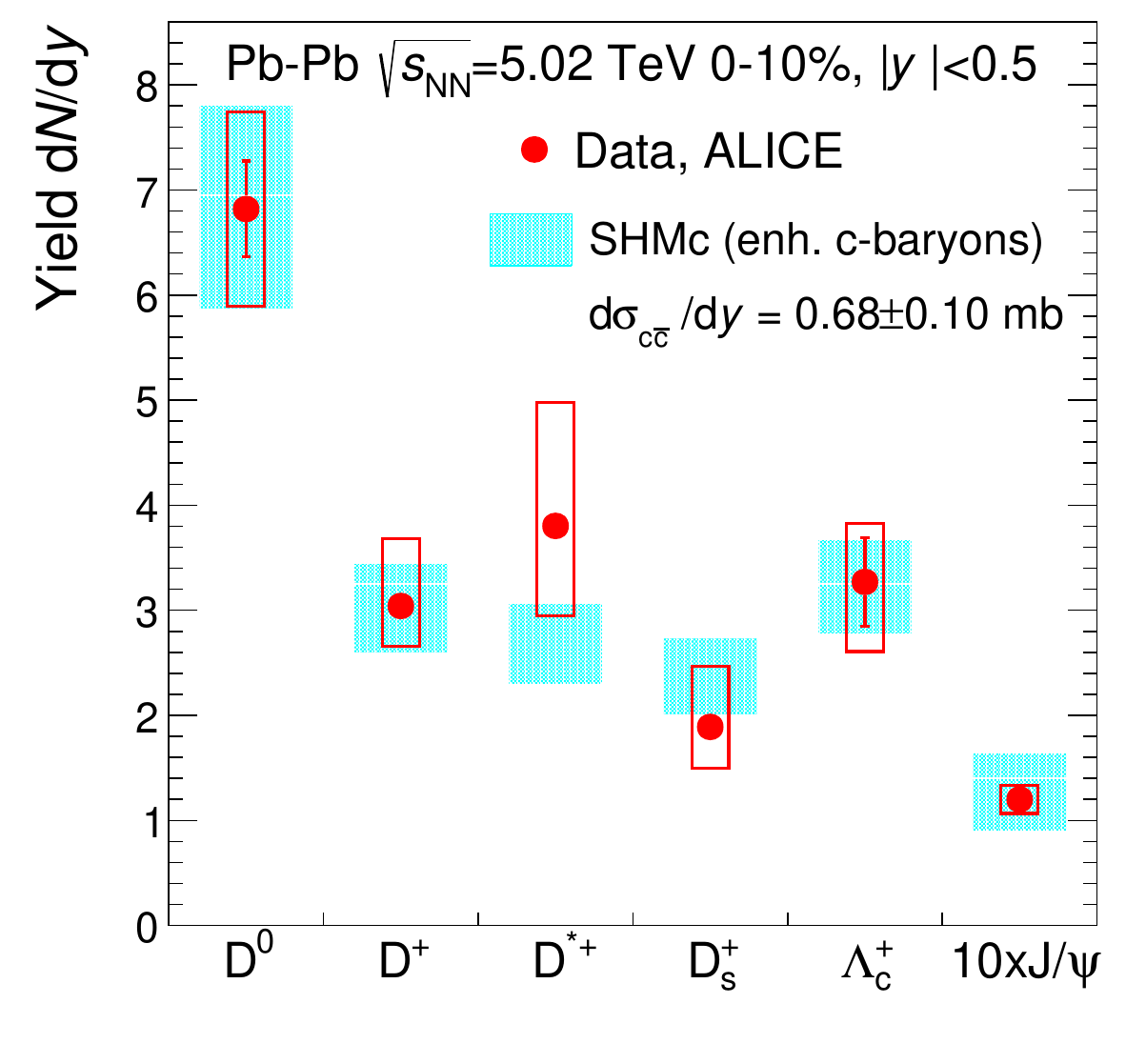}}
\end{minipage} \end{tabular}
\caption{Left panel: the SHMc predictions for production yields of a broad range of charm hadron species for central (0-10\%) Pb--Pb collisions (for the case of enhanced charmed-baryon spectrum, see text; $\frac{\ud N_{c\bar{c}}}{\ud y}=16.3\pm2.4$, $g_c$=31.5). The yields without the contribution from feed-down from decays of resonances are shown for the most-relevant species. Right panel: the comparison of SHMc predictions to experimental data from ALICE~(\cite{ALICE:2021rxa,ALICE:2021kfc,ALICE:2021bib,ALICE:2023gco}).
}\label{fig:charm}
\end{figure}

The model is developed to predict the full suite of charm hadron species as seen in Fig.~\ref{fig:charm} (left), from the common mesons and baryons to exotic charmonia like the $\chi_{c1}(3872)$, the pentaquark $P_c(4320)$, and the multiple-charm baryons, all key ALICE physics goals for LHC Runs 4-5. Interestingly, also nuclei containing charm quarks come into reach at the high luminosity phase of the LHC. The predicted (see ~\cite{Andronic:2021erx}) production cross sections are included in Fig.~\ref{fig:charm} for the charm-deuteron (c-d) and charm-triton (c-t), should they be bound. Not shown in the plot, the $T_{cc}^+$ state is predicted with a yield of about 90\% that of $\chi_{c1}(3872)$, while the $X(6900)$ (\ccbar\ccbar) state is predicted with a yield of about $10^{-8}$ per collision.

In ~\cite{Andronic:2019wva} it is demonstrated that, using SHMc, the measured yield for J/$\psi$ mesons is very well reproduced for all collision centralities along with the yield of all light-flavor hadrons. The uncertainty in the prediction is mainly caused by the uncertainty in the total charm cross section in Pb--Pb collisions. We note here that the excellent agreement of measured charmed hadron yields with those computed with the SHMc, Fig,~\ref{fig:charm} (right), implies that charm quarks, and consequently charmonia, are unbound inside the QGP; in fact their yields at full LHC energy exhibit enhancement compared to expectations using collision scaling from pp collisions and nuclear effects~(\cite{Klasen:2023uqj}), contrary to the original predictions based on ~\cite{Matsui:1986dk}. For a detailed discussion see ~\cite{Andronic:2017pug}.

To describe the measured yields of charmonia, feeding from excited charmonia can be neglected because these large-mass states are strongly Boltzmann suppressed. For the measured yields of open charm mesons and baryons, this is clearly not the case and feeding from excited $D^*$, $\Lambda_c^*$, and $\Sigma_c^*$ hadrons  is an essential ingredient for the successful description of open charm hadrons ~(\cite{Andronic:2021erx}), see Fig~\ref{fig:charm} (left). Even though the experimental mass spectrum of excited open charm hadrons is presumably not complete, in particular in the baryon sector, the prediction in the meson sector of yields of D-mesons compares very well with the measurements, both concerning transverse momentum $p_T$ and centrality dependence.
For $\Lambda_c$ baryons this is not the case and one has to augment the currently measured charm baryon spectrum to account for a large number of additional states predicted by LQCD~(\cite{Bazavov:2014yba}) and the quark model (as shown by~\cite{He:2019tik}) to achieve agreement with experimental data~(\cite{Andronic:2023ioz}), see Fig~\ref{fig:charm} (left). This leads to only a relatively small increase of the total charm production cross section to 0.68$\pm$0.10.

\begin{figure}[htb]
    \centering    \includegraphics[width=.47\linewidth,clip=true]{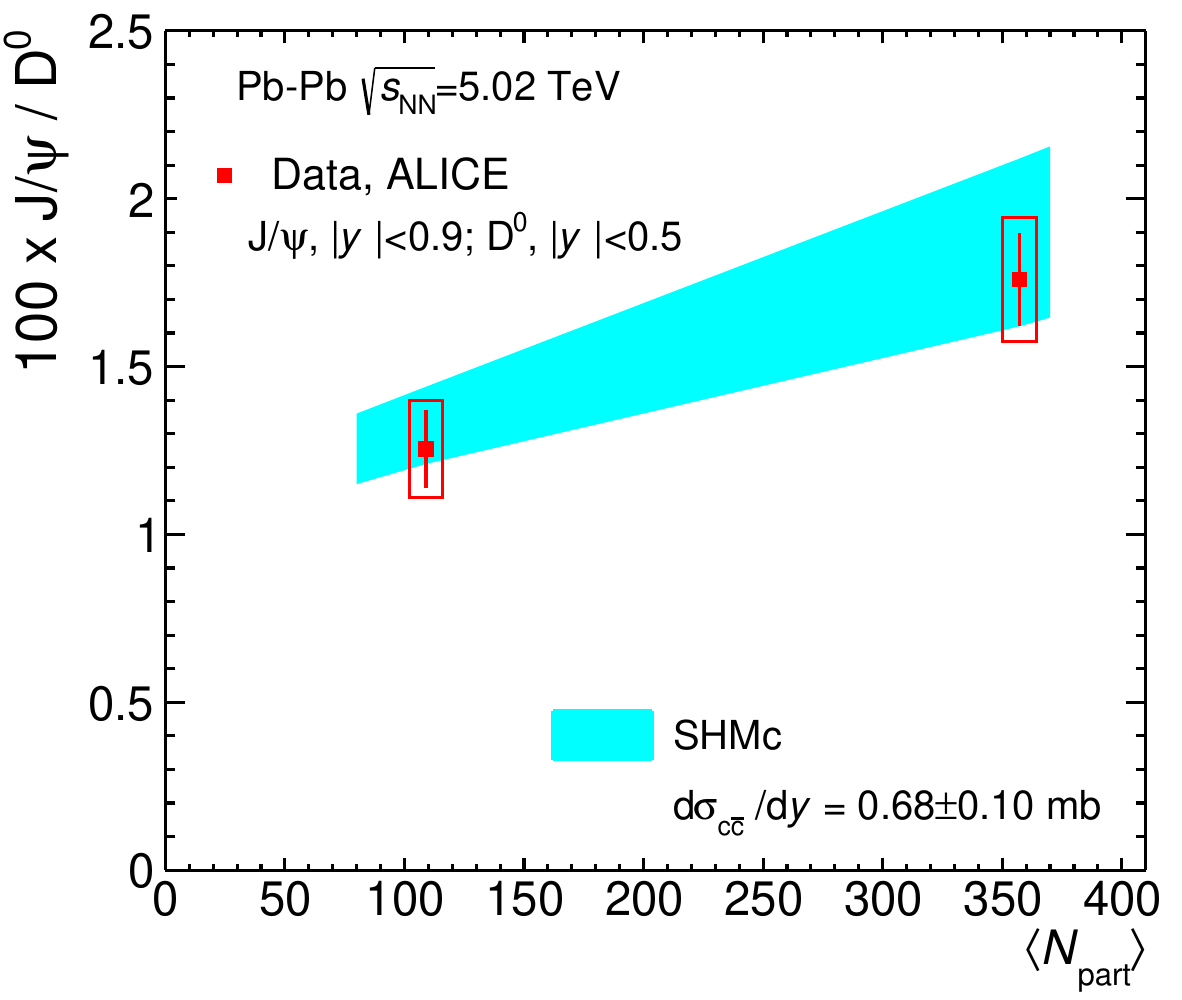} 
    \caption{The ratio of J/$\psi$ to $D^{0}$ midrapidity densities (in percent) measured in Pb--Pb collisions at the LHC by ALICE~(\cite{ALICE:2023gco,ALICE:2021rxa}) is compared to the prediction of SHMc ~\cite{Andronic:2021erx}. Note that the SHMc uncertainties, arising from the charm production cross section, are fully correlated over the \meanNp\ range.}    \label{fig:JpsitoD0}
\end{figure}

It is also interesting to make a precision study of charm yields as function of centrality or of the dependence on the number of nucleons participating in the collision, $\langle N_{part}\rangle$ . This is done in Fig.~\ref{fig:JpsitoD0} by looking at the data for Pb--Pb collisions. In this figure the $\langle N_{part}\rangle$ dependence of the measured rapidity density ratio $(J/\psi)/D^0$ is compared to the results of predictions of the SHMc. Recently, both the $D^0$ and $J/\psi$ production cross sections have been well measured down to $p_T$ = 0. The yield ratio $(J/\psi)/D_0$ is reproduced with very good precision for both measured centralities, as can be seen in Fig.~\ref{fig:JpsitoD0}. This ratio of hidden to open charm hadron yields clearly depends on the charm quarks fugacity $g_c$ and the excellent agreement between data and SHMc model predictions lends strong support to the assumption that both open and hidden charm states are produced from freely moving  charm and anticharm quarks that get bound into hadrons by statistical hadronization at the QCD phase boundary. Further comparisons between SHMc and data for open charm hadrons are presented  in~(\cite{Andronic:2021erx,ALICE:2022wpn}).

One may draw a number of important conclusions from this successful comparison of measured yields for the production of (u,d,s) as well as open and hidden charm hadrons obtained from  SHM or SHMc predictions. The temperature parameter here is obtained from a fit of the SHM model predictions to measured yields of (u,d,s) hadrons and agrees within uncertainties with LQCD predictions for the chiral pseudo-critical temperature ~(\cite{Bazavov:2018mes,Borsanyi:2020fev}) . 
\begin{itemize}
\item 
From the analysis of hadron production in relativistic nuclear collisions  on concludes that the data are  described quantitatively by the chemical freeze-out parameters  ($T_{CF}, \mu_{B}$). The fireball volume appearing in the partition function is determined by normalization to the measured number of primary charged particles. For collision energies $\sqrt{s_{NN}}\geq$ 10 GeV these freeze-out parameters agree with good precision with the results from LQCD for the location of the chiral cross over transition. Unfortunately no LQCD predictions exist for small energies. 

These results imply that, within experimental precision,  hadronization is independent of particle species and only dependent on the values of $T$ and $\mu_B$ at the phase boundary. At LHC energy, the chemical potentials vanish, and only $T = T_{c}$ is needed to describe hadronization. In the thermal limit, hadronization is universal.
\item
The mechanism implemented in the SHMc for the production of charmed hadrons implies that these particles are produced from uncorrelated, thermalized charm quarks as  expected for a strongly coupled, deconfined QGP (see also the discussion in ~\cite{Andronic:2021erx}).  At full LHC energy, where chemical freeze-out takes place for central Pb--Pb collisions in a volume per unit rapidity of $V \approx 5000 $ fm$^3$, this implies that charm quarks can travel over linear distances of order 10 fm (see ~\cite{Andronic:2017pug,Andronic:2021erx} for more detail). This result provides strong evidence for deconfinement in the charm sector.
Some recent theoretical studies have been interpreted as an indication
for the presence of a confined (gluon-less) phase at temperatures above the hadron-resonance gas phase but below the QGP phase~(\cite{Cohen:2023hbq,Fujimoto:2025sxx}). The above observation of deconfined charm quarks close to $T_c = 157$ MeV does not lend support for the existence of such a gluon-less  confined phase. Also, the J/$\psi$ meson cannot be formed from colored charm and anti-charm quarks in absence of gluons.
\end{itemize}

Future measurement campaigns at the LHC will yield detailed information on the production cross sections of hadrons with multiple charm quarks as well as excited charmonia. The predictions from the SHMc for the relevant cross sections exhibit a rather dramatic hie\-rarchy of enhancements, see~\cite{Andronic:2021erx} for such processes. Expe\-ri\-men\-tal tests of these predictions would lead to a fundamental understanding of confinement/deconfinement and hadronization. The vision is to obtain, from the measured charmonium spectrum compared to SHMc, a deconfinement temperature similar in spirit to the above cited freeze-out temperature for nuclei.

The SHM was applied to the bottom sector too, see ~(\cite{Andronic:2006ky,Andronic:2022ucg}), although in this case incomplete thermalization of bottom quarks needs to be considered. 

Very recent investigations have shown that the process underlying the SHMc can be successfully applied also in the beauty-quark sector ~\cite{Wu:2025lcj}. In this SHMb approach, generation (formation) of bottom hadrons takes in the QGP or  at the phase boundary. Further investigations are needed to understand how to deal with the apparent partial equilibration of b-quarks.

\section{Experimental plans and opportunities for the next decades} \label{sect:future}

A vigorous research program with ultra-relativistic nuclear collisions will continue at least into the 2040ties. The LHC at CERN will continue to run up to 2041, and with the upgrades underway for all 4 large experiments (ALICE, ATLAS, CMS, LHCb) a world-wide unique physics program will be conducted there by colliding nuclear beams at the highest energy available today. 
At the same time, there are exciting new opportunities for quark matter research focusing on much lower energy collisions at new high intensity accelerators such as FAIR at the German GSI Helmholtz-Center and the NICA Facility at JINR/Dubna in Russia. 
Also, research with relativistic nuclear beams will continue at the SPS facility, one of the injectors for the LHC at CERN. In addition, new facilities for high energy nuclear collisions are coming on-line: HIAF at Huizhou,  China and nuclear beams at J-PARC, the Japan Proton Accelerator Research Complex at Tokai, Japan. 
We look forward to rich new physics programs with relativistic nuclear collisions at these facilities.

\section{Summary} \label{sect:summary}

Experiments with ultra-relativistic nuclear collisions have provided a rich harvest of insights into the physics of the QGP, as well as important information on the interaction between hadrons, on the parton structure of nucleons and nuclei, and recently also on nuclear structure. We have focused in the present review on the significant improvement of our understanding of QGP properties from the detailed and increasingly precise measurement of hadron production in collisions between heavy atomic nuclei at center-of-mass energies ranging from a few GeV to 5 TeV per colliding nucleon pair. At all but possibly the lowest energies a hot fireball is formed with energy densities significantly exceeding predictions from first-principles lattice QCD calculations for a temperature near the phase boundary.

Analysis of the data over the full energy range in terms of the statistical hadronization model SHM has led to the establishment of a 'chemical freeze-out' line in the $T - \mub$ phase diagram. This freeze-out line closely coincides with the QGP phase boundary established from lattice QCD calculations for center-of-mass energies $\sqrt{s_{NN}}$ above 10~GeV, lending strong support to the interpretation that this LQCD prediction for the phase boundary is experimentally confirmed.

Focusing at the highest energies investigated at the LHC, where the chemical potential \mub, a measure of the asymmetry between produced matter and anti-matter vanishes, the experimentally determined pseudo-critical temperature $T_{pc}^{exp} = 156.6 \pm 1.7$ MeV. The uncertainty here is determined by the uncertainties of the measurements.

The excellent agreement between SHM calculations and the measured yields of more than 20 hadron species at LHC energy also implies that the thermal approach underlying the SHM accurately describes full hadronization pattern with one single parameter, the temperature $T_{pc}^{exp}$,  lending support to the notion that at the thermal limit hadronization is universal (independent of the quark flavor content of the hadrons).
Moreover, the remarkable description of the absolute yields over the whole range of collision energies implies a universal hadron production mechanism at chemical freeze-
out, despite vastly different fireball properties at the lowest and highest ends of the collision range.
Furthermore, SHMc analysis of the yields of closed-charm and open-charm hadrons provides first evidence for the deconfinement of charm quarks inside the hot LHC fireball.

There are also a number of open questions. One such question is whether there could be a flavor dependence of the chemical freeze-out temperature. For hadrons composed of (u,d,s) quarks no such dependence has been observed. It would hence be very important to establish a chemical freeze-out temperature exclusively from an analysis of open and hidden-charm hadron yields. This could not be achieved today since it involves the measurement of a number of such hadrons down to  vanishing transverse momentum. This has been possible up to now only for J/$\psi$, $\psi'$ and D$^0$ mesons but new vertex detectors for ALICE Run 4 and Run 5 promise progress here. 

Another open question is the mechanism of the formation of loosely bound nuclei such as the deuteron or the iconical hyper-triton, where the size of the bound state is equal to or even larger than the size of the entire fireball formed in the collision. There are pragmatic approaches based on simple criteria for bound state formation (such as the coalescence model) or direct use of the (parameter-free) SHM. Both work well for deuteron production but the case of hyper-triton is not conclusive yet. Progress in the theoretical understanding of bound state formation processes and, in particular, bound state formation times  would be very important.

At lower energies, the Beam Energy Scan program at the RHIC accelerator at Brookhaven has yielded exciting new data for Au--Au collisions in the energy range $3 < \sqrt{ s_{NN}} < 60$ GeV. The most important results of this experimental campaign are fluctuation measurements for net proton distributions, which were measured up to the sixth order and demonstrated the presence of unexpected structure as function of energy. Unfortunately, the precision of the data is not sufficient to make conclusive statements regarding a possible critical endpoint in the QCD phase diagram. Meanwhile, RHIC  was just discontinued to make room for an entirely new accelerator, the electron-ion collider EIC. But we look forward to the full analysis of the most recent data from the STAR experiment with potentially new insights.

At the EIC, high energy electrons will be  brought into collision with high energy protons or atomic nuclei to study the structure of the quarks and gluons inside nuclei. The very important topic of the existence of a critical endpoint will hence have to be taken up by the CBM experiment at FAIR which will cover just the right energy range for such a search and start running at the end of the present decade. 

\vspace{.5cm}
\begin{ack}[Acknowledgments]

  K.R. acknowledges support from the National Science Centre (NCN), Poland, under OPUS Grant No. 2022/45/B/ST2/01527, and from the Polish Ministry of Science and Higher Education. This work is part of and supported by the DFG Collaborative Research Centre, SFB1225/\ ISOQUANT.

\end{ack}

\bibliography{areview}

\begin{thebibliography*}{162}
\providecommand{\bibtype}[1]{}
\providecommand{\natexlab}[1]{#1}
{\catcode`\|=0\catcode`\#=12\catcode`\@=11\catcode`\\=12
|immediate|write|@auxout{\expandafter\ifx\csname
  natexlab\endcsname\relax\gdef\natexlab#1{#1}\fi}}
\renewcommand{\url}[1]{{\tt #1}}
\providecommand{\urlprefix}{URL }
\expandafter\ifx\csname urlstyle\endcsname\relax
  \providecommand{\doi}[1]{doi:\discretionary{}{}{}#1}\else
  \providecommand{\doi}{doi:\discretionary{}{}{}\begingroup
  \urlstyle{rm}\Url}\fi
\providecommand{\bibinfo}[2]{#2}
\providecommand{\eprint}[2][]{\url{#2}}

\bibtype{Article}%
\bibitem[Aarts et al.(2017)]{Aarts:2016hap}
\bibinfo{author}{Aarts G} and  et al. (\bibinfo{year}{2017}).
\bibinfo{title}{{Heavy-flavor production and medium properties in high-energy
  nuclear collisions - What next?}}
\bibinfo{journal}{{\em Eur. Phys. J. A}} \bibinfo{volume}{53}
  (\bibinfo{number}{5}): \bibinfo{pages}{93}.
  \bibinfo{doi}{\doi{10.1140/epja/i2017-12282-9}}.
\eprint{1612.08032}.

\bibtype{Article}%
\bibitem[Abbas et al.(2013)]{ALICE:2013xna}
\bibinfo{author}{Abbas E} and  et al. (\bibinfo{collaboration}{ALICE})
  (\bibinfo{year}{2013}).
\bibinfo{title}{{J/$\psi$ elliptic flow in Pb--Pb Collisions at $\sqrt{s_{_{\rm
  NN}}}$=2.76 TeV}}.
\bibinfo{journal}{{\em Phys. Rev. Lett.}} \bibinfo{volume}{111}:
  \bibinfo{pages}{162301}. \bibinfo{doi}{\doi{10.1103/PhysRevLett.111.162301}}.
\eprint{1303.5880}.

\bibtype{Article}%
\bibitem[Abdallah et al.(2022{\natexlab{a}})]{STAR:2021orx}
\bibinfo{author}{Abdallah M} and  et al. (\bibinfo{collaboration}{STAR})
  (\bibinfo{year}{2022}{\natexlab{a}}).
\bibinfo{title}{{Measurements of $H_\Lambda^3$ and $H_\Lambda^4$ Lifetimes and
  Yields in Au+Au Collisions in the High Baryon Density Region}}.
\bibinfo{journal}{{\em Phys. Rev. Lett.}} \bibinfo{volume}{128}
  (\bibinfo{number}{20}): \bibinfo{pages}{202301}.
  \bibinfo{doi}{\doi{10.1103/PhysRevLett.128.202301}}.
\eprint{2110.09513}.

\bibtype{Article}%
\bibitem[Abdallah et al.(2022{\natexlab{b}})]{STAR:2021hyx}
\bibinfo{author}{Abdallah MS} and  et al. (\bibinfo{collaboration}{STAR})
  (\bibinfo{year}{2022}{\natexlab{b}}).
\bibinfo{title}{{Probing strangeness canonical ensemble with K{\ensuremath{-}},
  {\ensuremath{\phi}}(1020) and {\ensuremath{\Xi}}{\ensuremath{-}} production
  in Au+Au collisions at sNN=3 GeV}}.
\bibinfo{journal}{{\em Phys. Lett. B}} \bibinfo{volume}{831}:
  \bibinfo{pages}{137152}. \bibinfo{doi}{\doi{10.1016/j.physletb.2022.137152}}.
\eprint{2108.00924}.

\bibtype{Article}%
\bibitem[Abdulhamid et al.(2024{\natexlab{a}})]{STAR:2023uxk}
\bibinfo{author}{Abdulhamid M} and  et al. (\bibinfo{collaboration}{STAR})
  (\bibinfo{year}{2024}{\natexlab{a}}).
\bibinfo{title}{{Production of protons and light nuclei in Au+Au collisions at
  $\sqrt{s_{\textrm{NN}}}$=3 GeV with the STAR detector}}.
\bibinfo{journal}{{\em Phys. Rev. C}} \bibinfo{volume}{110}
  (\bibinfo{number}{5}): \bibinfo{pages}{054911}.
  \bibinfo{doi}{\doi{10.1103/PhysRevC.110.054911}}.
\eprint{2311.11020}.

\bibtype{Article}%
\bibitem[Abdulhamid et al.(2024{\natexlab{b}})]{STAR:2024znc}
\bibinfo{author}{Abdulhamid MI} and  et al. (\bibinfo{collaboration}{STAR})
  (\bibinfo{year}{2024}{\natexlab{b}}).
\bibinfo{title}{{Strangeness production in $ \sqrt{s_{\textrm{NN}}}$ = 3 GeV
  Au+Au collisions at RHIC}}.
\bibinfo{journal}{{\em JHEP}} \bibinfo{volume}{10}: \bibinfo{pages}{139}.
  \bibinfo{doi}{\doi{10.1007/JHEP10(2024)139}}.
\eprint{2407.10110}.

\bibtype{Article}%
\bibitem[Abe et al.(1988)]{Abe:1988hq}
\bibinfo{author}{Abe K} and  et al. (\bibinfo{year}{1988}).
\bibinfo{title}{{Leading Particle Distributions in 200-{GeV}/c $P$ + a
  Interactions}}.
\bibinfo{journal}{{\em Phys. Lett. B}} \bibinfo{volume}{200}:
  \bibinfo{pages}{266--271}. \bibinfo{doi}{\doi{10.1016/0370-2693(88)90769-1}}.

\bibtype{Article}%
\bibitem[Abelev et al.(2009)]{Abelev:2008ab}
\bibinfo{author}{Abelev B} and  et al. (\bibinfo{collaboration}{STAR})
  (\bibinfo{year}{2009}).
\bibinfo{title}{{Systematic Measurements of Identified Particle Spectra in $p
  p, d^+$ Au and Au+Au Collisions from STAR}}.
\bibinfo{journal}{{\em Phys. Rev. C}} \bibinfo{volume}{79}:
  \bibinfo{pages}{034909}. \bibinfo{doi}{\doi{10.1103/PhysRevC.79.034909}}.
\eprint{0808.2041}.

\bibtype{Article}%
\bibitem[Abelev et al.(2010)]{Abelev:2009bw}
\bibinfo{author}{Abelev B} and  et al. (\bibinfo{collaboration}{STAR})
  (\bibinfo{year}{2010}).
\bibinfo{title}{{Identified particle production, azimuthal anisotropy, and
  interferometry measurements in Au+Au collisions at $\sqrt{s_{_{\rm NN}}}=9.2$
  GeV}}.
\bibinfo{journal}{{\em Phys. Rev. C}} \bibinfo{volume}{81}:
  \bibinfo{pages}{024911}. \bibinfo{doi}{\doi{10.1103/PhysRevC.81.024911}}.
\eprint{0909.4131}.

\bibtype{Article}%
\bibitem[Abelev et al.(2012)]{Abelev:2012wca}
\bibinfo{author}{Abelev B} and  et al. (\bibinfo{collaboration}{ALICE})
  (\bibinfo{year}{2012}).
\bibinfo{title}{{Pion, Kaon, and Proton Production in Central Pb--Pb Collisions
  at $\sqrt{s_{_{\rm NN}}}=2.76$ TeV}}.
\bibinfo{journal}{{\em Phys. Rev. Lett.}} \bibinfo{volume}{109}:
  \bibinfo{pages}{252301}. \bibinfo{doi}{\doi{10.1103/PhysRevLett.109.252301}}.
\eprint{1208.1974}.

\bibtype{Article}%
\bibitem[Abelev et al.(2013{\natexlab{a}})]{Abelev:2013vea}
\bibinfo{author}{Abelev B} and  et al. (\bibinfo{collaboration}{ALICE})
  (\bibinfo{year}{2013}{\natexlab{a}}).
\bibinfo{title}{{Centrality dependence of $\pi$, K, p production in Pb-Pb
  collisions at $\sqrt{s_{_{\rm NN}}}=2.76$ TeV}}.
\bibinfo{journal}{{\em Phys. Rev. C}} \bibinfo{volume}{88}:
  \bibinfo{pages}{044910}. \bibinfo{doi}{\doi{10.1103/PhysRevC.88.044910}}.
\eprint{1303.0737}.

\bibtype{Article}%
\bibitem[Abelev et al.(2013{\natexlab{b}})]{Abelev:2013xaa}
\bibinfo{author}{Abelev BB} and  et al. (\bibinfo{collaboration}{ALICE})
  (\bibinfo{year}{2013}{\natexlab{b}}).
\bibinfo{title}{{$K^0_S$ and $\Lambda$ production in Pb-Pb collisions at
  $\sqrt{s_{_{\rm NN}}}=2.76$ TeV}}.
\bibinfo{journal}{{\em Phys. Rev. Lett.}} \bibinfo{volume}{111}:
  \bibinfo{pages}{222301}. \bibinfo{doi}{\doi{10.1103/PhysRevLett.111.222301}}.
\eprint{1307.5530}.

\bibtype{Article}%
\bibitem[Abelev et al.(2014)]{ABELEV:2013zaa}
\bibinfo{author}{Abelev BB} and  et al. (\bibinfo{collaboration}{ALICE})
  (\bibinfo{year}{2014}).
\bibinfo{title}{{Multi-strange baryon production at mid-rapidity in Pb-Pb
  collisions at $\sqrt{s_{_{\rm NN}}}=2.76$ TeV}}.
\bibinfo{journal}{{\em Phys. Lett. B}} \bibinfo{volume}{728}:
  \bibinfo{pages}{216--227}.
  \bibinfo{doi}{\doi{10.1016/j.physletb.2013.11.048}}.
\eprint{1307.5543}.

\bibtype{Article}%
\bibitem[Abelev et al.(2015)]{Abelev:2014uua}
\bibinfo{author}{Abelev BB} and  et al. (\bibinfo{collaboration}{ALICE})
  (\bibinfo{year}{2015}).
\bibinfo{title}{{$K^*(892)^0$ and $\phi(1020)$ production in Pb-Pb collisions
  at $\sqrt{s_{\rm NN}}$ = 2.76 TeV}}.
\bibinfo{journal}{{\em Phys. Rev. C}} \bibinfo{volume}{91}:
  \bibinfo{pages}{024609}. \bibinfo{doi}{\doi{10.1103/PhysRevC.91.024609}}.
\eprint{1404.0495}.

\bibtype{Article}%
\bibitem[Aboona et al.(2026)]{STAR:2025xxf}
\bibinfo{author}{Aboona BE} and  et al. (\bibinfo{collaboration}{STAR})
  (\bibinfo{year}{2026}).
\bibinfo{title}{{Identified charged hadron production in Au+Au collisions at
  $\sqrt{s_{\textrm{NN}}}$=54.4 GeV with the STAR detector}}.
\bibinfo{journal}{{\em Phys. Rev. C}} \bibinfo{volume}{113}
  (\bibinfo{number}{5}): \bibinfo{pages}{054907}.
  \bibinfo{doi}{\doi{10.1103/4161-dflc}}.
\eprint{2512.06415}.

\bibtype{Article}%
\bibitem[Abou~Yassine et al.(2026)]{HADES:2025qum}
\bibinfo{author}{Abou~Yassine R} and  et al. (\bibinfo{collaboration}{HADES})
  (\bibinfo{year}{2026}).
\bibinfo{title}{{Formation and lifetime measurements of light hypernuclei in
  Ag+Ag collisions at sNN = 2.55 GeV}}.
\bibinfo{journal}{{\em Phys. Lett. B}} \bibinfo{volume}{880}:
  \bibinfo{pages}{140745}. \bibinfo{doi}{\doi{10.1016/j.physletb.2026.140745}}.
\eprint{2512.12454}.

\bibtype{Article}%
\bibitem[Acharya et al.(2018)]{Acharya:2017bso}
\bibinfo{author}{Acharya S} and  et al. (\bibinfo{collaboration}{ALICE})
  (\bibinfo{year}{2018}).
\bibinfo{title}{{Production of $^{4}$He and $^{4}\overline{\textrm{He}}$ in
  Pb-Pb collisions at $\sqrt{s_{\mathrm{NN}}}$ = 2.76 TeV at the LHC}}.
\bibinfo{journal}{{\em Nucl. Phys. A}} \bibinfo{volume}{971}:
  \bibinfo{pages}{1--20}. \bibinfo{doi}{\doi{10.1016/j.nuclphysa.2017.12.004}}.
\eprint{1710.07531}.

\bibtype{Article}%
\bibitem[Acharya et al.(2022{\natexlab{a}})]{ALICE:2021kfc}
\bibinfo{author}{Acharya S} and  et al. (\bibinfo{collaboration}{ALICE})
  (\bibinfo{year}{2022}{\natexlab{a}}).
\bibinfo{title}{{Measurement of prompt $D_s^+$-meson production and azimuthal
  anisotropy in Pb{\textendash}Pb collisions at $\sqrt {s_{NN}}$=5.02TeV}}.
\bibinfo{journal}{{\em Phys. Lett. B}} \bibinfo{volume}{827}:
  \bibinfo{pages}{136986}. \bibinfo{doi}{\doi{10.1016/j.physletb.2022.136986}}.
\eprint{2110.10006}.

\bibtype{Article}%
\bibitem[Acharya et al.(2022{\natexlab{b}})]{ALICE:2021rxa}
\bibinfo{author}{Acharya S} and  et al. (\bibinfo{collaboration}{ALICE})
  (\bibinfo{year}{2022}{\natexlab{b}}).
\bibinfo{title}{{Prompt D$^{0}$, D$^{+}$, and D$^{*+}$ production in
  Pb{\textendash}Pb collisions at $ \sqrt{s_{\mathrm{NN}}} $ = 5.02 TeV}}.
\bibinfo{journal}{{\em JHEP}} \bibinfo{volume}{01}: \bibinfo{pages}{174}.
  \bibinfo{doi}{\doi{10.1007/JHEP01(2022)174}}.
\eprint{2110.09420}.

\bibtype{Article}%
\bibitem[Acharya et al.(2023)]{ALICE:2021bib}
\bibinfo{author}{Acharya S} and  et al. (\bibinfo{collaboration}{ALICE})
  (\bibinfo{year}{2023}).
\bibinfo{title}{{Constraining hadronization mechanisms with
  {\ensuremath{\Lambda}}c+/D0 production ratios in Pb{\textendash}Pb collisions
  at sNN=5.02 TeV}}.
\bibinfo{journal}{{\em Phys. Lett. B}} \bibinfo{volume}{839}:
  \bibinfo{pages}{137796}. \bibinfo{doi}{\doi{10.1016/j.physletb.2023.137796}}.
\eprint{2112.08156}.

\bibtype{Article}%
\bibitem[Acharya et al.(2024{\natexlab{a}})]{ALICE:2023ulv}
\bibinfo{author}{Acharya S} and  et al. (\bibinfo{collaboration}{ALICE})
  (\bibinfo{year}{2024}{\natexlab{a}}).
\bibinfo{title}{{Measurements of Chemical Potentials in Pb-Pb Collisions at
  sNN=5.02{\,}{\,}TeV}}.
\bibinfo{journal}{{\em Phys. Rev. Lett.}} \bibinfo{volume}{133}
  (\bibinfo{number}{9}): \bibinfo{pages}{092301}.
  \bibinfo{doi}{\doi{10.1103/PhysRevLett.133.092301}}.
\eprint{2311.13332}.

\bibtype{Article}%
\bibitem[Acharya et al.(2024{\natexlab{b}})]{ALICE:2023gco}
\bibinfo{author}{Acharya S} and  et al. (\bibinfo{collaboration}{ALICE})
  (\bibinfo{year}{2024}{\natexlab{b}}).
\bibinfo{title}{{Measurements of inclusive J/{\ensuremath{\psi}} production at
  midrapidity and forward rapidity in Pb{\textendash}Pb collisions at sNN =
  5.02 TeV}}.
\bibinfo{journal}{{\em Phys. Lett. B}} \bibinfo{volume}{849}:
  \bibinfo{pages}{138451}. \bibinfo{doi}{\doi{10.1016/j.physletb.2024.138451}}.
\eprint{2303.13361}.

\bibtype{Article}%
\bibitem[Acharya et al.(2024{\natexlab{c}})]{ALICE:2022wpn}
\bibinfo{author}{Acharya S} and  et al. (\bibinfo{collaboration}{ALICE})
  (\bibinfo{year}{2024}{\natexlab{c}}).
\bibinfo{title}{{The ALICE experiment: a journey through QCD}}.
\bibinfo{journal}{{\em Eur. Phys. J. C}} \bibinfo{volume}{84}
  (\bibinfo{number}{8}): \bibinfo{pages}{813}.
  \bibinfo{doi}{\doi{10.1140/epjc/s10052-024-12935-y}}.
\eprint{2211.04384}.

\bibtype{Article}%
\bibitem[Acharya et al.(2025{\natexlab{a}})]{ALICE:2024djx}
\bibinfo{author}{Acharya S} and  et al. (\bibinfo{collaboration}{ALICE})
  (\bibinfo{year}{2025}{\natexlab{a}}).
\bibinfo{title}{{First Measurement of $A$ = 4 Hypernuclei and Antihypernuclei
  at the LH}}.
\bibinfo{journal}{{\em Phys. Rev. Lett.}} \bibinfo{volume}{134}
  (\bibinfo{number}{16}): \bibinfo{pages}{162301}.
  \bibinfo{doi}{\doi{10.1103/PhysRevLett.134.162301}}.
\eprint{2410.17769}.

\bibtype{Article}%
\bibitem[Acharya et al.(2025{\natexlab{b}})]{ALICE:2024koa}
\bibinfo{author}{Acharya S} and  et al. (\bibinfo{collaboration}{ALICE})
  (\bibinfo{year}{2025}{\natexlab{b}}).
\bibinfo{title}{{Measurement of H{\ensuremath{\Lambda}}3 production in
  Pb{\textendash}Pb collisions at sNN=5.02 TeV}}.
\bibinfo{journal}{{\em Phys. Lett. B}} \bibinfo{volume}{860}:
  \bibinfo{pages}{139066}. \bibinfo{doi}{\doi{10.1016/j.physletb.2024.139066}}.
\eprint{2405.19839}.

\bibtype{Article}%
\bibitem[Adam et al.(2016{\natexlab{a}})]{Adam:2015yta}
\bibinfo{author}{Adam J} and  et al. (\bibinfo{collaboration}{ALICE})
  (\bibinfo{year}{2016}{\natexlab{a}}).
\bibinfo{title}{{$^{3}_{\Lambda}\mathrm H$ and $^{3}_{\bar{\Lambda}}
  \overline{\mathrm H}$ production in Pb-Pb collisions at $\sqrt{s_{\rm NN}} =$
  2.76 TeV}}.
\bibinfo{journal}{{\em Phys. Lett. B}} \bibinfo{volume}{754}:
  \bibinfo{pages}{360--372}.
  \bibinfo{doi}{\doi{10.1016/j.physletb.2016.01.040}}.
\eprint{1506.08453}.

\bibtype{Article}%
\bibitem[Adam et al.(2016{\natexlab{b}})]{ALICE:2015oer}
\bibinfo{author}{Adam J} and  et al. (\bibinfo{collaboration}{ALICE})
  (\bibinfo{year}{2016}{\natexlab{b}}).
\bibinfo{title}{{$^{3}_{\Lambda}\mathrm H$ and $^{3}_{\bar{\Lambda}}
  \overline{\mathrm H}$ production in Pb-Pb collisions at $\sqrt{s_{\rm NN}} =$
  2.76 TeV}}.
\bibinfo{journal}{{\em Phys. Lett. B}} \bibinfo{volume}{754}:
  \bibinfo{pages}{360--372}.
  \bibinfo{doi}{\doi{10.1016/j.physletb.2016.01.040}}.
\eprint{1506.08453}.

\bibtype{Article}%
\bibitem[Adam et al.(2016{\natexlab{c}})]{ALICE:2016igk}
\bibinfo{author}{Adam J} and  et al. (\bibinfo{collaboration}{ALICE})
  (\bibinfo{year}{2016}{\natexlab{c}}).
\bibinfo{title}{{Measurement of transverse energy at midrapidity in Pb-Pb
  collisions at $\sqrt{s_{\rm NN}} = 2.76$ TeV}}.
\bibinfo{journal}{{\em Phys. Rev. C}} \bibinfo{volume}{94}
  (\bibinfo{number}{3}): \bibinfo{pages}{034903}.
  \bibinfo{doi}{\doi{10.1103/PhysRevC.94.034903}}.
\eprint{1603.04775}.

\bibtype{Article}%
\bibitem[Adam et al.(2016{\natexlab{d}})]{Adam:2015vda}
\bibinfo{author}{Adam J} and  et al. (\bibinfo{collaboration}{ALICE})
  (\bibinfo{year}{2016}{\natexlab{d}}).
\bibinfo{title}{{Production of light nuclei and anti-nuclei in pp and Pb-Pb
  collisions at energies available at the CERN Large Hadron Collider}}.
\bibinfo{journal}{{\em Phys. Rev. C}} \bibinfo{volume}{93}
  (\bibinfo{number}{2}): \bibinfo{pages}{024917}.
  \bibinfo{doi}{\doi{10.1103/PhysRevC.93.024917}}.
\eprint{1506.08951}.

\bibtype{Article}%
\bibitem[Adam et al.(2017{\natexlab{a}})]{ALICE:2016fbt}
\bibinfo{author}{Adam J} and  et al. (\bibinfo{collaboration}{ALICE})
  (\bibinfo{year}{2017}{\natexlab{a}}).
\bibinfo{title}{{Centrality dependence of the pseudorapidity density
  distribution for charged particles in Pb-Pb collisions at $\sqrt{s_{\rm
  NN}}=5.02$ TeV}}.
\bibinfo{journal}{{\em Phys. Lett. B}} \bibinfo{volume}{772}:
  \bibinfo{pages}{567--577}.
  \bibinfo{doi}{\doi{10.1016/j.physletb.2017.07.017}}.
\eprint{1612.08966}.

\bibtype{Article}%
\bibitem[Adam et al.(2017{\natexlab{b}})]{ALICE:2017jyt}
\bibinfo{author}{Adam J} and  et al. (\bibinfo{collaboration}{ALICE})
  (\bibinfo{year}{2017}{\natexlab{b}}).
\bibinfo{title}{{Enhanced production of multi-strange hadrons in
  high-multiplicity proton-proton collisions}}.
\bibinfo{journal}{{\em Nature Phys.}} \bibinfo{volume}{13}:
  \bibinfo{pages}{535--539}. \bibinfo{doi}{\doi{10.1038/nphys4111}}.
\eprint{1606.07424}.

\bibtype{Article}%
\bibitem[Adam et al.(2019)]{STAR:2019sjh}
\bibinfo{author}{Adam J} and  et al. (\bibinfo{collaboration}{STAR})
  (\bibinfo{year}{2019}).
\bibinfo{title}{{Beam energy dependence of (anti-)deuteron production in Au +
  Au collisions at the BNL Relativistic Heavy Ion Collider}}.
\bibinfo{journal}{{\em Phys. Rev. C}} \bibinfo{volume}{99}
  (\bibinfo{number}{6}): \bibinfo{pages}{064905}.
  \bibinfo{doi}{\doi{10.1103/PhysRevC.99.064905}}.
\eprint{1903.11778}.

\bibtype{Article}%
\bibitem[Adam et al.(2020)]{STAR:2019bjj}
\bibinfo{author}{Adam J} and  et al. (\bibinfo{collaboration}{STAR})
  (\bibinfo{year}{2020}).
\bibinfo{title}{{Strange hadron production in Au+Au collisions at
  $\sqrt{s_{_{\mathrm{NN}}}}$ = 7.7, 11.5, 19.6, 27, and 39 GeV}}.
\bibinfo{journal}{{\em Phys. Rev. C}} \bibinfo{volume}{102}
  (\bibinfo{number}{3}): \bibinfo{pages}{034909}.
  \bibinfo{doi}{\doi{10.1103/PhysRevC.102.034909}}.
\eprint{1906.03732}.

\bibtype{Article}%
\bibitem[Adamczewski-Musch et al.(2018)]{HADES:2017jgz}
\bibinfo{author}{Adamczewski-Musch J} and  et al.
  (\bibinfo{collaboration}{HADES}) (\bibinfo{year}{2018}).
\bibinfo{title}{{Deep sub-threshold $\phi$ production in Au+Au collisions}}.
\bibinfo{journal}{{\em Phys. Lett. B}} \bibinfo{volume}{778}:
  \bibinfo{pages}{403--407}.
  \bibinfo{doi}{\doi{10.1016/j.physletb.2018.01.048}}.
\eprint{1703.08418}.

\bibtype{Article}%
\bibitem[Adamczyk et al.(2017)]{Adamczyk:2017iwn}
\bibinfo{author}{Adamczyk L} and  et al. (\bibinfo{collaboration}{STAR})
  (\bibinfo{year}{2017}).
\bibinfo{title}{{Bulk Properties of the Medium Produced in Relativistic
  Heavy-Ion Collisions from the Beam Energy Scan Program}}.
\bibinfo{journal}{{\em Phys. Rev. C}} \bibinfo{volume}{96}
  (\bibinfo{number}{4}): \bibinfo{pages}{044904}.
  \bibinfo{doi}{\doi{10.1103/PhysRevC.96.044904}}.
\eprint{1701.07065}.

\bibtype{Article}%
\bibitem[Adams et al.(2007)]{Adams:2006ke}
\bibinfo{author}{Adams J} and  et al. (\bibinfo{collaboration}{STAR})
  (\bibinfo{year}{2007}).
\bibinfo{title}{{Scaling Properties of Hyperon Production in Au+Au Collisions
  at $\sqrt{s_{_{\rm NN}}} = 200$ GeV}}.
\bibinfo{journal}{{\em Phys. Rev. Lett.}} \bibinfo{volume}{98}:
  \bibinfo{pages}{062301}. \bibinfo{doi}{\doi{10.1103/PhysRevLett.98.062301}}.
\eprint{nucl-ex/0606014}.

\bibtype{Article}%
\bibitem[Adare et al.(2016)]{PHENIX:2015tbb}
\bibinfo{author}{Adare A} and  et al. (\bibinfo{collaboration}{PHENIX})
  (\bibinfo{year}{2016}).
\bibinfo{title}{{Transverse energy production and charged-particle multiplicity
  at midrapidity in various systems from $\sqrt{s_{NN}}=7.7$ to 200 GeV}}.
\bibinfo{journal}{{\em Phys. Rev. C}} \bibinfo{volume}{93}
  (\bibinfo{number}{2}): \bibinfo{pages}{024901}.
  \bibinfo{doi}{\doi{10.1103/PhysRevC.93.024901}}.
\eprint{1509.06727}.

\bibtype{Article}%
\bibitem[Adler et al.(2002)]{Adler:2002uv}
\bibinfo{author}{Adler C} and  et al. (\bibinfo{collaboration}{STAR})
  (\bibinfo{year}{2002}).
\bibinfo{title}{{Midrapidity $\Lambda$ and $\bar{\Lambda}$ production in Au +
  Au collisions at $\sqrt{s_{_{\rm NN}}}= 130$ GeV}}.
\bibinfo{journal}{{\em Phys. Rev. Lett.}} \bibinfo{volume}{89}:
  \bibinfo{pages}{092301}. \bibinfo{doi}{\doi{10.1103/PhysRevLett.89.092301}}.
\eprint{nucl-ex/0203016}.

\bibtype{Article}%
\bibitem[Adler et al.(2004)]{Adler:2003cb}
\bibinfo{author}{Adler S} and  et al. (\bibinfo{collaboration}{PHENIX})
  (\bibinfo{year}{2004}).
\bibinfo{title}{{Identified charged particle spectra and yields in Au+Au
  collisions at $\sqrt{s_{_{\rm NN}}}= 200$ GeV}}.
\bibinfo{journal}{{\em Phys. Rev. C}} \bibinfo{volume}{69}:
  \bibinfo{pages}{034909}. \bibinfo{doi}{\doi{10.1103/PhysRevC.69.034909}}.
\eprint{nucl-ex/0307022}.

\bibtype{Article}%
\bibitem[Afanasiev et al.(2002)]{Afanasiev:2002mx}
\bibinfo{author}{Afanasiev S} and  et al. (\bibinfo{collaboration}{NA49})
  (\bibinfo{year}{2002}).
\bibinfo{title}{{Energy dependence of pion and kaon production in central Pb +
  Pb collisions}}.
\bibinfo{journal}{{\em Phys. Rev. C}} \bibinfo{volume}{66}:
  \bibinfo{pages}{054902}. \bibinfo{doi}{\doi{10.1103/PhysRevC.66.054902}}.
\eprint{nucl-ex/0205002}.

\bibtype{Article}%
\bibitem[Aggarwal et al.(2001)]{WA98:2000mvt}
\bibinfo{author}{Aggarwal MM} and  et al. (\bibinfo{collaboration}{WA98})
  (\bibinfo{year}{2001}).
\bibinfo{title}{{Scaling of particle and transverse energy production in Pb-208
  + Pb-208 collisions at 158-A-GeV}}.
\bibinfo{journal}{{\em Eur. Phys. J. C}} \bibinfo{volume}{18}:
  \bibinfo{pages}{651--663}. \bibinfo{doi}{\doi{10.1007/s100520100578}}.
\eprint{nucl-ex/0008004}.

\bibtype{Article}%
\bibitem[Aggarwal et al.(2011)]{Aggarwal:2010ig}
\bibinfo{author}{Aggarwal M} and  et al. (\bibinfo{collaboration}{STAR})
  (\bibinfo{year}{2011}).
\bibinfo{title}{{Strange and Multi-strange Particle Production in Au+Au
  Collisions at $\sqrt{s_{_{\rm NN}}}= 62.4$ GeV}}.
\bibinfo{journal}{{\em Phys. Rev. C}} \bibinfo{volume}{83}:
  \bibinfo{pages}{024901}. \bibinfo{doi}{\doi{10.1103/PhysRevC.83.024901}}.
\eprint{1010.0142}.

\bibtype{Article}%
\bibitem[Ahle et al.(2000{\natexlab{a}})]{Ahle:2000wq}
\bibinfo{author}{Ahle L} and  et al. (\bibinfo{collaboration}{E866, E917})
  (\bibinfo{year}{2000}{\natexlab{a}}).
\bibinfo{title}{{An Excitation function of K- and K+ production in Au + Au
  reactions at the AGS}}.
\bibinfo{journal}{{\em Phys. Lett. B}} \bibinfo{volume}{490}:
  \bibinfo{pages}{53--60}. \bibinfo{doi}{\doi{10.1016/S0370-2693(00)00916-3}}.
\eprint{nucl-ex/0008010}.

\bibtype{Article}%
\bibitem[Ahle et al.(2000{\natexlab{b}})]{Ahle:1999uy}
\bibinfo{author}{Ahle L} and  et al. (\bibinfo{collaboration}{E866, E917})
  (\bibinfo{year}{2000}{\natexlab{b}}).
\bibinfo{title}{{Excitation function of K+ and pi+ production in Au + Au
  reactions at 2/A GeV to 10/A GeV}}.
\bibinfo{journal}{{\em Phys. Lett. B}} \bibinfo{volume}{476}:
  \bibinfo{pages}{1--8}. \bibinfo{doi}{\doi{10.1016/S0370-2693(00)00037-X}}.
\eprint{nucl-ex/9910008}.

\bibtype{Article}%
\bibitem[Ahmad et al.(1998)]{Ahmad:1998sg}
\bibinfo{author}{Ahmad S}, \bibinfo{author}{Bonner B}, \bibinfo{author}{Efremov
  S}, \bibinfo{author}{Mutchler G}, \bibinfo{author}{Platner E} and  et al.
  (\bibinfo{year}{1998}).
\bibinfo{title}{{Nuclear matter expansion parameters from the measurement of
  differential multiplicities for lambda production in central Au+Au collisions
  at AGS}}.
\bibinfo{journal}{{\em Nucl. Phys. A}} \bibinfo{volume}{636}:
  \bibinfo{pages}{507--524}.
  \bibinfo{doi}{\doi{10.1016/S0375-9474(98)00218-8}}.
\eprint{nucl-ex/9803006}.

\bibtype{Article}%
\bibitem[Ali Hassan~Abdallah et al.(2026)]{ALICE:2026zcz}
\bibinfo{author}{Ali Hassan~Abdallah D} and  et al.
  (\bibinfo{collaboration}{ALICE}) (\bibinfo{year}{2026}), \bibinfo{month}{3}.
\bibinfo{title}{{Evidence of different $\Lambda_{\rm c}$-baryon and D-meson
  elliptic flow in Pb$-$Pb collisions at $\sqrt{s_{\rm NN}}$ = 5.36 TeV with
  ALICE at the LHC}} \eprint{2603.18966}.

\bibtype{Article}%
\bibitem[Alt et al.(2006)]{Alt:2005gr}
\bibinfo{author}{Alt C} and  et al. (\bibinfo{collaboration}{NA49})
  (\bibinfo{year}{2006}).
\bibinfo{title}{{Energy and centrality dependence of antiproton and proton
  production in relativistic Pb + Pb collisions at the CERN SPS}}.
\bibinfo{journal}{{\em Phys. Rev. C}} \bibinfo{volume}{73}:
  \bibinfo{pages}{044910}. \bibinfo{doi}{\doi{10.1103/PhysRevC.73.044910}}.
\eprint{nucl-ex/0512033}.

\bibtype{Article}%
\bibitem[Alt et al.(2008{\natexlab{a}})]{Alt:2008qm}
\bibinfo{author}{Alt C} and  et al. (\bibinfo{collaboration}{NA49})
  (\bibinfo{year}{2008}{\natexlab{a}}).
\bibinfo{title}{{Energy dependence of Lambda and Xi production in central Pb+Pb
  collisions at A20, A30, A40, A80, and A158 GeV measured at the CERN Super
  Proton Synchrotron}}.
\bibinfo{journal}{{\em Phys. Rev. C}} \bibinfo{volume}{78}:
  \bibinfo{pages}{034918}. \bibinfo{doi}{\doi{10.1103/PhysRevC.78.034918}}.
\eprint{0804.3770}.

\bibtype{Article}%
\bibitem[Alt et al.(2008{\natexlab{b}})]{Alt:2007aa}
\bibinfo{author}{Alt C} and  et al. (\bibinfo{collaboration}{NA49})
  (\bibinfo{year}{2008}{\natexlab{b}}).
\bibinfo{title}{{Pion and kaon production in central Pb + Pb collisions at 20A
  and 30A GeV: Evidence for the onset of deconfinement}}.
\bibinfo{journal}{{\em Phys. Rev. C}} \bibinfo{volume}{77}:
  \bibinfo{pages}{024903}. \bibinfo{doi}{\doi{10.1103/PhysRevC.77.024903}}.
\eprint{0710.0118}.

\bibtype{Article}%
\bibitem[Altenkort et al.(2021)]{Altenkort:2020fgs}
\bibinfo{author}{Altenkort L}, \bibinfo{author}{Eller AM},
  \bibinfo{author}{Kaczmarek O}, \bibinfo{author}{Mazur L},
  \bibinfo{author}{Moore GD} and  \bibinfo{author}{Shu HT}
  (\bibinfo{year}{2021}).
\bibinfo{title}{{Heavy quark momentum diffusion from the lattice using gradient
  flow}}.
\bibinfo{journal}{{\em Phys. Rev. D}} \bibinfo{volume}{103}
  (\bibinfo{number}{1}): \bibinfo{pages}{014511}.
  \bibinfo{doi}{\doi{10.1103/PhysRevD.103.014511}}.
\eprint{2009.13553}.

\bibtype{Article}%
\bibitem[Altenkort et al.(2023)]{Altenkort:2023oms}
\bibinfo{author}{Altenkort L}, \bibinfo{author}{Kaczmarek O},
  \bibinfo{author}{Larsen R}, \bibinfo{author}{Mukherjee S},
  \bibinfo{author}{Petreczky P}, \bibinfo{author}{Shu HT} and
  \bibinfo{author}{Stendebach S} (\bibinfo{collaboration}{HotQCD})
  (\bibinfo{year}{2023}).
\bibinfo{title}{{Heavy Quark Diffusion from 2+1 Flavor Lattice QCD with 320~MeV
  Pion Mass}}.
\bibinfo{journal}{{\em Phys. Rev. Lett.}} \bibinfo{volume}{130}
  (\bibinfo{number}{23}): \bibinfo{pages}{231902}.
  \bibinfo{doi}{\doi{10.1103/PhysRevLett.130.231902}}.
\eprint{2302.08501}.

\bibtype{Article}%
\bibitem[Andronic(2014)]{Andronic:2014zha}
\bibinfo{author}{Andronic A} (\bibinfo{year}{2014}).
\bibinfo{title}{{An overview of the experimental study of quark-gluon matter in
  high-energy nucleus-nucleus collisions}}.
\bibinfo{journal}{{\em Int. J. Mod. Phys. A}} \bibinfo{volume}{29}:
  \bibinfo{pages}{1430047}. \bibinfo{doi}{\doi{10.1142/S0217751X14300476}}.
\eprint{1407.5003}.

\bibtype{Article}%
\bibitem[Andronic and Arnaldi(2025)]{Andronic:2025jbp}
\bibinfo{author}{Andronic A} and  \bibinfo{author}{Arnaldi R}
  (\bibinfo{year}{2025}).
\bibinfo{title}{{Quarkonia and Deconfined Quark{\textendash}Gluon Matter in
  Heavy-Ion Collisions}}.
\bibinfo{journal}{{\em Ann. Rev. Nucl. Part. Sci.}} \bibinfo{volume}{75}
  (\bibinfo{number}{1}): \bibinfo{pages}{351--375}.
  \bibinfo{doi}{\doi{10.1146/annurev-nucl-121423-101041}}.
\eprint{2501.08290}.

\bibtype{Article}%
\bibitem[Andronic et al.(2003)]{Andronic:2003zv}
\bibinfo{author}{Andronic A}, \bibinfo{author}{Braun-Munzinger P},
  \bibinfo{author}{Redlich K} and  \bibinfo{author}{Stachel J}
  (\bibinfo{year}{2003}).
\bibinfo{title}{{Statistical hadronization of charm in heavy ion collisions at
  SPS, RHIC and LHC}}.
\bibinfo{journal}{{\em Phys. Lett. B}} \bibinfo{volume}{571}:
  \bibinfo{pages}{36--44}. \bibinfo{doi}{\doi{10.1016/j.physletb.2003.07.066}}.
\eprint{nucl-th/0303036}.

\bibtype{Article}%
\bibitem[Andronic et al.(2006)]{Andronic:2005yp}
\bibinfo{author}{Andronic A}, \bibinfo{author}{Braun-Munzinger P} and
  \bibinfo{author}{Stachel J} (\bibinfo{year}{2006}).
\bibinfo{title}{{Hadron production in central nucleus-nucleus collisions at
  chemical freeze-out}}.
\bibinfo{journal}{{\em Nucl. Phys. A}} \bibinfo{volume}{772}:
  \bibinfo{pages}{167--199}.
  \bibinfo{doi}{\doi{10.1016/j.nuclphysa.2006.03.012}}.
\eprint{nucl-th/0511071}.

\bibtype{Article}%
\bibitem[Andronic et al.(2007)]{Andronic:2006ky}
\bibinfo{author}{Andronic A}, \bibinfo{author}{Braun-Munzinger P},
  \bibinfo{author}{Redlich K} and  \bibinfo{author}{Stachel J}
  (\bibinfo{year}{2007}).
\bibinfo{title}{{Statistical hadronization of heavy quarks in
  ultra-relativistic nucleus-nucleus collisions}}.
\bibinfo{journal}{{\em Nucl. Phys. A}} \bibinfo{volume}{789}:
  \bibinfo{pages}{334--356}.
  \bibinfo{doi}{\doi{10.1016/j.nuclphysa.2007.02.013}}.
\eprint{nucl-th/0611023}.

\bibtype{Article}%
\bibitem[Andronic et al.(2009)]{Andronic:2008gu}
\bibinfo{author}{Andronic A}, \bibinfo{author}{Braun-Munzinger P} and
  \bibinfo{author}{Stachel J} (\bibinfo{year}{2009}).
\bibinfo{title}{{Thermal hadron production in relativistic nuclear collisions:
  the hadron mass spectrum, the horn, and the QCD phase transition}}.
\bibinfo{journal}{{\em Phys. Lett. B}} \bibinfo{volume}{673}:
  \bibinfo{pages}{142--145}. \bibinfo{doi}{\doi{10.1016/j.physletb.2009.06.021,
  10.1016/j.physletb.2009.02.014}}.
\eprint{0812.1186}.

\bibtype{Article}%
\bibitem[Andronic et al.(2011)]{Andronic:2010qu}
\bibinfo{author}{Andronic A}, \bibinfo{author}{Braun-Munzinger P},
  \bibinfo{author}{Stachel J} and  \bibinfo{author}{St{\"o}cker H}
  (\bibinfo{year}{2011}).
\bibinfo{title}{{Production of light nuclei, hypernuclei and their
  antiparticles in relativistic nuclear collisions}}.
\bibinfo{journal}{{\em Phys. Lett. B}} \bibinfo{volume}{697}:
  \bibinfo{pages}{203--207}.
  \bibinfo{doi}{\doi{10.1016/j.physletb.2011.01.053}}.
\eprint{1010.2995}.

\bibtype{Article}%
\bibitem[Andronic et al.(2018)]{Andronic:2017pug}
\bibinfo{author}{Andronic A}, \bibinfo{author}{Braun-Munzinger P},
  \bibinfo{author}{Redlich K} and  \bibinfo{author}{Stachel J}
  (\bibinfo{year}{2018}).
\bibinfo{title}{{Decoding the phase structure of QCD via particle production at
  high energy}}.
\bibinfo{journal}{{\em Nature}} \bibinfo{volume}{561} (\bibinfo{number}{7723}):
  \bibinfo{pages}{321--330}. \bibinfo{doi}{\doi{10.1038/s41586-018-0491-6}}.
\eprint{1710.09425}.

\bibtype{Article}%
\bibitem[Andronic et al.(2019{\natexlab{a}})]{Andronic:2018qqt}
\bibinfo{author}{Andronic A}, \bibinfo{author}{Braun-Munzinger P},
  \bibinfo{author}{Friman B}, \bibinfo{author}{Lo PM}, \bibinfo{author}{Redlich
  K} and  \bibinfo{author}{Stachel J} (\bibinfo{year}{2019}{\natexlab{a}}).
\bibinfo{title}{{The thermal proton yield anomaly in Pb-Pb collisions at the
  LHC and its resolution}}.
\bibinfo{journal}{{\em Phys. Lett. B}} \bibinfo{volume}{792}:
  \bibinfo{pages}{304--309}.
  \bibinfo{doi}{\doi{10.1016/j.physletb.2019.03.052}}.
\eprint{1808.03102}.

\bibtype{Article}%
\bibitem[Andronic et al.(2019{\natexlab{b}})]{Andronic:2019wva}
\bibinfo{author}{Andronic A}, \bibinfo{author}{Braun-Munzinger P},
  \bibinfo{author}{K\"ohler MK}, \bibinfo{author}{Redlich K} and
  \bibinfo{author}{Stachel J} (\bibinfo{year}{2019}{\natexlab{b}}).
\bibinfo{title}{{Transverse momentum distributions of charmonium states with
  the statistical hadronization model}}.
\bibinfo{journal}{{\em Phys. Lett. B}} \bibinfo{volume}{797}:
  \bibinfo{pages}{134836}. \bibinfo{doi}{\doi{10.1016/j.physletb.2019.134836}}.
\eprint{1901.09200}.

\bibtype{Article}%
\bibitem[Andronic et al.(2021{\natexlab{a}})]{Andronic:2020iyg}
\bibinfo{author}{Andronic A}, \bibinfo{author}{Braun-Munzinger P},
  \bibinfo{author}{G{\"u}nd{\"u}z D}, \bibinfo{author}{Kirchhoff Y},
  \bibinfo{author}{K{\"o}hler MK}, \bibinfo{author}{Stachel J} and
  \bibinfo{author}{Winn M} (\bibinfo{year}{2021}{\natexlab{a}}).
\bibinfo{title}{{Influence of modified light-flavor hadron spectra on particle
  yields in the statistical hadronization model}}.
\bibinfo{journal}{{\em Nucl. Phys. A}} \bibinfo{volume}{1010}:
  \bibinfo{pages}{122176}.
  \bibinfo{doi}{\doi{10.1016/j.nuclphysa.2021.122176}}.
\eprint{2011.03826}.

\bibtype{Article}%
\bibitem[Andronic et al.(2021{\natexlab{b}})]{Andronic:2021erx}
\bibinfo{author}{Andronic A}, \bibinfo{author}{Braun-Munzinger P},
  \bibinfo{author}{K{\"o}hler MK}, \bibinfo{author}{Mazeliauskas A},
  \bibinfo{author}{Redlich K}, \bibinfo{author}{Stachel J} and
  \bibinfo{author}{Vislavicius V} (\bibinfo{year}{2021}{\natexlab{b}}).
\bibinfo{title}{{The multiple-charm hierarchy in the statistical hadronization
  model}}.
\bibinfo{journal}{{\em JHEP}} \bibinfo{volume}{07}: \bibinfo{pages}{035}.
  \bibinfo{doi}{\doi{10.1007/JHEP07(2021)035}}.
\eprint{2104.12754}.

\bibtype{Article}%
\bibitem[Andronic et al.(2023)]{Andronic:2022ucg}
\bibinfo{author}{Andronic A}, \bibinfo{author}{Braun-Munzinger P},
  \bibinfo{author}{Redlich K} and  \bibinfo{author}{Stachel J}
  (\bibinfo{year}{2023}).
\bibinfo{title}{{Statistical Hadronization of $b$-quarks in Pb\textendash{}Pb
  Collisions at LHC Energy: A Case for Partial Equilibration of $b$-quarks?}}
\bibinfo{journal}{{\em Acta Phys. Polon. Supp.}} \bibinfo{volume}{16}
  (\bibinfo{number}{1}): \bibinfo{pages}{1--A107}.
  \bibinfo{doi}{\doi{10.5506/APhysPolBSupp.16.1-A107}}.
\eprint{2209.14562}.

\bibtype{Article}%
\bibitem[Andronic et al.(2024)]{Andronic:2023ioz}
\bibinfo{author}{Andronic A}, \bibinfo{author}{Braun-Munzinger P},
  \bibinfo{author}{Brun{\ss}en H}, \bibinfo{author}{Crkovsk{\'a} J},
  \bibinfo{author}{Stachel J}, \bibinfo{author}{Vislavicius V} and
  \bibinfo{author}{V{\"o}lkl M} (\bibinfo{year}{2024}).
\bibinfo{title}{{Transverse dynamics of charmed hadrons in ultra-relativistic
  nuclear collisions}}.
\bibinfo{journal}{{\em JHEP}} \bibinfo{volume}{10}: \bibinfo{pages}{229}.
  \bibinfo{doi}{\doi{10.1007/JHEP10(2024)229}}.
\eprint{2308.14821}.

\bibtype{Article}%
\bibitem[Antinori et al.(2004)]{Antinori:2004ee}
\bibinfo{author}{Antinori F} and  et al. (\bibinfo{collaboration}{NA57})
  (\bibinfo{year}{2004}).
\bibinfo{title}{{Energy dependence of hyperon production in nucleus nucleus
  collisions at SPS}}.
\bibinfo{journal}{{\em Phys. Lett. B}} \bibinfo{volume}{595}:
  \bibinfo{pages}{68--74}. \bibinfo{doi}{\doi{10.1016/j.physletb.2004.05.025}}.
\eprint{nucl-ex/0403022}.

\bibtype{Article}%
\bibitem[Aoki et al.(2006)]{Aoki:2006we}
\bibinfo{author}{Aoki Y}, \bibinfo{author}{Endrodi G}, \bibinfo{author}{Fodor
  Z}, \bibinfo{author}{Katz S} and  \bibinfo{author}{Szabo K}
  (\bibinfo{year}{2006}).
\bibinfo{title}{{The order of the quantum chromodynamics transition predicted
  by the standard model of particle physics}}.
\bibinfo{journal}{{\em Nature}} \bibinfo{volume}{443}:
  \bibinfo{pages}{675--678}. \bibinfo{doi}{\doi{10.1038/nature05120}}.
\eprint{hep-lat/0611014}.

\bibtype{Article}%
\bibitem[Apolin{\'a}rio et al.(2022)]{Apolinario:2022vzg}
\bibinfo{author}{Apolin{\'a}rio L}, \bibinfo{author}{Lee YJ} and
  \bibinfo{author}{Winn M} (\bibinfo{year}{2022}).
\bibinfo{title}{{Heavy quarks and jets as probes of the QGP}}.
\bibinfo{journal}{{\em Prog. Part. Nucl. Phys.}} \bibinfo{volume}{127}:
  \bibinfo{pages}{103990}. \bibinfo{doi}{\doi{10.1016/j.ppnp.2022.103990}}.
\eprint{2203.16352}.

\bibtype{Article}%
\bibitem[Arsene et al.(2005)]{Arsene:2005mr}
\bibinfo{author}{Arsene I} and  et al. (\bibinfo{collaboration}{BRAHMS})
  (\bibinfo{year}{2005}).
\bibinfo{title}{{Centrality dependent particle production at $y=0$ and $y \sim
  1$ in Au + Au collisions at $\sqrt{s_{_{\rm NN}}}= 200$ GeV}}.
\bibinfo{journal}{{\em Phys. Rev. C}} \bibinfo{volume}{72}:
  \bibinfo{pages}{014908}. \bibinfo{doi}{\doi{10.1103/PhysRevC.72.014908}}.
\eprint{nucl-ex/0503010}.

\bibtype{Article}%
\bibitem[Bailhache and Appelsh{\"a}user(2025)]{Bailhache:2025kwa}
\bibinfo{author}{Bailhache R} and  \bibinfo{author}{Appelsh{\"a}user H}
  (\bibinfo{year}{2025}).
\bibinfo{title}{{Dileptons at Colliders as Probes of the
  Quark{\textendash}Gluon Plasma}}.
\bibinfo{journal}{{\em Ann. Rev. Nucl. Part. Sci.}} \bibinfo{volume}{75}
  (\bibinfo{number}{1}): \bibinfo{pages}{463--486}.
  \bibinfo{doi}{\doi{10.1146/annurev-nucl-121423-100858}}.
\eprint{2512.10597}.

\bibtype{Article}%
\bibitem[Barrette et al.(1993)]{E814E877:1993rlr}
\bibinfo{author}{Barrette J} and  et al. (\bibinfo{collaboration}{E814/E877})
  (\bibinfo{year}{1993}).
\bibinfo{title}{{Measurement of transverse energy production with Si and Au
  beams at relativistic energy: Towards hot and dense hadronic matter}}.
\bibinfo{journal}{{\em Phys. Rev. Lett.}} \bibinfo{volume}{70}:
  \bibinfo{pages}{2996--2999}.
  \bibinfo{doi}{\doi{10.1103/PhysRevLett.70.2996}}.

\bibtype{Article}%
\bibitem[Bauswein et al.(2019)]{Bauswein:2018bma}
\bibinfo{author}{Bauswein A}, \bibinfo{author}{Bastian NUF},
  \bibinfo{author}{Blaschke DB}, \bibinfo{author}{Chatziioannou K},
  \bibinfo{author}{Clark JA}, \bibinfo{author}{Fischer T} and
  \bibinfo{author}{Oertel M} (\bibinfo{year}{2019}).
\bibinfo{title}{{Identifying a first-order phase transition in neutron star
  mergers through gravitational waves}}.
\bibinfo{journal}{{\em Phys. Rev. Lett.}} \bibinfo{volume}{122}
  (\bibinfo{number}{6}): \bibinfo{pages}{061102}.
  \bibinfo{doi}{\doi{10.1103/PhysRevLett.122.061102}}.
\eprint{1809.01116}.

\bibtype{Article}%
\bibitem[Baym et al.(2019)]{Baym:2019iky}
\bibinfo{author}{Baym G}, \bibinfo{author}{Furusawa S},
  \bibinfo{author}{Hatsuda T}, \bibinfo{author}{Kojo T} and
  \bibinfo{author}{Togashi H} (\bibinfo{year}{2019}).
\bibinfo{title}{{New Neutron Star Equation of State with Quark-Hadron
  Crossover}}.
\bibinfo{journal}{{\em Astrophys. J.}} \bibinfo{volume}{885}:
  \bibinfo{pages}{42}. \bibinfo{doi}{\doi{10.3847/1538-4357/ab441e}}.
\eprint{1903.08963}.

\bibtype{Article}%
\bibitem[Bazavov et al.(2014)]{Bazavov:2014yba}
\bibinfo{author}{Bazavov A}, \bibinfo{author}{Ding HT}, \bibinfo{author}{Hegde
  P}, \bibinfo{author}{Kaczmarek O}, \bibinfo{author}{Karsch F} and  et al.
  (\bibinfo{year}{2014}).
\bibinfo{title}{{The melting and abundance of open charm hadrons}}.
\bibinfo{journal}{{\em Phys. Lett. B}} \bibinfo{volume}{737}:
  \bibinfo{pages}{210}. \bibinfo{doi}{\doi{10.1016/j.physletb.2014.08.034}}.
\eprint{1404.4043}.

\bibtype{Article}%
\bibitem[Bazavov et al.(2017)]{Bazavov:2017dus}
\bibinfo{author}{Bazavov A} and  et al. (\bibinfo{year}{2017}).
\bibinfo{title}{{The QCD Equation of State to $\mathcal{O}(\mu_B^6)$ from
  Lattice QCD}}.
\bibinfo{journal}{{\em Phys. Rev. D}} \bibinfo{volume}{95}
  (\bibinfo{number}{5}): \bibinfo{pages}{054504}.
  \bibinfo{doi}{\doi{10.1103/PhysRevD.95.054504}}.
\eprint{1701.04325}.

\bibtype{Article}%
\bibitem[Bazavov et al.(2019)]{Bazavov:2018mes}
\bibinfo{author}{Bazavov A} and  et al. (\bibinfo{collaboration}{HotQCD})
  (\bibinfo{year}{2019}).
\bibinfo{title}{{Chiral crossover in QCD at zero and non-zero chemical
  potentials}}.
\bibinfo{journal}{{\em Phys. Lett. B}} \bibinfo{volume}{795}:
  \bibinfo{pages}{15--21}. \bibinfo{doi}{\doi{10.1016/j.physletb.2019.05.013}}.
\eprint{1812.08235}.

\bibtype{Article}%
\bibitem[Bearden et al.(2002)]{Bearden:2002ib}
\bibinfo{author}{Bearden I} and  et al. (\bibinfo{collaboration}{NA44})
  (\bibinfo{year}{2002}).
\bibinfo{title}{{Particle production in central Pb + Pb collisions at 158A
  GeV/c}}.
\bibinfo{journal}{{\em Phys. Rev. C}} \bibinfo{volume}{66}:
  \bibinfo{pages}{044907}. \bibinfo{doi}{\doi{10.1103/PhysRevC.66.044907}}.
\eprint{nucl-ex/0202019}.

\bibtype{Article}%
\bibitem[Becattini et al.(2017)]{Becattini:2016xct}
\bibinfo{author}{Becattini F}, \bibinfo{author}{Steinheimer J},
  \bibinfo{author}{Stock R} and  \bibinfo{author}{Bleicher M}
  (\bibinfo{year}{2017}).
\bibinfo{title}{{Hadronization conditions in relativistic nuclear collisions
  and the QCD pseudo-critical line}}.
\bibinfo{journal}{{\em Phys. Lett. B}} \bibinfo{volume}{764}:
  \bibinfo{pages}{241--246}.
  \bibinfo{doi}{\doi{10.1016/j.physletb.2016.11.033}}.
\eprint{1605.09694}.

\bibtype{Article}%
\bibitem[Benecke et al.(1969)]{Benecke:1969sh}
\bibinfo{author}{Benecke J}, \bibinfo{author}{Chou TT}, \bibinfo{author}{Yang
  CN} and  \bibinfo{author}{Yen E} (\bibinfo{year}{1969}).
\bibinfo{title}{{Hypothesis of Limiting Fragmentation in High-Energy
  Collisions}}.
\bibinfo{journal}{{\em Phys. Rev.}} \bibinfo{volume}{188}:
  \bibinfo{pages}{2159--2169}. \bibinfo{doi}{\doi{10.1103/PhysRev.188.2159}}.

\bibtype{Article}%
\bibitem[Bernhard et al.(2019)]{Bernhard:2019bmu}
\bibinfo{author}{Bernhard JE}, \bibinfo{author}{Moreland JS} and
  \bibinfo{author}{Bass SA} (\bibinfo{year}{2019}).
\bibinfo{title}{{Bayesian estimation of the specific shear and bulk viscosity
  of quark{\textendash}gluon plasma}}.
\bibinfo{journal}{{\em Nature Phys.}} \bibinfo{volume}{15}
  (\bibinfo{number}{11}): \bibinfo{pages}{1113--1117}.
  \bibinfo{doi}{\doi{10.1038/s41567-019-0611-8}}.

\bibtype{Article}%
\bibitem[Beth and Uhlenbeck(1937)]{Beth:1937zz}
\bibinfo{author}{Beth E} and  \bibinfo{author}{Uhlenbeck G}
  (\bibinfo{year}{1937}).
\bibinfo{title}{{The quantum theory of the non-ideal gas. II. Behaviour at low
  temperatures}}.
\bibinfo{journal}{{\em Physica}} \bibinfo{volume}{4}:
  \bibinfo{pages}{915--924}.
  \bibinfo{doi}{\doi{10.1016/S0031-8914(37)80189-5}}.

\bibtype{Article}%
\bibitem[Bjorken(1983)]{Bjorken:1982qr}
\bibinfo{author}{Bjorken J} (\bibinfo{year}{1983}).
\bibinfo{title}{{Highly Relativistic Nucleus-Nucleus Collisions: The Central
  Rapidity Region}}.
\bibinfo{journal}{{\em Phys. Rev. D}} \bibinfo{volume}{27}:
  \bibinfo{pages}{140--151}. \bibinfo{doi}{\doi{10.1103/PhysRevD.27.140}}.

\bibtype{Article}%
\bibitem[Blaschke et al.(2025)]{Blaschke:2025qvv}
\bibinfo{author}{Blaschke D}, \bibinfo{author}{Ivanytskyi O} and
  \bibinfo{author}{R{\"o}pke G} (\bibinfo{year}{2025}).
\bibinfo{title}{{Generalized Beth-Uhlenbeck approach to the thermodynamics of
  quark-hadron matter}}.
\bibinfo{journal}{{\em PoS}} \bibinfo{volume}{QCHSC24}: \bibinfo{pages}{247}.
  \bibinfo{doi}{\doi{10.22323/1.483.0247}}.
\eprint{2507.10497}.

\bibtype{Article}%
\bibitem[Bonati et al.(2018)]{Bonati:2018nut}
\bibinfo{author}{Bonati C}, \bibinfo{author}{D'Elia M}, \bibinfo{author}{Negro
  F}, \bibinfo{author}{Sanfilippo F} and  \bibinfo{author}{Zambello K}
  (\bibinfo{year}{2018}).
\bibinfo{title}{{Curvature of the pseudocritical line in QCD: Taylor expansion
  matches analytic continuation}}.
\bibinfo{journal}{{\em Phys. Rev. D}} \bibinfo{volume}{98}
  (\bibinfo{number}{5}): \bibinfo{pages}{054510}.
  \bibinfo{doi}{\doi{10.1103/PhysRevD.98.054510}}.
\eprint{1805.02960}.

\bibtype{Article}%
\bibitem[Borsanyi and Parotto(2025)]{Borsanyi:2025ttb}
\bibinfo{author}{Borsanyi S} and  \bibinfo{author}{Parotto P}
  (\bibinfo{year}{2025}), \bibinfo{month}{12}.
\bibinfo{title}{{The QCD phase diagram}} \eprint{2512.08843}.

\bibtype{Article}%
\bibitem[Borsanyi et al.(2020)]{Borsanyi:2020fev}
\bibinfo{author}{Borsanyi S}, \bibinfo{author}{Fodor Z},
  \bibinfo{author}{Guenther JN}, \bibinfo{author}{Kara R},
  \bibinfo{author}{Katz SD}, \bibinfo{author}{Parotto P},
  \bibinfo{author}{Pasztor A}, \bibinfo{author}{Ratti C} and
  \bibinfo{author}{Szabo KK} (\bibinfo{year}{2020}).
\bibinfo{title}{{QCD Crossover at Finite Chemical Potential from Lattice
  Simulations}}.
\bibinfo{journal}{{\em Phys. Rev. Lett.}} \bibinfo{volume}{125}
  (\bibinfo{number}{5}): \bibinfo{pages}{052001}.
  \bibinfo{doi}{\doi{10.1103/PhysRevLett.125.052001}}.
\eprint{2002.02821}.

\bibtype{Article}%
\bibitem[Borsanyi et al.(2024)]{Borsanyi:2024xrx}
\bibinfo{author}{Borsanyi S}, \bibinfo{author}{Fodor Z},
  \bibinfo{author}{Guenther JN}, \bibinfo{author}{Parotto P},
  \bibinfo{author}{Pasztor A}, \bibinfo{author}{Pirelli L},
  \bibinfo{author}{Szabo KK} and  \bibinfo{author}{Wong CH}
  (\bibinfo{year}{2024}).
\bibinfo{title}{{QCD deconfinement transition line up to
  {\ensuremath{\mu}}B=400{\,}{\,}MeV from finite volume lattice simulations}}.
\bibinfo{journal}{{\em Phys. Rev. D}} \bibinfo{volume}{110}
  (\bibinfo{number}{11}): \bibinfo{pages}{114507}.
  \bibinfo{doi}{\doi{10.1103/PhysRevD.110.114507}}.
\eprint{2410.06216}.

\bibtype{Article}%
\bibitem[Bors{\'a}nyi et al.(2025)]{Borsanyi:2025lim}
\bibinfo{author}{Bors{\'a}nyi S}, \bibinfo{author}{Fodor Z},
  \bibinfo{author}{Guenther JN}, \bibinfo{author}{Kara R},
  \bibinfo{author}{Parotto P}, \bibinfo{author}{P{\'a}sztor A},
  \bibinfo{author}{Pirelli L} and  \bibinfo{author}{Wong CH}
  (\bibinfo{year}{2025}).
\bibinfo{title}{{Chiral versus deconfinement properties of the QCD crossover:
  Differences in the volume and chemical potential dependence from the
  lattice}}.
\bibinfo{journal}{{\em Phys. Rev. D}} \bibinfo{volume}{111}
  (\bibinfo{number}{1}): \bibinfo{pages}{014506}.
  \bibinfo{doi}{\doi{10.1103/PhysRevD.111.014506}}.

\bibtype{Article}%
\bibitem[Boyanovsky et al.(2006)]{Boyanovsky:2006bf}
\bibinfo{author}{Boyanovsky D}, \bibinfo{author}{de~Vega H} and
  \bibinfo{author}{Schwarz D} (\bibinfo{year}{2006}).
\bibinfo{title}{{Phase transitions in the early and the present universe}}.
\bibinfo{journal}{{\em Ann. Rev. Nucl. Part. Sci.}} \bibinfo{volume}{56}:
  \bibinfo{pages}{441--500}.
  \bibinfo{doi}{\doi{10.1146/annurev.nucl.56.080805.140539}}.
\eprint{hep-ph/0602002}.

\bibtype{Article}%
\bibitem[Braun-Munzinger and D\"onigus(2019)]{Braun-Munzinger:2018hat}
\bibinfo{author}{Braun-Munzinger P} and  \bibinfo{author}{D\"onigus B}
  (\bibinfo{year}{2019}).
\bibinfo{title}{{Loosely-bound objects produced in nuclear collisions at the
  LHC}}.
\bibinfo{journal}{{\em Nucl. Phys. A}} \bibinfo{volume}{987}:
  \bibinfo{pages}{144--201}.
  \bibinfo{doi}{\doi{10.1016/j.nuclphysa.2019.02.006}}.
\eprint{1809.04681}.

\bibtype{Article}%
\bibitem[Braun-Munzinger and Redlich(2000)]{Braun-Munzinger:2000uqj}
\bibinfo{author}{Braun-Munzinger P} and  \bibinfo{author}{Redlich K}
  (\bibinfo{year}{2000}).
\bibinfo{title}{{Charmonium production from the secondary collisions at LHC
  energy}}.
\bibinfo{journal}{{\em Eur. Phys. J. C}} \bibinfo{volume}{16}:
  \bibinfo{pages}{519--525}. \bibinfo{doi}{\doi{10.1007/s100520000356}}.
\eprint{hep-ph/0001008}.

\bibtype{Article}%
\bibitem[Braun-Munzinger and Stachel(1998)]{BraunMunzinger:1998cg}
\bibinfo{author}{Braun-Munzinger P} and  \bibinfo{author}{Stachel J}
  (\bibinfo{year}{1998}).
\bibinfo{title}{{Dynamics of ultrarelativistic nuclear collisions with heavy
  beams: An Experimental overview}}.
\bibinfo{journal}{{\em Nucl. Phys. A}} \bibinfo{volume}{638}:
  \bibinfo{pages}{3--18}. \bibinfo{doi}{\doi{10.1016/S0375-9474(98)00342-X}}.
\eprint{nucl-ex/9803015}.

\bibtype{Article}%
\bibitem[Braun-Munzinger and Stachel(2000)]{Braun-Munzinger:2000csl}
\bibinfo{author}{Braun-Munzinger P} and  \bibinfo{author}{Stachel J}
  (\bibinfo{year}{2000}).
\bibinfo{title}{{(Non)thermal aspects of charmonium production and a new look
  at J / psi suppression}}.
\bibinfo{journal}{{\em Phys. Lett. B}} \bibinfo{volume}{490}:
  \bibinfo{pages}{196--202}.
  \bibinfo{doi}{\doi{10.1016/S0370-2693(00)00991-6}}.
\eprint{nucl-th/0007059}.

\bibtype{Article}%
\bibitem[Braun-Munzinger and Wambach(2009)]{BraunMunzinger:2008tz}
\bibinfo{author}{Braun-Munzinger P} and  \bibinfo{author}{Wambach J}
  (\bibinfo{year}{2009}).
\bibinfo{title}{{The Phase Diagram of Strongly-Interacting Matter}}.
\bibinfo{journal}{{\em Rev. Mod. Phys.}} \bibinfo{volume}{81}:
  \bibinfo{pages}{1031--1050}. \bibinfo{doi}{\doi{10.1103/RevModPhys.81.1031}}.
\eprint{0801.4256}.

\bibtype{Article}%
\bibitem[Braun-Munzinger et al.(2003)]{Braun-Munzinger:2003pwq}
\bibinfo{author}{Braun-Munzinger P}, \bibinfo{author}{Redlich K} and
  \bibinfo{author}{Stachel J} (\bibinfo{year}{2003}).
\bibinfo{title}{{Particle production in heavy ion collisions. In {\it Quark
  Gluon Plasma 3}, eds. R. C. Hwa and Xin-Nian Wang, World Scientific
  Publishing, 491--599}} \eprint{nucl-th/0304013}.

\bibtype{Article}%
\bibitem[Braun-Munzinger et al.(2004)]{BraunMunzinger:2003zz}
\bibinfo{author}{Braun-Munzinger P}, \bibinfo{author}{Stachel J} and
  \bibinfo{author}{Wetterich C} (\bibinfo{year}{2004}).
\bibinfo{title}{{Chemical freezeout and the QCD phase transition temperature}}.
\bibinfo{journal}{{\em Phys. Lett. B}} \bibinfo{volume}{596}:
  \bibinfo{pages}{61--69}. \bibinfo{doi}{\doi{10.1016/j.physletb.2004.05.081}}.
\eprint{nucl-th/0311005}.

\bibtype{Article}%
\bibitem[Braun-Munzinger et al.(2016)]{Braun-Munzinger:2015hba}
\bibinfo{author}{Braun-Munzinger P}, \bibinfo{author}{Koch V},
  \bibinfo{author}{Sch{\"a}fer T} and  \bibinfo{author}{Stachel J}
  (\bibinfo{year}{2016}).
\bibinfo{title}{{Properties of hot and dense matter from relativistic heavy ion
  collisions}}.
\bibinfo{journal}{{\em Phys. Rept.}} \bibinfo{volume}{621}:
  \bibinfo{pages}{76--126}. \bibinfo{doi}{\doi{10.1016/j.physrep.2015.12.003}}.
\eprint{1510.00442}.

\bibtype{Article}%
\bibitem[Braun-Munzinger et al.(2021)]{Braun-Munzinger:2020jbk}
\bibinfo{author}{Braun-Munzinger P}, \bibinfo{author}{Friman B},
  \bibinfo{author}{Redlich K}, \bibinfo{author}{Rustamov A} and
  \bibinfo{author}{Stachel J} (\bibinfo{year}{2021}).
\bibinfo{title}{{Relativistic nuclear collisions: Establishing a non-critical
  baseline for fluctuation measurements}}.
\bibinfo{journal}{{\em Nucl. Phys. A}} \bibinfo{volume}{1008}:
  \bibinfo{pages}{122141}.
  \bibinfo{doi}{\doi{10.1016/j.nuclphysa.2021.122141}}.
\eprint{2007.02463}.

\bibtype{Article}%
\bibitem[Braun-Munzinger et al.(2025{\natexlab{a}})]{Braun-Munzinger:2024ybd}
\bibinfo{author}{Braun-Munzinger P}, \bibinfo{author}{Redlich K},
  \bibinfo{author}{Sharma N} and  \bibinfo{author}{Stachel J}
  (\bibinfo{year}{2025}{\natexlab{a}}).
\bibinfo{title}{{Emergence of new systematics for open charm production in high
  energy collisions}}.
\bibinfo{journal}{{\em JHEP}} \bibinfo{volume}{04}: \bibinfo{pages}{058}.
  \bibinfo{doi}{\doi{10.1007/JHEP04(2025)058}}.
\eprint{2408.07496}.

\bibtype{Inproceedings}%
\bibitem[Braun-Munzinger et al.(2025{\natexlab{b}})]{Braun-Munzinger:2025mud}
\bibinfo{author}{Braun-Munzinger P}, \bibinfo{author}{Redlich K} and
  \bibinfo{author}{Stachel J} (\bibinfo{year}{2025}{\natexlab{b}}),
  \bibinfo{month}{6}, \bibinfo{title}{{The quark-gluon plasma: diagnosis with
  thermal hadron production from the early history until detailed
  characterization at high energy colliders}}, \eprint{2506.04733}.

\bibtype{Article}%
\bibitem[Braun-Munzinger et al.(2026)]{Braun-Munzinger:2026krf}
\bibinfo{author}{Braun-Munzinger P}, \bibinfo{author}{Rustamov A} and
  \bibinfo{author}{Xu N} (\bibinfo{year}{2026}), \bibinfo{month}{1}.
\bibinfo{title}{{The phase structure of QCD: Fluctuations and Correlations}}
  \bibinfo{doi}{\doi{10.1146/annurev-nucl-100324-014902}}.
\eprint{2601.18666}.

\bibtype{Article}%
\bibitem[Busza et al.(2018)]{Busza:2018rrf}
\bibinfo{author}{Busza W}, \bibinfo{author}{Rajagopal K} and
  \bibinfo{author}{van~der Schee W} (\bibinfo{year}{2018}).
\bibinfo{title}{{Heavy Ion Collisions: The Big Picture, and the Big
  Questions}}.
\bibinfo{journal}{{\em Ann. Rev. Nucl. Part. Sci.}} \bibinfo{volume}{68}:
  \bibinfo{pages}{339--376}.
  \bibinfo{doi}{\doi{10.1146/annurev-nucl-101917-020852}}.
\eprint{1802.04801}.

\bibtype{Article}%
\bibitem[Cabibbo and Parisi(1975)]{Cabibbo:1975ig}
\bibinfo{author}{Cabibbo N} and  \bibinfo{author}{Parisi G}
  (\bibinfo{year}{1975}).
\bibinfo{title}{{Exponential Hadronic Spectrum and Quark Liberation}}.
\bibinfo{journal}{{\em Phys. Lett. B}} \bibinfo{volume}{59}:
  \bibinfo{pages}{67--69}. \bibinfo{doi}{\doi{10.1016/0370-2693(75)90158-6}}.

\bibtype{Article}%
\bibitem[Cacciari et al.(2012)]{Cacciari:2012ny}
\bibinfo{author}{Cacciari M}, \bibinfo{author}{Frixione S},
  \bibinfo{author}{Houdeau N}, \bibinfo{author}{Mangano ML},
  \bibinfo{author}{Nason P} and  \bibinfo{author}{Ridolfi G}
  (\bibinfo{year}{2012}).
\bibinfo{title}{{Theoretical predictions for charm and bottom production at the
  LHC}}.
\bibinfo{journal}{{\em JHEP}} \bibinfo{volume}{10}: \bibinfo{pages}{137}.
  \bibinfo{doi}{\doi{10.1007/JHEP10(2012)137}}.
\eprint{1205.6344}.

\bibtype{Article}%
\bibitem[Chapline and Nauenberg(1977)]{Chapline:1976gy}
\bibinfo{author}{Chapline G} and  \bibinfo{author}{Nauenberg M}
  (\bibinfo{year}{1977}).
\bibinfo{title}{{Asymptotic Freedom and the Baryon-Quark Phase Transition}}.
\bibinfo{journal}{{\em Phys. Rev. D}} \bibinfo{volume}{16}:
  \bibinfo{pages}{450}. \bibinfo{doi}{\doi{10.1103/PhysRevD.16.450}}.

\bibtype{Article}%
\bibitem[Chatrchyan et al.(2012)]{CMS:2012krf}
\bibinfo{author}{Chatrchyan S} and  et al. (\bibinfo{collaboration}{CMS})
  (\bibinfo{year}{2012}).
\bibinfo{title}{{Measurement of the Pseudorapidity and Centrality Dependence of
  the Transverse Energy Density in PbPb Collisions at $\sqrt{s_{NN}}=2.76$
  TeV}}.
\bibinfo{journal}{{\em Phys. Rev. Lett.}} \bibinfo{volume}{109}:
  \bibinfo{pages}{152303}. \bibinfo{doi}{\doi{10.1103/PhysRevLett.109.152303}}.
\eprint{1205.2488}.

\bibtype{Article}%
\bibitem[Cho et al.(2017)]{ExHIC:2017smd}
\bibinfo{author}{Cho S} and  et al. (\bibinfo{collaboration}{ExHIC})
  (\bibinfo{year}{2017}).
\bibinfo{title}{{Exotic hadrons from heavy ion collisions}}.
\bibinfo{journal}{{\em Prog. Part. Nucl. Phys.}} \bibinfo{volume}{95}:
  \bibinfo{pages}{279--322}. \bibinfo{doi}{\doi{10.1016/j.ppnp.2017.02.002}}.
\eprint{1702.00486}.

\bibtype{Article}%
\bibitem[Cho et al.(2020)]{Cho:2019lxb}
\bibinfo{author}{Cho S}, \bibinfo{author}{Sun KJ}, \bibinfo{author}{Ko CM},
  \bibinfo{author}{Lee SH} and  \bibinfo{author}{Oh Y} (\bibinfo{year}{2020}).
\bibinfo{title}{{Charmed hadron production in an improved quark coalescence
  model}}.
\bibinfo{journal}{{\em Phys. Rev. C}} \bibinfo{volume}{101}
  (\bibinfo{number}{2}): \bibinfo{pages}{024909}.
  \bibinfo{doi}{\doi{10.1103/PhysRevC.101.024909}}.
\eprint{1905.09774}.

\bibtype{Article}%
\bibitem[Cleymans et al.(1991)]{Cleymans:1990mn}
\bibinfo{author}{Cleymans J}, \bibinfo{author}{Redlich K} and
  \bibinfo{author}{Suhonen E} (\bibinfo{year}{1991}).
\bibinfo{title}{{Canonical description of strangeness conservation and particle
  production}}.
\bibinfo{journal}{{\em Z. Phys. C}} \bibinfo{volume}{51}:
  \bibinfo{pages}{137--141}. \bibinfo{doi}{\doi{10.1007/BF01579571}}.

\bibtype{Article}%
\bibitem[Cleymans et al.(1999)]{Cleymans:1998yb}
\bibinfo{author}{Cleymans J}, \bibinfo{author}{Oeschler H} and
  \bibinfo{author}{Redlich K} (\bibinfo{year}{1999}).
\bibinfo{title}{{Influence of impact parameter on thermal description of
  relativistic heavy ion collisions at (1-2) A-GeV}}.
\bibinfo{journal}{{\em Phys. Rev. C}} \bibinfo{volume}{59}:
  \bibinfo{pages}{1663}. \bibinfo{doi}{\doi{10.1103/PhysRevC.59.1663}}.
\eprint{nucl-th/9809027}.

\bibtype{Article}%
\bibitem[Cleymans et al.(2021)]{Cleymans:2020fsc}
\bibinfo{author}{Cleymans J}, \bibinfo{author}{Lo PM}, \bibinfo{author}{Redlich
  K} and  \bibinfo{author}{Sharma N} (\bibinfo{year}{2021}).
\bibinfo{title}{{Multiplicity dependence of (multi)strange baryons in the
  canonical ensemble with phase shift corrections}}.
\bibinfo{journal}{{\em Phys. Rev. C}} \bibinfo{volume}{103}
  (\bibinfo{number}{1}): \bibinfo{pages}{014904}.
  \bibinfo{doi}{\doi{10.1103/PhysRevC.103.014904}}.
\eprint{2009.04844}.

\bibtype{Article}%
\bibitem[Cohen and Glozman(2024)]{Cohen:2023hbq}
\bibinfo{author}{Cohen TD} and  \bibinfo{author}{Glozman LY}
  (\bibinfo{year}{2024}).
\bibinfo{title}{{Large $N_c$ QCD phase diagram at $\mu _B=0$}}.
\bibinfo{journal}{{\em Eur. Phys. J. A}} \bibinfo{volume}{60}
  (\bibinfo{number}{9}): \bibinfo{pages}{171}.
  \bibinfo{doi}{\doi{10.1140/epja/s10050-024-01400-9}}.
\eprint{2311.07333}.

\bibtype{Article}%
\bibitem[Collins and Perry(1975)]{Collins:1974ky}
\bibinfo{author}{Collins JC} and  \bibinfo{author}{Perry M}
  (\bibinfo{year}{1975}).
\bibinfo{title}{{Superdense Matter: Neutrons Or Asymptotically Free Quarks?}}
\bibinfo{journal}{{\em Phys. Rev. Lett.}} \bibinfo{volume}{34}:
  \bibinfo{pages}{1353}. \bibinfo{doi}{\doi{10.1103/PhysRevLett.34.1353}}.

\bibtype{Article}%
\bibitem[d'Enterria and Loizides(2021)]{dEnterria:2020dwq}
\bibinfo{author}{d'Enterria D} and  \bibinfo{author}{Loizides C}
  (\bibinfo{year}{2021}).
\bibinfo{title}{{Progress in the Glauber Model at Collider Energies}}.
\bibinfo{journal}{{\em Ann. Rev. Nucl. Part. Sci.}} \bibinfo{volume}{71}:
  \bibinfo{pages}{315--344}.
  \bibinfo{doi}{\doi{10.1146/annurev-nucl-102419-060007}}.
\eprint{2011.14909}.

\bibtype{Article}%
\bibitem[Fischer and Pawlowski(2026)]{Fischer:2026uni}
\bibinfo{author}{Fischer CS} and  \bibinfo{author}{Pawlowski JM}
  (\bibinfo{year}{2026}), \bibinfo{month}{3}.
\bibinfo{title}{{Phase structure and observables at high densities from first
  principles QCD}} \eprint{2603.11135}.

\bibtype{Article}%
\bibitem[Floerchinger and Wetterich(2012)]{Floerchinger:2012xd}
\bibinfo{author}{Floerchinger S} and  \bibinfo{author}{Wetterich C}
  (\bibinfo{year}{2012}).
\bibinfo{title}{{Chemical freeze-out in heavy ion collisions at large baryon
  densities}}.
\bibinfo{journal}{{\em Nucl. Phys. A}} \bibinfo{volume}{890-891}:
  \bibinfo{pages}{11--24}.
  \bibinfo{doi}{\doi{10.1016/j.nuclphysa.2012.07.009}}.
\eprint{1202.1671}.

\bibtype{Article}%
\bibitem[Fujimoto et al.(2025)]{Fujimoto:2025sxx}
\bibinfo{author}{Fujimoto Y}, \bibinfo{author}{Fukushima K},
  \bibinfo{author}{Hidaka Y} and  \bibinfo{author}{McLerran L}
  (\bibinfo{year}{2025}).
\bibinfo{title}{{New state of matter between the hadronic phase and the
  quark-gluon plasma?}}
\bibinfo{journal}{{\em Phys. Rev. D}} \bibinfo{volume}{112}
  (\bibinfo{number}{7}): \bibinfo{pages}{074006}.
  \bibinfo{doi}{\doi{10.1103/h71y-km92}}.
\eprint{2506.00237}.

\bibtype{Article}%
\bibitem[Fukushima(2025)]{Fukushima:2025ujk}
\bibinfo{author}{Fukushima K} (\bibinfo{year}{2025}).
\bibinfo{title}{{QCD phase diagram and astrophysical implications}}.
\bibinfo{journal}{{\em J. Subatomic Part. Cosmol.}} \bibinfo{volume}{3}:
  \bibinfo{pages}{100066}. \bibinfo{doi}{\doi{10.1016/j.jspc.2025.100066}}.
\eprint{2501.01907}.

\bibtype{Article}%
\bibitem[Fukushima and Hatsuda(2011)]{Fukushima:2010bq}
\bibinfo{author}{Fukushima K} and  \bibinfo{author}{Hatsuda T}
  (\bibinfo{year}{2011}).
\bibinfo{title}{{The phase diagram of dense QCD}}.
\bibinfo{journal}{{\em Rept. Prog. Phys.}} \bibinfo{volume}{74}:
  \bibinfo{pages}{014001}. \bibinfo{doi}{\doi{10.1088/0034-4885/74/1/014001}}.
\eprint{1005.4814}.

\bibtype{Article}%
\bibitem[Gardim et al.(2020)]{Gardim:2019xjs}
\bibinfo{author}{Gardim FG}, \bibinfo{author}{Giacalone G},
  \bibinfo{author}{Luzum M} and  \bibinfo{author}{Ollitrault JY}
  (\bibinfo{year}{2020}).
\bibinfo{title}{{Thermodynamics of hot strong-interaction matter from
  ultrarelativistic nuclear collisions}}.
\bibinfo{journal}{{\em Nature Phys.}} \bibinfo{volume}{16}
  (\bibinfo{number}{6}): \bibinfo{pages}{615--619}.
  \bibinfo{doi}{\doi{10.1038/s41567-020-0846-4}}.
\eprint{1908.09728}.

\bibtype{Article}%
\bibitem[Gazdzicki and Gorenstein(1999)]{Gazdzicki:1998vd}
\bibinfo{author}{Gazdzicki M} and  \bibinfo{author}{Gorenstein MI}
  (\bibinfo{year}{1999}).
\bibinfo{title}{{On the early stage of nucleus-nucleus collisions}}.
\bibinfo{journal}{{\em Acta Phys. Polon. B}} \bibinfo{volume}{30}:
  \bibinfo{pages}{2705}.
\eprint{hep-ph/9803462}.

\bibtype{Article}%
\bibitem[Glozman(2023)]{Glozman:2022zpy}
\bibinfo{author}{Glozman LY} (\bibinfo{year}{2023}).
\bibinfo{title}{{Chiral spin symmetry and hot/dense QCD}}.
\bibinfo{journal}{{\em Prog. Part. Nucl. Phys.}} \bibinfo{volume}{131}:
  \bibinfo{pages}{104049}. \bibinfo{doi}{\doi{10.1016/j.ppnp.2023.104049}}.
\eprint{2209.10235}.

\bibtype{Article}%
\bibitem[Gorda et al.(2023)]{Gorda:2022jvk}
\bibinfo{author}{Gorda T}, \bibinfo{author}{Komoltsev O} and
  \bibinfo{author}{Kurkela A} (\bibinfo{year}{2023}).
\bibinfo{title}{{Ab-initio QCD Calculations Impact the Inference of the
  Neutron-star-matter Equation of State}}.
\bibinfo{journal}{{\em Astrophys. J.}} \bibinfo{volume}{950}
  (\bibinfo{number}{2}): \bibinfo{pages}{107}.
  \bibinfo{doi}{\doi{10.3847/1538-4357/acce3a}}.
\eprint{2204.11877}.

\bibtype{Article}%
\bibitem[Gorenstein et al.(2001)]{Gorenstein:2000ck}
\bibinfo{author}{Gorenstein MI}, \bibinfo{author}{Kostyuk A},
  \bibinfo{author}{Stoecker H} and  \bibinfo{author}{Greiner W}
  (\bibinfo{year}{2001}).
\bibinfo{title}{{Statistical coalescence model with exact charm conservation}}.
\bibinfo{journal}{{\em Phys. Lett. B}} \bibinfo{volume}{509}:
  \bibinfo{pages}{277--282}.
  \bibinfo{doi}{\doi{10.1016/S0370-2693(01)00516-0}}.
\eprint{hep-ph/0010148}.

\bibtype{Article}%
\bibitem[Greco et al.(2004)]{Greco:2003vf}
\bibinfo{author}{Greco V}, \bibinfo{author}{Ko C} and  \bibinfo{author}{Rapp R}
  (\bibinfo{year}{2004}).
\bibinfo{title}{{Quark coalescence for charmed mesons in ultrarelativistic
  heavy ion collisions}}.
\bibinfo{journal}{{\em Phys. Lett. B}} \bibinfo{volume}{595}:
  \bibinfo{pages}{202--208}.
  \bibinfo{doi}{\doi{10.1016/j.physletb.2004.06.064}}.
\eprint{nucl-th/0312100}.

\bibtype{Article}%
\bibitem[Hagedorn(1965)]{Hagedorn:1965st}
\bibinfo{author}{Hagedorn R} (\bibinfo{year}{1965}).
\bibinfo{title}{{Statistical thermodynamics of strong interactions at
  high-energies}}.
\bibinfo{journal}{{\em Nuovo Cim. Suppl.}} \bibinfo{volume}{3}:
  \bibinfo{pages}{147--186}.

\bibtype{Article}%
\bibitem[Hagedorn and Redlich(1985)]{Hagedorn:1984uy}
\bibinfo{author}{Hagedorn R} and  \bibinfo{author}{Redlich K}
  (\bibinfo{year}{1985}).
\bibinfo{title}{{Statistical Thermodynamics in Relativistic Particle and Ion
  Physics: Canonical or Grand Canonical?}}
\bibinfo{journal}{{\em Z. Phys. C}} \bibinfo{volume}{27}: \bibinfo{pages}{541}.
  \bibinfo{doi}{\doi{10.1007/BF01436508}}.

\bibtype{Article}%
\bibitem[Hamieh et al.(2000)]{Hamieh:2000tk}
\bibinfo{author}{Hamieh S}, \bibinfo{author}{Redlich K} and
  \bibinfo{author}{Tounsi A} (\bibinfo{year}{2000}).
\bibinfo{title}{{Canonical description of strangeness enhancement from p-A to
  Pb Pb collisions}}.
\bibinfo{journal}{{\em Phys. Lett.}} \bibinfo{volume}{B486}:
  \bibinfo{pages}{61--66}. \bibinfo{doi}{\doi{10.1016/S0370-2693(00)00762-0}}.
\eprint{hep-ph/0006024}.

\bibtype{Article}%
\bibitem[Harris and M{\"u}ller(2024)]{Harris:2023tti}
\bibinfo{author}{Harris JW} and  \bibinfo{author}{M{\"u}ller B}
  (\bibinfo{year}{2024}).
\bibinfo{title}{{''QGP Signatures'' Revisited}}.
\bibinfo{journal}{{\em Eur. Phys. J. C}} \bibinfo{volume}{84}
  (\bibinfo{number}{3}): \bibinfo{pages}{247}.
  \bibinfo{doi}{\doi{10.1140/epjc/s10052-024-12533-y}}.
\eprint{2308.05743}.

\bibtype{Article}%
\bibitem[He and Rapp(2019)]{He:2019tik}
\bibinfo{author}{He M} and  \bibinfo{author}{Rapp R} (\bibinfo{year}{2019}).
\bibinfo{title}{{Charm-Baryon Production in Proton-Proton Collisions}}.
\bibinfo{journal}{{\em Phys. Lett. B}} \bibinfo{volume}{795}:
  \bibinfo{pages}{117--121}.
  \bibinfo{doi}{\doi{10.1016/j.physletb.2019.06.004}}.
\eprint{1902.08889}.

\bibtype{Article}%
\bibitem[He et al.(2022)]{He:2021zej}
\bibinfo{author}{He M}, \bibinfo{author}{Wu B} and  \bibinfo{author}{Rapp R}
  (\bibinfo{year}{2022}).
\bibinfo{title}{{Collectivity of J/{\ensuremath{\psi}} Mesons in Heavy-Ion
  Collisions}}.
\bibinfo{journal}{{\em Phys. Rev. Lett.}} \bibinfo{volume}{128}
  (\bibinfo{number}{16}): \bibinfo{pages}{162301}.
  \bibinfo{doi}{\doi{10.1103/PhysRevLett.128.162301}}.
\eprint{2111.13528}.

\bibtype{Article}%
\bibitem[He et al.(2023)]{He:2022ywp}
\bibinfo{author}{He M}, \bibinfo{author}{van Hees H} and  \bibinfo{author}{Rapp
  R} (\bibinfo{year}{2023}).
\bibinfo{title}{{Heavy-quark diffusion in the quark{\textendash}gluon plasma}}.
\bibinfo{journal}{{\em Prog. Part. Nucl. Phys.}} \bibinfo{volume}{130}:
  \bibinfo{pages}{104020}. \bibinfo{doi}{\doi{10.1016/j.ppnp.2023.104020}}.
\eprint{2204.09299}.

\bibtype{Article}%
\bibitem[Itoh(1970)]{Itoh:1970uw}
\bibinfo{author}{Itoh N} (\bibinfo{year}{1970}).
\bibinfo{title}{{Hydrostatic Equilibrium of Hypothetical Quark Stars}}.
\bibinfo{journal}{{\em Prog. Theor. Phys.}} \bibinfo{volume}{44}:
  \bibinfo{pages}{291}. \bibinfo{doi}{\doi{10.1143/PTP.44.291}}.

\bibtype{Article}%
\bibitem[Kaiser and Weise(2026)]{Kaiser:2026msy}
\bibinfo{author}{Kaiser N} and  \bibinfo{author}{Weise W}
  (\bibinfo{year}{2026}), \bibinfo{month}{2}.
\bibinfo{title}{{Liquid-gas phase transition of nuclear matter}}
  \eprint{2602.09916}.

\bibtype{Article}%
\bibitem[Karsch(2022)]{Karsch:2022opd}
\bibinfo{author}{Karsch F} (\bibinfo{year}{2022}), \bibinfo{month}{12}.
\bibinfo{title}{{Lattice QCD at non-zero temperature and density}}
  \eprint{2212.03015}.

\bibtype{Article}%
\bibitem[Klasen and Paukkunen(2024)]{Klasen:2023uqj}
\bibinfo{author}{Klasen M} and  \bibinfo{author}{Paukkunen H}
  (\bibinfo{year}{2024}).
\bibinfo{title}{{Nuclear Parton Distribution Functions After the First Decade
  of LHC Data}}.
\bibinfo{journal}{{\em Ann. Rev. Nucl. Part. Sci.}} \bibinfo{volume}{74}
  (\bibinfo{number}{1}): \bibinfo{pages}{49--87}.
  \bibinfo{doi}{\doi{10.1146/annurev-nucl-102122-022747}}.
\eprint{2311.00450}.

\bibtype{Article}%
\bibitem[Klay et al.(2002)]{Klay:2001tf}
\bibinfo{author}{Klay J} and  et al. (\bibinfo{collaboration}{E895})
  (\bibinfo{year}{2002}).
\bibinfo{title}{{Longitudinal flow from 2-A-GeV to 8-A-GeV Au+Au collisions at
  the Brookhaven AGS}}.
\bibinfo{journal}{{\em Phys. Rev. Lett.}} \bibinfo{volume}{88}:
  \bibinfo{pages}{102301}. \bibinfo{doi}{\doi{10.1103/PhysRevLett.88.102301}}.
\eprint{nucl-ex/0111006}.

\bibtype{Article}%
\bibitem[Klay et al.(2003)]{Klay:2003zf}
\bibinfo{author}{Klay J} and  et al. (\bibinfo{collaboration}{E895})
  (\bibinfo{year}{2003}).
\bibinfo{title}{{Charged pion production in 2 to 8 AGev central au+au
  collisions}}.
\bibinfo{journal}{{\em Phys. Rev. C}} \bibinfo{volume}{68}:
  \bibinfo{pages}{054905}. \bibinfo{doi}{\doi{10.1103/PhysRevC.68.054905}}.
\eprint{nucl-ex/0306033}.

\bibtype{Article}%
\bibitem[Koch(1997)]{Koch:1997ei}
\bibinfo{author}{Koch V} (\bibinfo{year}{1997}).
\bibinfo{title}{{Aspects of chiral symmetry}}.
\bibinfo{journal}{{\em Int. J. Mod. Phys. E}} \bibinfo{volume}{6}:
  \bibinfo{pages}{203--250}. \bibinfo{doi}{\doi{10.1142/S0218301397000147}}.
\eprint{nucl-th/9706075}.

\bibtype{Inproceedings}%
\bibitem[Maiani and Pilloni(2022)]{Maiani:2022psl}
\bibinfo{author}{Maiani L} and  \bibinfo{author}{Pilloni A}
  (\bibinfo{year}{2022}), \bibinfo{month}{7}, \bibinfo{title}{{GGI Lectures on
  Exotic Hadrons}}, \eprint{2207.05141}.

\bibtype{Article}%
\bibitem[Matsui and Satz(1986)]{Matsui:1986dk}
\bibinfo{author}{Matsui T} and  \bibinfo{author}{Satz H}
  (\bibinfo{year}{1986}).
\bibinfo{title}{{J/$\psi$ suppression by quark-gluon plasma formation}}.
\bibinfo{journal}{{\em Phys. Lett. B}} \bibinfo{volume}{178}:
  \bibinfo{pages}{416}. \bibinfo{doi}{\doi{10.1016/0370-2693(86)91404-8}}.

\bibtype{Article}%
\bibitem[McLerran(2026)]{McLerran:2026dio}
\bibinfo{author}{McLerran L} (\bibinfo{year}{2026}).
\bibinfo{title}{{Two Lectures on the Phase Diagram of QCD}}.
\bibinfo{journal}{{\em Acta Phys. Polon. B}} \bibinfo{volume}{57}
  (\bibinfo{number}{4}): \bibinfo{pages}{4--A2}.
  \bibinfo{doi}{\doi{10.5506/APhysPolB.57.4-A2}}.
\eprint{2604.03849}.

\bibtype{Article}%
\bibitem[Minissale et al.(2021)]{Minissale:2020bif}
\bibinfo{author}{Minissale V}, \bibinfo{author}{Plumari S} and
  \bibinfo{author}{Greco V} (\bibinfo{year}{2021}).
\bibinfo{title}{{Charm hadrons in pp collisions at LHC energy within a
  coalescence plus fragmentation approach}}.
\bibinfo{journal}{{\em Phys. Lett. B}} \bibinfo{volume}{821}:
  \bibinfo{pages}{136622}. \bibinfo{doi}{\doi{10.1016/j.physletb.2021.136622}}.
\eprint{2012.12001}.

\bibtype{Article}%
\bibitem[Navas et al.(2024)]{ParticleDataGroup:2024cfk}
\bibinfo{author}{Navas S} and  et al. (\bibinfo{collaboration}{Particle Data
  Group}) (\bibinfo{year}{2024}).
\bibinfo{title}{{Review of particle physics}}.
\bibinfo{journal}{{\em Phys. Rev. D}} \bibinfo{volume}{110}
  (\bibinfo{number}{3}): \bibinfo{pages}{030001}.
  \bibinfo{doi}{\doi{10.1103/PhysRevD.110.030001}}.

\bibtype{Article}%
\bibitem[Pinkenburg et al.(2002)]{Pinkenburg:2001fj}
\bibinfo{author}{Pinkenburg C} and  et al. (\bibinfo{collaboration}{E895})
  (\bibinfo{year}{2002}).
\bibinfo{title}{{Production and collective behavior of strange particles in Au
  + Au collisions at 2-AGeV - 8-AGeV}}.
\bibinfo{journal}{{\em Nucl. Phys. A}} \bibinfo{volume}{698}:
  \bibinfo{pages}{495--498}.
  \bibinfo{doi}{\doi{10.1016/S0375-9474(01)01412-9}}.
\eprint{nucl-ex/0104025}.

\bibtype{Article}%
\bibitem[Rafelski et al.(2023)]{Rafelski:2023emw}
\bibinfo{author}{Rafelski J}, \bibinfo{author}{Birrell J},
  \bibinfo{author}{Steinmetz A} and  \bibinfo{author}{Yang CT}
  (\bibinfo{year}{2023}).
\bibinfo{title}{{A Short Survey of Matter-Antimatter Evolution in the
  Primordial Universe}}.
\bibinfo{journal}{{\em Universe}} \bibinfo{volume}{9} (\bibinfo{number}{7}):
  \bibinfo{pages}{309}. \bibinfo{doi}{\doi{10.3390/universe9070309}}.
\eprint{2305.09055}.

\bibtype{Article}%
\bibitem[Rapp and Wambach(2000)]{Rapp:1999ej}
\bibinfo{author}{Rapp R} and  \bibinfo{author}{Wambach J}
  (\bibinfo{year}{2000}).
\bibinfo{title}{{Chiral symmetry restoration and dileptons in relativistic
  heavy ion collisions}}.
\bibinfo{journal}{{\em Adv. Nucl. Phys.}} \bibinfo{volume}{25}:
  \bibinfo{pages}{1}. \bibinfo{doi}{\doi{10.1007/0-306-47101-9_1}}.
\eprint{hep-ph/9909229}.

\bibtype{Article}%
\bibitem[Reisdorf et al.(2010)]{FOPI:2010xrt}
\bibinfo{author}{Reisdorf W} and  et al. (\bibinfo{collaboration}{FOPI})
  (\bibinfo{year}{2010}).
\bibinfo{title}{{Systematics of central heavy ion collisions in the 1A GeV
  regime}}.
\bibinfo{journal}{{\em Nucl. Phys. A}} \bibinfo{volume}{848}:
  \bibinfo{pages}{366--427}.
  \bibinfo{doi}{\doi{10.1016/j.nuclphysa.2010.09.008}}.
\eprint{1005.3418}.

\bibtype{Article}%
\bibitem[Rothkopf(2020)]{Rothkopf:2019ipj}
\bibinfo{author}{Rothkopf A} (\bibinfo{year}{2020}).
\bibinfo{title}{{Heavy Quarkonium in Extreme Conditions}}.
\bibinfo{journal}{{\em Phys. Rept.}} \bibinfo{volume}{858}:
  \bibinfo{pages}{1--117}. \bibinfo{doi}{\doi{10.1016/j.physrep.2020.02.006}}.
\eprint{1912.02253}.

\bibtype{Article}%
\bibitem[Salabura and Stroth(2021)]{Salabura:2020tou}
\bibinfo{author}{Salabura P} and  \bibinfo{author}{Stroth J}
  (\bibinfo{year}{2021}).
\bibinfo{title}{{Dilepton radiation from strongly interacting systems}}.
\bibinfo{journal}{{\em Prog. Part. Nucl. Phys.}} \bibinfo{volume}{120}:
  \bibinfo{pages}{103869}. \bibinfo{doi}{\doi{10.1016/j.ppnp.2021.103869}}.
\eprint{2005.14589}.

\bibtype{Article}%
\bibitem[Schenke et al.(2020)]{Schenke:2020mbo}
\bibinfo{author}{Schenke B}, \bibinfo{author}{Shen C} and
  \bibinfo{author}{Tribedy P} (\bibinfo{year}{2020}).
\bibinfo{title}{{Running the gamut of high energy nuclear collisions}}.
\bibinfo{journal}{{\em Phys. Rev. C}} \bibinfo{volume}{102}
  (\bibinfo{number}{4}): \bibinfo{pages}{044905}.
  \bibinfo{doi}{\doi{10.1103/PhysRevC.102.044905}}.
\eprint{2005.14682}.

\bibtype{Article}%
\bibitem[Shuryak(1978)]{Shuryak:1978ij}
\bibinfo{author}{Shuryak EV} (\bibinfo{year}{1978}).
\bibinfo{title}{{Quark-gluon plasma and hadronic production of leptons, photons
  and psions}}.
\bibinfo{journal}{{\em Phys. Lett. B}} \bibinfo{volume}{78}:
  \bibinfo{pages}{150}. \bibinfo{doi}{\doi{10.1016/0370-2693(78)90370-2}}.

\bibtype{Article}%
\bibitem[Song and Coci(2022)]{Song:2021mvc}
\bibinfo{author}{Song T} and  \bibinfo{author}{Coci G} (\bibinfo{year}{2022}).
\bibinfo{title}{{Prerequisites for heavy quark coalescence in heavy-ion
  collisions}}.
\bibinfo{journal}{{\em Nucl. Phys. A}} \bibinfo{volume}{1028}:
  \bibinfo{pages}{122539}.
  \bibinfo{doi}{\doi{10.1016/j.nuclphysa.2022.122539}}.
\eprint{2104.10987}.

\bibtype{Article}%
\bibitem[Sorensen et al.(2024)]{Sorensen:2023zkk}
\bibinfo{author}{Sorensen A} and  et al. (\bibinfo{year}{2024}).
\bibinfo{title}{{Dense nuclear matter equation of state from heavy-ion
  collisions}}.
\bibinfo{journal}{{\em Prog. Part. Nucl. Phys.}} \bibinfo{volume}{134}:
  \bibinfo{pages}{104080}. \bibinfo{doi}{\doi{10.1016/j.ppnp.2023.104080}}.
\eprint{2301.13253}.

\bibtype{Article}%
\bibitem[STA(2026)]{STAR:2026khp}
 (\bibinfo{year}{2026}), \bibinfo{month}{4}.
\bibinfo{title}{{Collision Energy Dependence of Hypertriton Production in Au+Au
  Collisions at RHIC}} \eprint{2604.18214}.

\bibtype{Article}%
\bibitem[Stock(1999)]{Stock:1999hm}
\bibinfo{author}{Stock R} (\bibinfo{year}{1999}).
\bibinfo{title}{{The parton to hadron phase transition observed in Pb+Pb
  collisions at 158-GeV per nucleon}}.
\bibinfo{journal}{{\em Phys. Lett. B}} \bibinfo{volume}{456}:
  \bibinfo{pages}{277--282}.
  \bibinfo{doi}{\doi{10.1016/S0370-2693(99)00482-7}}.
\eprint{hep-ph/9905247}.

\bibtype{Article}%
\bibitem[Vovchenko et al.(2016)]{Vovchenko:2015idt}
\bibinfo{author}{Vovchenko V}, \bibinfo{author}{Begun VV} and
  \bibinfo{author}{Gorenstein MI} (\bibinfo{year}{2016}).
\bibinfo{title}{{Hadron multiplicities and chemical freeze-out conditions in
  proton-proton and nucleus-nucleus collisions}}.
\bibinfo{journal}{{\em Phys. Rev. C}} \bibinfo{volume}{93}
  (\bibinfo{number}{6}): \bibinfo{pages}{064906}.
  \bibinfo{doi}{\doi{10.1103/PhysRevC.93.064906}}.
\eprint{1512.08025}.

\bibtype{Inproceedings}%
\bibitem[Wang and Wiedemann(2025)]{Wang:2025lct}
\bibinfo{author}{Wang XN} and  \bibinfo{author}{Wiedemann UA}
  (\bibinfo{year}{2025}), \bibinfo{month}{8}, \bibinfo{title}{{QGP@50: More
  than Four Decades of Jet Quenching}}, \eprint{2508.18794}.

\bibtype{Article}%
\bibitem[Weise(2012)]{Weise:2012yv}
\bibinfo{author}{Weise W} (\bibinfo{year}{2012}).
\bibinfo{title}{{Nuclear chiral dynamics and phases of QCD}}.
\bibinfo{journal}{{\em Prog. Part. Nucl. Phys.}} \bibinfo{volume}{67}:
  \bibinfo{pages}{299--311}. \bibinfo{doi}{\doi{10.1016/j.ppnp.2011.12.034}}.
\eprint{1201.0950}.

\bibtype{Article}%
\bibitem[Wu and Rapp(2026)]{Wu:2025lcj}
\bibinfo{author}{Wu B} and  \bibinfo{author}{Rapp R} (\bibinfo{year}{2026}).
\bibinfo{title}{{Bottomonium transport in a strongly coupled quark-gluon
  plasma}}.
\bibinfo{journal}{{\em Phys. Lett. B}} \bibinfo{volume}{873}:
  \bibinfo{pages}{140223}. \bibinfo{doi}{\doi{10.1016/j.physletb.2026.140223}}.
\eprint{2508.20995}.

\bibtype{Article}%
\bibitem[Yasui et al.(2026)]{Yasui:2026vve}
\bibinfo{author}{Yasui S}, \bibinfo{author}{Lee SH}, \bibinfo{author}{Lo PM}
  and  \bibinfo{author}{Sasaki C} (\bibinfo{year}{2026}), \bibinfo{month}{3}.
\bibinfo{title}{{New nonet scalar mesons and glueballs: the mass spectra and
  the production yields in relativistic heavy ion collisions}}
  \eprint{2603.13764}.

\bibtype{Article}%
\bibitem[Zhou et al.(2014)]{Zhou:2014kka}
\bibinfo{author}{Zhou K}, \bibinfo{author}{Xu N}, \bibinfo{author}{Xu Z} and
  \bibinfo{author}{Zhuang P} (\bibinfo{year}{2014}).
\bibinfo{title}{{Medium effects on charmonium production at ultrarelativistic
  energies available at the CERN Large Hadron Collider}}.
\bibinfo{journal}{{\em Phys. Rev. C}} \bibinfo{volume}{89}:
  \bibinfo{pages}{054911}. \bibinfo{doi}{\doi{10.1103/PhysRevC.89.054911}}.
\eprint{1401.5845}.

\end{thebibliography*}

\end{document}